\documentclass[a4paper,11pt]{article}
\pdfoutput=1 

\usepackage{jheppub} 
\usepackage{graphicx}
\usepackage{amsmath}
\usepackage{bm}
\usepackage{yhmath}
\usepackage{mathtools}
\usepackage{wasysym}
\usepackage{subfigure}
\usepackage{color}
\usepackage{cases}
\usepackage{times}
\usepackage{dcolumn,booktabs,bm}
\usepackage{slashed}
\usepackage{amsfonts,amssymb,stmaryrd,latexsym,amsmath}
\usepackage{textcomp}
\usepackage{multirow}
\usepackage{cancel}
\usepackage{array}
\usepackage{diagbox}

\title{\boldmath Two-particle scattering from finite-volume quantization conditions using the plane wave basis}

\author[a]{Lu Meng,}
\author[a]{E.~Epelbaum}

\affiliation[a]{Institut f\"ur Theoretische Physik II, Ruhr-Universit\"at Bochum,  D-44780 Bochum,
	Germany}

\emailAdd{lu.meng@rub.de}
\emailAdd{evgeny.epelbaum@rub.de}

\abstract{We propose an alternative approach to L\"uscher's formula for extracting two-body scattering phase shifts from finite volume spectra with no reliance on the partial wave expansion. We use an effective-field-theory-based Hamiltonian method in the plane wave basis and decompose the corresponding matrix elements of operators into irreducible representations of the relevant point groups. The proposed approach allows one to benefit from the knowledge of the long-range interaction and avoids complications from partial wave mixing in a finite volume. We consider spin-singlet channels in the two-nucleon system and pion-pion scattering in the $\rho$-meson channel in the rest and moving frames to illustrate the method for non-relativistic and relativistic systems, respectively. For the two-nucleon system, the long-range interaction due to the one-pion exchange is found to make the single-channel L\"uscher formula unreliable at the physical pion mass. For S-wave dominated states, the single-channel L\"uscher method suffers from significant finite-volume artifacts for a $L=3$~fm box, but it works well for boxes with $L>5$~fm. However, for P-wave dominated states, significant partial wave mixing effects prevent the application of the single-channel L\"uscher formula regardless of the box size (except for the near-threshold region). Using a toy model to generate synthetic
 data for finite-volume energies, we show that our effective-field-theory-based approach in the plane wave basis is capable of a reliable extraction of the phase shifts. For pion-pion scattering, we employ a phenomenological model to fit lattice QCD results at the physical pion mass. The extracted P-wave  phase shifts are found to be in a good agreement with the experimental results.
}

\begin{document} 
\maketitle
\flushbottom

\section{Introduction}
\label{sec:intro}

Three decades ago, L\"uscher proposed a model-independent approach for extracting two-particle elastic scattering phase shifts from energy levels in a finite box with periodic boundary conditions~\cite{Luscher:1985dn,Luscher:1986pf,Luscher:1990ux}.  This seminal work opened the way for investigating hadronic scattering observables using lattice QCD simulations. The L\"uscher formula has been generalized to moving systems in finite boxes~\cite{Rummukainen:1995vs,Kim:2005gf,Gockeler:2012yj}, higher partial waves~\cite{Luu:2011ep}, particles with different masses~\cite{Fu:2011xz,Leskovec:2012gb}, particles with nonzero spins~\cite{Briceno:2014oea}, multichannel scattering~\cite{Lage:2009zv,He:2005ey,Hansen:2012tf,Doring:2012eu} and to cases with asymmetric boxes~\cite{Feng:2004ua}, see also \cite{Briceno:2017max,Mai:2021lwb} for review articles.

The L\"uscher method assumes that the box size $L$ is much larger than the interaction range $R$, such that exponentially suppressed corrections $\sim e^{-L/R}$ are negligible. In currently feasible lattice simulations, the box size can usually not be chosen very large given the available computational resources.  On the other hand, in certain cases such as e.g.~the two-nucleon system, the long-range interaction is known and appears to play a very prominent role. Another example of a system featuring long-range dynamics is the $\bar{D}^*D / \bar{D}D^*$ system, in which the one-pion exchange (OPE) potential plays an important role for the formation of exotic hadrons $X(3872)$ and $Z_c(3900)$~\cite{Chen:2016qju,Guo:2017jvc,Brambilla:2019esw}. The accidental mass relation $m_{\bar{D}^*}\approx m_{\bar{D}}+M_\pi$ makes the exchanged pion in the OPE potential almost on-shell, which results in the interaction range much larger than $1/M_\pi$. In such situations, taking into account the long-range dynamics may allow one to reduce the cost of lattice QCD simulations by considering smaller volumes. It, however, requires the finite-volume L\"uscher quantization conditions to be generalized by taking into account exponentially suppressed effects due to the long-range interaction. 

In single-channel cases, each energy level in a finite volume (FV) can be related to the scattering phase shift at this energy using interaction-independent L\"uscher's formula. On the other hand, one has to deal with coupled-channel effects when taking into account inelastic channels or partial wave mixing. In the multichannel case, L\"uscher's formula becomes a determinant equation $\det [f(T,E)]=0$, which  relates the $T$-matrix and the FV energy levels. Though this result is still model-independent, there is no one-to-one correspondences between phase shifts and FV energy levels. It does not allow one to extract the full $T$-matrix, i.e.~phase shifts for every channel and the corresponding mixing angles.
In practice, the $T$-matrix can be parameterized within a particular theoretical framework, and several FV energies can be used as input to determine the parameters in the $T$-matrix. This may, however, introduce some model dependence.  
Another complication is related to rotational symmetry breaking in finite boxes, which results in the energy levels generally receiving contributions from multiple partial waves. 

In the literature, much effort has been devoted to cure or estimate the two drawbacks of the L\"uscher formula, namely exponentially suppressed corrections and mixing effects in multichannel scattering. In refs.~\cite{Bedaque:2006yi,Sato:2007ms}, the authors estimated exponentially suppressed corrections for $\pi\pi$ and NN scatterings in the FV. It was shown that for realistic pion masses, simulations in a box of the size $L>5$~fm would result in exponential corrections to the phase shifts less than one degree. In ref.~\cite{Jansen:2015lha}, finite volume corrections to the $X(3872)$ were analyzed in an effective field theory (EFT) with perturbative pions. It was shown that FV effects are significant for box lengths as large as $20$~fm.
Unitarized chiral perturbation theory in a finite volume was employed in refs.~\cite{Doring:2011vk,Doring:2012eu,Guo:2016zep,Guo:2018tjx}, in which some exponentially suppressed effects were included by adopting a fully relativistic propagator. Such exponentially suppressed differences with the L\"uscher formula for $\pi\pi$ scattering were calculated explicitly in refs.~\cite{Albaladejo:2012jr,Chen:2012rp}. In order to deal with coupled-channel effects in a FV, an effective Hamiltonian approach was developed in refs.~\cite{Hall:2013qba,Wu:2014vma,Liu:2015ktc,Li:2021mob}. Within different approaches, partial wave mixing effects were estimated in refs.~\cite{Luu:2011ep,Doring:2012eu,Morningstar:2017spu,Li:2019qvh}.  The sensitivity to the second lowest partial
wave in L\"uscher's quantization conditions was discussed in ref.~\cite{Lee:2021kfn}.

Starting from L\"uscher's seminal work, FV effects were usually investigated using the partial wave expansion technique. In the infinite volume, partial wave expansion allows one to simplify the scattering problem thanks to the rotational symmetry. 
However, in the finite volume, continuous rotational symmetry is broken and the angular momentum is no longer a good quantum number. To some extent, introducing the partial wave expansion in a finite volume complicates the problem. For small boxes and/or systems with coupled channels as well as in the presence of long-range interactions, the above mentioned drawbacks of the L\"uscher approach get amplified.
A natural alternative method is to employ vectors of discrete momenta. 
In refs.~\cite{Luu:2011ep,Li:2019qvh}, the authors used the discrete momentum basis to investigate partial wave mixing effects. Recently, a Fourier basis $\{\cos(2\pi n_i r_i/L),\sin(2\pi n_i r_i/L)\}$ in coordinate space was adopted to diagonalize the Hamiltonian in FV~\cite{Lee:2020fbo}. 

In this work, we employ a similar approach as that of ref.~\cite{Lee:2020fbo} but in momentum space. We consider the Lippmann-Schwinger equation (LSE) in the non-relativistic case or a reduced Bethe-Salpeter equation (BSE) in the relativistic case using the plane wave basis with discrete momenta. Next, we reduce the matrix LSE and BSE into the irreducible representations (irreps) of the cubic group. With the reduced matrix equation, we compute the energy levels in the FV corresponding to specific irreps. In such a framework, effects from  mixing with higher partial waves are embedded naturally. We also extend this approach to systems with nonvanishing total momentum. We consider NN and $\pi\pi$ scattering as examples of the non-relativistic and relativistic systems, respectively,  to illustrate the calculations and compare our results with those obtained using the single-channel L\"uscher approach. In the case of the NN system, we also study the role played by the long-range interaction due to the one-pion exchange.

This work is organized as follows. In section~\ref{sec:p-dis}, we review the discretization conditions of three-momenta and discuss the relevant symmetry groups for two particles in a cubic box with periodic boundary conditions. In section~\ref{sec:irrep}, we construct the representation spaces of the corresponding point groups with the plane wave basis and perform their reduction into irreps using the projection operator technique.  In section~\ref{sec:FVft-slv}, we introduce a general approach to obtain the FV energy levels and to fit  lattice QCD energy spectra in the FV. Next, in section~\ref{sec:NN}, we consider spin-singlet NN scattering as an explicit example of a non-relativistic system to demonstrate our method. In this section, we also discuss in detail partial wave mixing effects in the FV. In section~\ref{sec:pipiSystem}, we consider $\pi\pi$ scattering in the $\rho$-meson channel as an example of a relativistic system. Using a phenomenological model to parameterize the $\pi\pi$ scattering amplitude, we calculate the FV energy levels and fit the parameters of the model to the lattice QCD spectra. Finally, in section~\ref{sec:sum}, we summarize the main results of our study and discuss possible generalizations. Some basic information about the point groups used in this work and L\"uscher's quantization conditions is given in appendices \ref{app:pgrp} and \ref{app:lush}.

\section{Two particles in a finite volume}
\label{sec:p-dis}
Discretization of three-momenta of two particles in a cubic box with periodic boundary conditions is considered e.g.~in refs.~\cite{Rummukainen:1995vs,Leskovec:2012gb}. In the following, we briefly review the relevant findings for the sake of completeness.
Throughout this paper, we focus on systems of  two spinless particles of the same mass. It is straightforward to generalize our results to particles with arbitrary spin and different masses. 

\subsection{Non-interacting case}

In the box frame (BF), the momenta $\bm{p}_1$ and $\bm{p}_2$ of two non-interacting particles are discrete,
\begin{eqnarray}
&E=\sqrt{\bm{p}_{1}^{2}+m_{1}^{2}}+\sqrt{\bm{p}_{2}^{2}+m_{2}^{2}},\\
&\bm{p}_{1}+\bm{p}_{2}=\bm{P},\quad\bm{p}_{1}=\frac{2\pi}{L}\bm{n},\quad\bm{P}=\frac{2\pi}{L}\bm{d},\qquad\bm{n},\bm{d}\in {Z}^{3},\label{eq:dis_bf}
\end{eqnarray}
where $\bm{P}$ and $E$ are the total momentum and energy of the two-particle system, respectively. For the non-interacting case, each particle is on-shell with $E_{i}=\sqrt{\bm{p}_{i}^{2}+m_{i}^{2}}$. In the center-of-mass frame (CMF), the related momenta and energies read
\begin{equation}
\bm{P}^{*}=0,\quad\bm{p}^{*}\equiv\frac{1}{2}(\bm{p}_{1}^{*}-\bm{p}_{2}^{*})=\bm{p}_{1}^{*}=-\bm{p}_{2}^{*},\quad E^{*2}=E^{2}-\bm{P}^{2},~\label{eq:cm}
\end{equation}
where the quantities in the CMF are labeled by superscript ``$*$''. In eq.~\eqref{eq:cm}, the on-shell relation of each particle is not used. Thus, eq.~\eqref{eq:cm} is also valid for interacting systems. For non-interacting systems, the momentum and energy in the CMF are related to those in the BF by the Lorentz transformation
\begin{eqnarray}
\bm{p}_{1}^{*}&=&\left[(\gamma-1)\frac{\bm{P}\cdot\bm{p}_{1}}{\bm{P}^{2}}-\frac{E_{1}}{E^{*}}\right]\bm{P}+\bm{p}_{1},\label{eq:lrp}
\\E_{1}^{*}&=&\frac{EE_{1}-\bm{P}\cdot\bm{p}_{1}}{E^{*}}, \label{eq:lre}
\end{eqnarray}
where $\gamma=\frac{E}{E^{*}}$  is the Lorentz factor. Equations \eqref{eq:lrp} and \eqref{eq:lre} can be combined and simplified to
\begin{equation}
	\bm{p}_{1}^{*}=\gamma^{-1}\left(\bm{p}_{1\parallel}-\frac{1}{2}A\bm{P}\right)+\bm{p}_{1\perp},\label{eq:lrcomb}
\end{equation}
where
\begin{eqnarray}
	A\equiv1+\frac{m_{1}^{2}-m_{2}^{2}}{E^{*2}}. 
\end{eqnarray} 
For a given vector $\bm{u}$, $\bm{u}_{\parallel}$ and $\bm{u}_{\perp}$ are defined via $\bm{u}_{\parallel}=\frac{(\bm{u}\cdot\bm{P})\bm{P}}{\bm{P}^{2}}$ and $\bm{u}_{\perp}=\bm{u}-\bm{u}_{\parallel}$. 

In this paper we are interested in systems of two particles with equal masses (and thus $A=1$).   With eqs.~\eqref{eq:dis_bf} and \eqref{eq:lrcomb}, one obtains the discretization relations in the CMF
\begin{eqnarray}
&\bm{p}^{*}=\bm{n}^{*}\frac{2\pi}{L},\quad\bm{n}^{*}\in P_{d}, \nonumber \\
&P_{d}=\{\gamma^{-1}(\bm{n}_{\parallel}-\frac{1}{2}\bm{d})+\bm{n}_{\perp}\}, \quad \bm{n}\in {Z}^{3}. \label{eq:dis_cmf}    
\end{eqnarray}

\begin{figure}[tbp]
	\centering 
	\includegraphics[width=.80\textwidth]{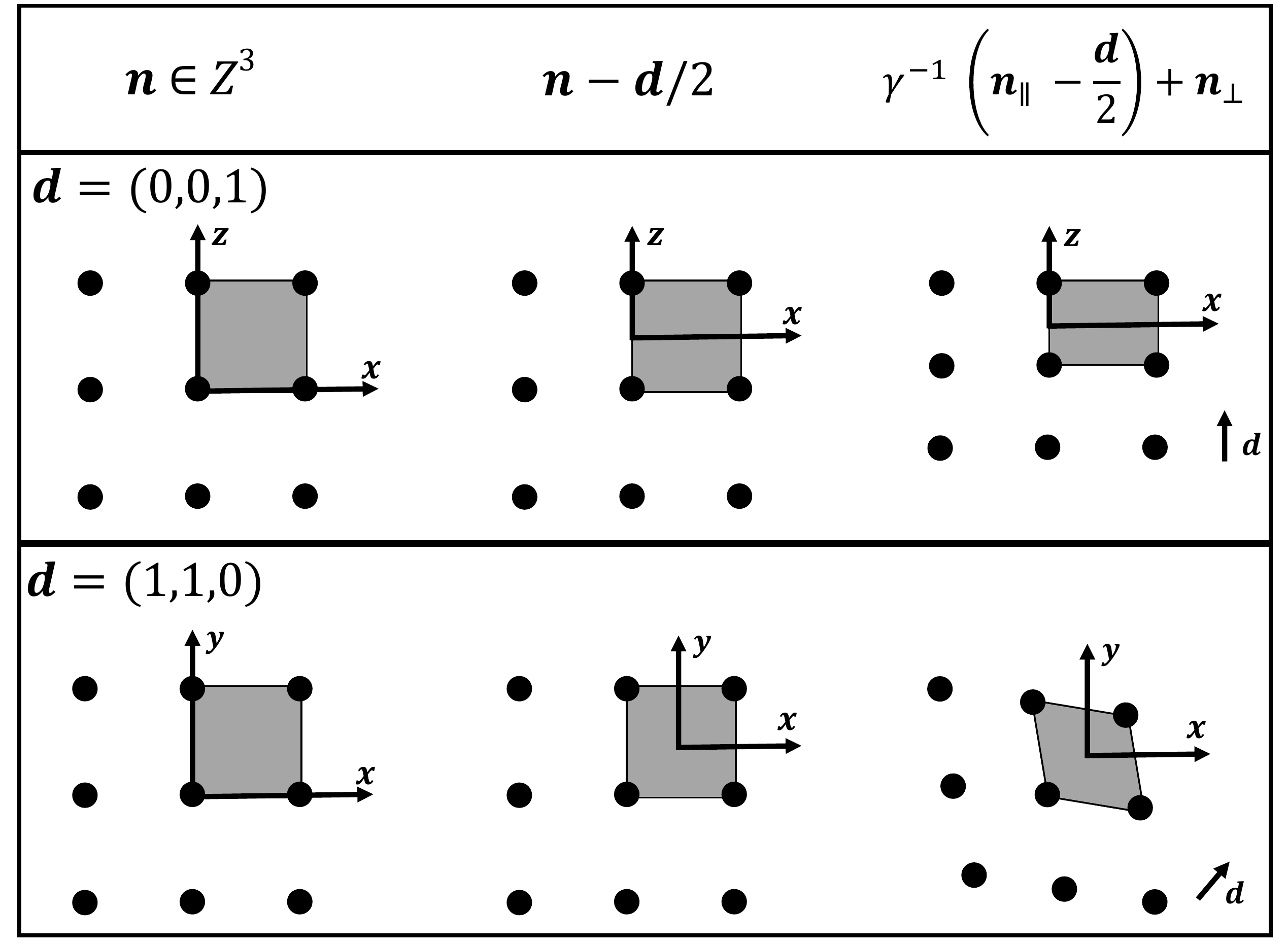}
	\caption{\label{fig:mesh} Meshes of the set of discrete momenta for a two-particle system with equal masses in the BF (left column) and CMF (right column).
          The second and third meshes in each row  illustrate 
the shift of the origin by $\bm{d}/2$ and the deformation in the direction of $\bm{d}$ to obtain the mesh in the CMF.  }
\end{figure}

In figure~\ref{fig:mesh}, we show a set of discrete momenta with the corresponding meshes in different frames. The first mesh in each row corresponds to the momenta in the BF. The second and third ones illustrate the steps needed to obtain the momenta in the CMF according to eq.~\eqref{eq:dis_cmf}, namely the shift of the origin by $\bm{d}/2$ and the deformation in the direction of $\bm{d}$. In figure~\ref{fig:mesh}, we take $\bm{d}=(0,0,1)$ and $\bm{d}=(1,1,0)$ as examples. The meshes show the symmetries of the system in the FV.  In the infinite volume, the momenta are continuous and satisfy the rotational symmetry described by the $SO(3)$ group. Considering the conservation of space inversion, the symmetric group becomes $O(3)=SO(3)\otimes \{E,I\}$, where $E$ and $I$ are the identity and the space inversion elements, respectively. In the BF corresponding to $\bm{d}=(0,0,0)$, the symmetry group is the cubic $O_h=O\otimes\{E,I\}$ one, which is a subgroup of the $O(3)$ group. When the two-particle system is moving in the box, the $O_{h}$ symmetry is broken by shifting the origin and a deformation in the direction of $\bm{d}$. The symmetric point groups corresponding to different values of $\bm{d}$ are presented in table~\ref{tab:grp}. In this work we are only interested in the $d^2=0,1,2,3,4$ cases and, therefore, consider only the $O_h$, $D_{4h}$, $D_{2h}$ and $D_{3d}$ groups and the corresponding subgroups $O$, $D_4$, $D_2$ and $D_3$ containing only proper rotations. It should be noticed that the symmetric groups considered here are only applicable to the case of two particles with equal masses. For particles with different masses ($A\neq 1$), the origin of the coordinate system  in figure~\ref{fig:mesh} is not in the center of its adjacent points anymore after shifting by ${1\over 2}A\bm{d}$, and the system is not invariant with respect to space inversions~\cite{Leskovec:2012gb}. In appendix~\ref{app:pgrp}, we list the group elements, conjugacy classes, and character tables for the relevant point groups.

\begin{table}[tbp]
	\centering
	\begin{tabular}{|c|c|c|c|}
		\hline 
		$\bm{d}$ & point group & $n_{G}$ & representation space\tabularnewline
		\hline 
		$(0,0,0)$ & $O_{h}$ & 48 & $\{n_{1},n_{2},n_{3}\}$\tabularnewline
		\hline 
		$(0,0,a)$ & $D_{4h}$ & 16 & $\{n_{1},n_{2};n_{3}\}$\tabularnewline
		\hline 
		$(a,a,0)$ & $D_{2h}$ & 8 & $\{n_{1},n_{2};n_{3}\}$\tabularnewline
		\hline 
		$(a,a,a)$ & $D_{3d}$ & 12 & $\{n_{1},n_{2},n_{3}\}$\tabularnewline
		\hline 
	\end{tabular}
	\caption{\label{tab:grp}The point groups, their orders $n_G$, and representation spaces of the two-particle systems in finite volume as defined in eqs.~(\ref{eq:space30}) and (\ref{eq:space21}).}
\end{table}

In the literature, two conventions to denote irreps of $D_{4h}$, $D_{2h}$ and $D_{3d}$ are used.
For example, the $D_{4h}$ group can be formed either as $D_{4h}=D_4\otimes \{E,I\}$ or as $D_{4h}=C_{4v}\otimes \{E,I\}$. The elements of $D_4$ are all proper rotations. The $C_{4v}$ group contains two improper rotations
and is
the symmetry group for  systems with particles of unequal masses. In the first convention, the irreps of $D_{4h}$ are denoted by the irreps of the $D_4$ group with parity~\cite{Rummukainen:1995vs,Briceno:2013lba}. In the second convention, the irreps of $D_{4h}$ are denoted by the irreps of the $C_{4v}$ group with parity~\cite{Gockeler:2012yj,Morningstar:2013bda}. We list the irreps of the two conventions  and their corresponding relations in table~\ref{tab:grp_twocv}. In this work, we adopt the first convention. 

\begin{table}[tbp]
	\centering
	\begin{tabular}{|c|c|c|ccccc|ccccc|}
		\hline 
		Point group & \multicolumn{2}{c|}{Two conventions} & \multicolumn{5}{c|}{irreps with even parity} & \multicolumn{5}{c|}{irreps with odd parity}\tabularnewline
		\hline 
		$O_{h}$ &  & $O\otimes\{I,E\}$ & $A_{1}^{+}$ & $A_{2}^{+}$ & $E^{+}$ & $T_{1}^{+}$ & $T_{2}^{+}$ & $A_{1}^{-}$ & $A_{2}^{-}$ & $E^{-}$ & $T_{1}^{-}$ & $T_{2}^{-}$\tabularnewline
		\hline 
		\multirow{2}{*}{$D_{4h}$} & I & $D_{4}\otimes\{I,E\}$ & $A_{1}^{+}$ & $A_{2}^{+}$ & $B_{1}^{+}$ & $B_{2}^{+}$ & $E^{+}$ & $A_{1}^{-}$ & $A_{2}^{-}$ & $B_{1}^{-}$ & $B_{2}^{-}$ & $E^{-}$\tabularnewline
		& II & $C_{4v}\otimes\{I,E\}$ & $A_{1}^{+}$ & $A_{2}^{+}$ & $B_{1}^{+}$ & $B_{2}^{+}$ & $E^{+}$ & $A_{2}^{-}$ & $A_{1}^{-}$ & $B_{2}^{-}$ & $B_{1}^{-}$ & $E^{-}$\tabularnewline
		\hline 
		\multirow{2}{*}{$D_{2h}$} & I & $D_{2}\otimes\{I,E\}$ & $A_{1}^{+}$ & $B_{1}^{+}$ & $B_{2}^{+}$ & $B_{3}^{+}$ &  & $A_{1}^{-}$ & $B_{1}^{-}$ & $B_{2}^{-}$ & $B_{3}^{-}$ & \tabularnewline
		& II & $C_{2v}\otimes\{I,E\}$ & $A_{1}^{+}$ & $A_{2}^{+}$ & $B_{1}^{+}$ & $B_{2}^{+}$ &  & $A_{2}^{-}$ & $A_{1}^{-}$ & $B_{2}^{-}$ & $B_{1}^{-}$ & \tabularnewline
		\hline 
		\multirow{2}{*}{$D_{3d}$} & I & $D_{3}\otimes\{I,E\}$ & $A_{1}^{+}$ & $A_{2}^{+}$ & $E^{+}$ &  &  & $A_{1}^{-}$ & $A_{2}^{-}$ & $E^{-}$ &  & \tabularnewline
		& II & $C_{3v}\otimes\{I,E\}$ & $A_{1}^{+}$ & $A_{2}^{+}$ & $E^{+}$ &  &  & $A_{2}^{-}$ & $A_{1}^{-}$ & $E^{+}$ &  & \tabularnewline
		\hline 
	\end{tabular}
	\caption{\label{tab:grp_twocv}Irreps of the relevant point groups using the two conventions as discussed in the text and their corresponding relations. The convention I is adopted in ref.~\cite{Rummukainen:1995vs,Briceno:2013lba} and in this paper. The convention II is employed in refs.~\cite{Gockeler:2012yj,Morningstar:2013bda}.}
\end{table}

For non-relativistic systems, the above discussion can be simplified since the Lorentz factor becomes $\gamma=1$ and   the Lorentz transformation reduces to the Galilean one. Thus, the second step in figure~\ref{fig:mesh}
is unnecessary. One consequence is that the meshes for non-relativistic systems might be more symmetric than the ones with the same $\bm{d}$ in a relativistic theory. For example, in a non-relativistic theory, the mesh for the $\bm{d}=(0,0,2)$ system is the same as that of the $\bm{d}=(0,0,0)$ system. The shift of origin by ${\bm{d}/2}=(0,0,1)$ in figure~\ref{fig:mesh} makes no
difference for momentum discretization. Thus, the discrete symmetry group of a non-relativistic system with $\bm{d}=(0,0,2)$ is $O_h$. However, the relativistic system with the same $\bm{d}$ only possesses the $D_{4h}$ symmetry. In this work, for calculational convenience, we adopt an unified approach for both non-relativistic and relativistic systems. E.g., for a given system with $\bm{d}=(0,0,2)$, we classify the energy levels according to the irreps of $D_{4h}$. Thus, for non-relativistic systems, we will end up with several degenerate states in different irreps reflecting the additional degeneracy of the more symmetric group $O_h$.

\subsection{Interacting case}~\label{subsec:intac}
To compute FV energy spectra in the interacting case, we need to perform boost transformations
for loop momenta with particles being off shell, i.e.~$E_i^{(*) 2}\neq m_i^2+\bm{p}_i^{(*)2}$.
In the literature, different schemes have been proposed to define
the quantization conditions for the cases with $\bm{P}\neq 0$. 
The one of ref.~\cite{Doring:2012eu} starts with the prescription
\begin{equation}
	E_{1,2}^{*}=\frac{E^{*2}+m_{1,2}^{2}-m_{2,1}^{2}}{2E^{*}} \,,
	\label{eq:erelation}
\end{equation}
which is the same as in the non-interacting case and ensures that $E_1^*+E_2^*=E^*$, and $E_1^*=E_2^*={E^*\over 2}$ for $m_1=m_2$. By substituting eq.~\eqref{eq:erelation} into eqs.~\eqref{eq:lrp} and ~\eqref{eq:lre}, one obtains the same quantization conditions as those in eqs.~\eqref{eq:lrcomb} and \eqref{eq:dis_cmf}.

Recently, an alternative scheme was proposed in ref.~\cite{Li:2021mob}. Their relation of $\bm{p}_1^*$ with $\bm{p}_1$ reads
\begin{equation}
	\bm{p}_{1}^{*}=\left[(\gamma-1)\frac{\bm{P}\cdot\bm{p}_{1}}{\bm{P}^{2}}-\frac{\omega_{1}}{\sqrt{(\omega_{1}+\omega_{2})^{2}-P^{2}}}\right]\bm{P}+\bm{p}_{1},\label{eq:transWu}
\end{equation}
where $\omega_i=\sqrt{m_1^2+\bm{p}_i^2}$. The Lorentz factor $\gamma$ is defined as
\begin{equation}
	\gamma=\frac{\omega_{1}+\omega_{2}}{\sqrt{(\omega_{1}+\omega_{2})^{2}-\bm{P}^{2}}} \,.
\end{equation}

In the first scheme, the Lorentz factor $\gamma$ is defined rigorously, and it only depends on the total energies of the two-particle system in the BF and CMF. 
In the second scheme, the Lorentz factor is taken as an approximation using the expression for the non-interacting case. Such an approximation makes the transformation independent of the energy, which simplifies the determination of poles of the  $T$-matrix. It was argued that the difference between the two schemes is exponentially suppressed~\cite{Li:2021mob}.

In the infinite volume, the LSE and BSE are usually written in the CMF. We need to replace the integration over the loop momentum in the LSE or BSE with a summation over the discrete momentum values. For a general case with $\bm{P}\neq 0$, the transformation reads
\begin{equation}
\int\frac{d^{3}\bm{q}^{*}}{(2\pi)^{3}}f(\bm{q}^{*})=\int {\cal J} \frac{d^{3}\bm{q}}{(2\pi)^{3}}f[\bm{q}^{*}(\bm{q})]\xrightarrow{FV}\frac{1}{L^{3}}\sum_{\bm{n}\in {Z}^{3}} {\cal J}  f[\bm{q}^{*}(\bm{q}_{\bm{n}})]=\frac{1}{L^{3}}\sum_{\bm{n}\in P_{d}} {\cal J}  f(\bm{q}_{\bm{n}}^{*}),\label{eq:tofv}	
\end{equation}
where $\cal J$ is the Jacobi determinant for the momentum transformation from the CMF to LF. The explicit expressions for $\cal J$ are different for the two schemes in eqs.~\eqref{eq:lrcomb} and \eqref{eq:transWu},
\begin{equation}
{\cal J}=\begin{cases}
	\gamma^{-1} & \text{scheme-I}\\
	\left.\frac{\omega_{1}\omega_{2}}{\omega_{1}+\omega_{2}}\right/\frac{\omega_{1}^{*}\omega_{2}^{*}}{\omega_{1}^{*}+\omega_{2}^{*}} & \text{scheme-II}
\end{cases}\,. \label{eq:jcob}
\end{equation}

The ambiguities in choosing the transformations in the interacting case are related to the need to define the interactions in the off-shell kinematics. However, for non-relativistic systems, the two schemes reduce to the same one, and the Jacobi determinant in eq.~\eqref{eq:tofv} becomes $\mathcal{J}=1$.

\section{Construction of irreducible representations using the plane wave basis }
\label{sec:irrep}
\subsection{Representation space}~\label{subsec:rep_space}
The rotation operator in the Hilbert space is $\hat{D}(g(\bm{n},\theta))=e^{-i\bm{n}\cdot\hat{\bm{J}}\theta}$, where $\hat{J}$ is the angular momentum operator and $g(\bm{n},\theta)$ is the rotation around $\bm{n}$ axis with angle $\theta$. 
For spinless systems in the infinite volume featuring the $SO(3)$ symmetry, the space spanned by $|lm\rangle,m=-l,...,l$,  forms an irrep. The matrix elements of the rotation operator in the irrep give rise to the Wigner $\cal{D}$-function defined via
\begin{eqnarray}
	\langle l'm'|\hat{D}(g)|lm\rangle \equiv {\cal D}_{m'm}^{l}(g)\delta_{ll'},
\end{eqnarray}
where $l$ and $m$ are the quantum numbers of angular momentum and its third component, respectively. In the finite volume, the Hilbert space with fixed $l$ also forms a representation space of $O$, $D_4$, $D_2$, and $D_3$ groups, but it becomes reducible. 
The procedure of reducing the Hilbert space with a fixed $l$ to the irrep ones of the corresponding point group is well established, see e.g. ref.~\cite{Bernard:2008ax}.

In the infinite volume, the Hilbert space spanned by the plane wave basis $|\bm{p}\rangle$  
also forms a representation space of the $O(3)$ group,
\begin{eqnarray}
\hat{D}(g)|\bm{p}\rangle=|g\bm{p}\rangle,\quad\langle\bm{p}'|\hat{D}(g)|\bm{p}\rangle=\delta_{\bm{p}',g\bm{p}}.\label{eq:Dpl_wv}
\end{eqnarray}
In the finite volume, a natural choice for the representation space of the cubic $O_h$ group is
\begin{equation}
\{n_{1},n_{2},n_{3}\} \equiv \{|n_{1},n_{2},n_{3}\rangle\text{ with permutations of }n_{1},n_{2},n_{3}\text{ and changing signs}\},\label{eq:space30}
\end{equation}
where $n_i$ is a multiple of ${2\pi \over L}$ for each component of the momentum. We use the notation $|n_i,n_j,n_k\rangle$ to denote the plane wave basis in a finite cubic box. It is obvious that $n_i$ are integer numbers for the $O_h$ group when $\bm{d}=(0,0,0)$. The dimensions of $\{n_1,n_2,n_3\}$, which we refer to as patterns throughout this paper, depend on the number of zeros in three components and the number of equality relations among them. 
\begin{table}[tbp]
	\centering
	\begin{tabular}{|c|c|c|c|c|}
		\hline 
		$\{n_{1},n_{2},n_{3}\}_{D}$ & \multicolumn{2}{c|}{$O_{h}:\{A_{1},A_{2},E,T_{1},T_{2}\}^{+|-}$} & \multicolumn{2}{c|}{$D_{3d}:\{A_{1},A_{2},E\}^{+|-}$}\tabularnewline
		\hline 
		$\{0,0,0\}_{1}$ & \{1,0,0,0,0\} & \{0,0,0,0,0\} & \{1,0,0\} & \{0,0,0\}\tabularnewline
		$\{0,0,a\}_{6}$ & \{1,0,1,0,0\} & \{0,0,0,1,0\} & \{1,0,1\} & \{0,1,1\}\tabularnewline
		$\{0,a,a\}_{12}$ & \{1,0,1,0,1\} & \{0,0,0,1,1\} & \{2,0,2\} & \{1,1,2\}\tabularnewline
		$\{0,a,b\}_{24}$ & \{1,1,2,1,1\} & \{0,0,0,2,2\} & \{2,2,4\} & \{2,2,4\}\tabularnewline
		$\{a,a,a\}_{8}$ & \{1,0,0,0,1\} & \{0,1,0,1,0\} & \{2,0,1\} & \{0,2,1\}\tabularnewline
		$\{a,a,b\}_{24}$ & \{1,0,1,1,2\} & \{0,1,1,2,1\} & \{3,1,4\} & \{1,3,4\}\tabularnewline
		$\{a,b,c\}_{48}$ & \{1,1,2,3,3\} & \{1,1,2,3,3\} & \{4,4,8\} & \{4,4,8\}\tabularnewline
		\hline 
	\end{tabular}
	\caption{\label{tab:30}Different patterns for the representation spaces $\{n_1,n_2,n_3\}$ and the multiplicity of irreps in such patterns for the $O_h$ and $D_{3d}$ groups. The subscript $D$ denotes the dimension of the representation space, while $a$, $b$ and $c$ are different non-zero numbers. The numbers in the last four columns are the multiplicities of the corresponding irreps in different representation spaces.   }
\end{table}
In table~\ref{tab:30}, we list seven different patterns and their dimensions. The group representations in the $\{n_1,n_2,n_3\}$ space are reducible. Fortunately, the reduction only depends on the patterns of this space. For the $O_h$ group, we only need to deal with the reduction for seven patterns as shown in table \ref{tab:30}.
Notice that the $A_i$ and $B_i$ irreps are one-dimensional ones, while the $E$ and $T_i$ representations are two- and three-dimensional irreps, respectively.

For moving systems in a box with $\bm{d}=(a,a,a)$, the corresponding point group is $D_{3d}$ and we can choose a similar representation space as in eq.~\eqref{eq:space30}. In this case, $n_i$ are unlikely to be integer numbers according to the quantization condition in eq.~\eqref{eq:dis_cmf}. In order to obtain the irreps, we need
to deal with at most seven different patterns shown in figure~\ref{tab:30}. If $a$ is odd, the patterns with zero components will not appear.

For moving systems with $\bm{d}=(0,0,a)$ or $\bm{d}=(a,a,0)$, it is more convenient to employ the representation spaces
\begin{equation}
	\{n_{1},n_{2};n_{3}\} \equiv \{|n_{1},n_{2},n_{3}\rangle\text{ with permutations of} \ n_{1}\text{and} \ n_{2}\text{ and changing signs}\}.\label{eq:space21}
\end{equation}
We use a semicolon to separate the third component from the first two components to signify permutations of $n_1$ and $n_2$ only. The $\{n_{1},n_{2};n_{3}\}$ representation spaces have eight different patterns as listed in table~\ref{tab:21}.

With the representation spaces in eqs.~\eqref{eq:space30} and \eqref{eq:space21} we can obtain the representation matrices for each pattern using eq.~\eqref{eq:Dpl_wv}. Next, we are going to reduce the representations into the direct sums of irreps.

\begin{table}
	\centering
	\begin{tabular}{|c|c|c|c|c|}
		\hline 
		$\{n_{1},n_{2};n_{3}\}_{D}$ & \multicolumn{2}{c|}{$D_{4h}:\{A_{1},A_{2},B_{1},B_{2},E\}^{+|-}$} & \multicolumn{2}{c|}{$D_{2h}:\{A_{1},B_{1},B_{2},B_{3}\}^{+|-}$}\tabularnewline
		\hline 
		$\{0,0;0\}_{1}$ & \{1,0,0,0,0\} & \{0,0,0,0,0\} & \{1,0,0,0\} & \{0,0,0,0\}\tabularnewline
		$\{0,0;c\}_{2}$ & \{1,0,0,0,0\} & \{0,1,0,0,0\} & \{1,0,0,0\} & \{0,0,0,1\}\tabularnewline
		$\{0,a;0\}_{4}$ & \{1,0,1,0,0\} & \{0,0,0,0,1\} & \{1,0,0,1\} & \{0,1,1,0\}\tabularnewline
		$\{0,a;c\}_{8}$ & \{1,0,1,0,1\} & \{0,1,0,1,1\} & \{1,1,1,1\} & \{1,1,1,1\}\tabularnewline
		$\{a,a;0\}_{4}$ & \{1,0,0,1,0\} & \{0,0,0,0,1\} & \{2,0,0,0\} & \{0,1,1,0\}\tabularnewline
		$\{a,a;c\}_{8}$ & \{1,0,0,1,1\} & \{0,1,1,0,1\} & \{2,1,1,0\} & \{0,1,1,2\}\tabularnewline
		$\{a,b;0\}_{8}$ & \{1,1,1,1,0\} & \{0,0,0,0,2\} & \{2,0,0,2\} & \{0,2,2,0\}\tabularnewline
		$\{a,b;c\}_{16}$ & \{1,1,1,1,2\} & \{1,1,1,1,2\} & \{2,2,2,2\} & \{2,2,2,2\}\tabularnewline
		\hline 
	\end{tabular}
        \caption{\label{tab:21}Different patterns for the representation spaces $\{n_1,n_2;n_3\}$ and the multiplicity of irreps in such patterns for the $D_{4h}$ and $D_{2h}$ groups. $a$, $b$ and $c$ are non-zero numbers with $a \neq b$. 
          The numbers in the last four columns are the multiplicities of the corresponding irreps in different representation spaces.}
\end{table}

\subsection{Reduction of the representations }
The procedure of reducing the representations with the help of projection operators is well established~\cite{dresselhaus2007group}. It has been used to reduce representations in the partial wave basis~\cite{Bernard:2008ax} and in the Fourier basis~\cite{Lee:2020fbo}. For a related discussion see ref.~\cite{Doring:2018xxx}.  We will employ the same technique to reduce representations in the plane wave basis. The projection operator is defined as
\begin{equation}
\hat{P}_{\alpha\beta}^{\Gamma_{a}}\equiv\sum_{g_{i} \in G}\frac{N(\Gamma_{a})}{n_{G}}R_{\alpha\beta}^{\Gamma_{a}}(g_{i})^{*}\hat{D}(g_{i}),~\label{eq:prjop}
\end{equation}
where $\hat{D}(g_{i})$ is the (im)proper rotation operator  in the Hilbert space corresponding to the group element $g_i$ as discussed in section~\ref{subsec:rep_space}, $\Gamma_{a}$ denotes the corresponding irrep, while $N(\Gamma_{a})$ and $n_G$ refer to the dimensionality of the irrep and the order of the group, respectively. Further, $R^{\Gamma_a}$ is the unitary matrix irrep of $\Gamma_a$. For the point groups relevant for this study, the unitary irreps can be taken from the literature~\cite{Bernard:2008ax,Gockeler:2012yj,Morningstar:2013bda}. We will specify the procedure of constructing  $R^{\Gamma_a}$ later. Here, we list some  properties of the projection operator without proof~\cite{dresselhaus2007group},
\begin{equation}
	\hat{P}_{\alpha\alpha'}^{\Gamma_{a}}\hat{P}_{\beta'\beta}^{\Gamma_{b}}=\delta_{\Gamma_{a}\Gamma_{b}}\delta_{\alpha'\beta'}\hat{P}_{\alpha\beta}^{\Gamma_{a}},\quad\hat{P}_{\alpha\beta}^{\Gamma_{a}}|\Gamma_{b},\beta'\rangle=\delta_{\Gamma_{a}\Gamma_{b}}\delta_{\beta\beta'}|\Gamma_{a},\alpha\rangle.
\end{equation}
When acting on a general state $|\psi \rangle$ defined as
\begin{equation}
|\psi\rangle\equiv\sum_{\Gamma_{b}}\sum_{\beta}a_{\beta}^{\Gamma_{b}}|\Gamma_{b},\beta\rangle,
\end{equation}
the projection operator projects the state onto a single component of a fixed irrep,
\begin{equation}
\hat{P}_{\alpha\alpha'}^{\Gamma_{a}}|\psi\rangle=a_{\alpha'}^{\Gamma_{a}}|\Gamma_{a},\alpha\rangle.
\end{equation}
For the case at hand, we choose $|\psi\rangle$ as the basis of the representation space $|n_1,n_2,n_3\rangle$, fix $\alpha'$ and vary $|\alpha\rangle$ to obtain the basis states $|\Gamma_a,\alpha\rangle $. With the transformation from $|n_1,n_2,n_2\rangle$ to $|\Gamma_a,\alpha \rangle$, we reduce the original representation to a direct sum of irreps.

The unitary matrix irreps $R^{\Gamma_a}$ in eq.~\eqref{eq:prjop} can be constructed with the character projection operator, which is defined as
\begin{eqnarray}
	\hat{P}^{\Gamma_{a}}	\equiv\sum_{\alpha}\hat{P}_{\alpha\alpha}^{\Gamma_{a}}=\sum_{g_{i}\in G}\frac{N(\Gamma_{a})}{n_{G}}\chi^{\Gamma_{a}}(g_{i})\hat{D}(g_{i})\,,
\end{eqnarray}
where $\chi^{\Gamma_{a}}(g_{i})=\sum_{\alpha}R_{\alpha\alpha}^{\Gamma_{a}}(g_{i})$ is the character of the irrep $\Gamma_a$. 
Acting with the character projection operator on $|\psi \rangle$, we obtain
\begin{equation}
\hat{P}^{\Gamma_{a}}|\psi\rangle	=\sum_{\alpha}a_{\alpha}^{\Gamma_{a}}|\Gamma_{a},\alpha\rangle.
\end{equation}
One can vary $|\psi\rangle$ to obtain a sufficient number of vectors, which span the irrep space after orthogonalization. In particular, in order to obtain the  $R^{\Gamma_a}$, one can start with the faithful representation. 
For point groups,  $|\psi \rangle$ and $\hat{D}(g_i)$ are replaced with the vector $(x,y,z)$ and (im)proper rotation matrices in the Euclidean space, respectively. For point groups, the character table is unique but the unitary irrep matrices are not. One can choose a convention to make $R^{\Gamma_a}$ real symmetric matrices~\cite{Bernard:2008ax,Gockeler:2012yj}.

In summary, the general procedure to reduce a representation to irreps is as follows:
\begin{enumerate}
	\item Identify the symmetric group and its elements and character table,
	\item Construct the unitary irrep matrices with character projection operation,
	\item Reduce the representation to a direct sum of irreps.
\end{enumerate}


\section{Finite volume energy levels: determination and fitting}
\label{sec:FVft-slv}
In general, FV energy levels can be obtained by finding the poles of the $T$-matrix in the box, calculated as described in sections \ref{sec:NN} and \ref{sec:pipiSystem}. Using the technique described in section~\ref{sec:irrep}, one ends up with a determinant equation for a specific irrep $\Gamma$
\begin{equation}
	\det[\mathbb{M}_\Gamma(E)]=0.~\label{eq:det}
\end{equation}
The specific expressions for the energy-dependent matrix $\mathbb{M}$ will be presented for the non-relativistic and relativistic systems in the next two sections. In general, one can adopt the root-finding algorithm to search for the roots of eq.~\eqref{eq:det}. To this end we define $\Omega $~\cite{Morningstar:2017spu}
\begin{equation}
	\Omega_\Gamma(E;\mu)\equiv\prod\frac{\lambda_{\Gamma,i}(E)}{\sqrt{\lambda_{\Gamma,i}(E)^{2}+\mu^{2}}}, \quad \quad \det\left[\mathbb{M}_\Gamma(E)\right]=\prod_{i}\lambda_{\Gamma,i}(E), \label{eq:omega}
\end{equation}
where $\lambda_{\Gamma,i}$ denotes the eigenvalue of the matrix $\mathbb{M}_\Gamma$. The root of eq.~\eqref{eq:det} corresponds to at least one $\lambda_{\Gamma,i}=0$. Thus, the zeros of $\Omega_\Gamma(E)$ are the roots of eq.~\eqref{eq:det}. The nonzero $\mu$ can be chosen to optimize the root-finding procedure. The quantity $\Omega_\Gamma(E;\mu)$ is bounded between $-1$ and $1$. Introducing $\Omega_\Gamma(E;\mu)$ thus allows one to avoid the numerical complication when finding the roots  with the determinant in eq.~\eqref{eq:det} becoming very large for large-dimensional matrices. 

Meanwhile, one could use eq.~\eqref{eq:det} to determine the parameters of the interaction within a given theoretical framework
by matching to the FV energy levels from lattice QCD. To this end, we have two options, the spectrum method and the determinant residual method~\cite{Morningstar:2017spu}. In the spectrum method, one compares the FV energy levels from the calculation with the lattice QCD spectrum and minimizes the residuals. In the fitting procedure, the FV energy levels need to be obtained by the root-finding algorithm repeatedly, which is time-consuming, see also ref.~\cite{Woss:2020cmp} for a discussion of further challenges. Alternatively, in the determinant residual method, the determinant $\det(\mathbb{M}_\Gamma)$ that becomes zero when the lattice QCD energy level is the solution of eq.~\eqref{eq:det}, is regarded as a residual. As already pointed out above, it is more practical to employ the quantity $\Omega_\Gamma(E;\mu)$ defined in eq.~\eqref{eq:omega} as a residual. Then, the $\chi^2$-function to be minimized is defined via
\begin{equation}
	\chi^{2}=\sum_{\Gamma,i}\frac{\Omega_{\Gamma}(E_{\Gamma,i})^{2}}{\sigma[\Omega_{\Gamma}(E_{\Gamma,i})]^{2}},\label{eq:chi2}
\end{equation}
where $E_{\Gamma,i}$ is the $i$th lattice QCD energy level in the irrep $\Gamma$ while $\sigma[\Omega_{\Gamma}(E_{\Gamma,i})]$ is the error of $\Omega_\Gamma$ propagated from the error of $E_{\Gamma,i}$. In this exploratory study, we neglect possible correlations among the different FV energy levels. Here and in what follows, we employ
the determinant residual approach.

\section{Application I: Spin-singlet two-nucleon scattering}
\label{sec:NN}

We now consider nucleon-nucleon scattering in spin-singlet channels as an example of a non-relativistic system.
This is a particularly interesting case for several reasons. First, the formalism described in sections~\ref{sec:p-dis} and \ref{sec:irrep}
for spinless particles of equal masses is well suited for NN scattering in spin-zero channels assuming the exact isospin symmetry.
Secondly, because of the appearance of the strong long-range interaction due to the OPE, the NN system is expected to feature severe
FV artifacts
due to partial wave mixing, which is one of
the central focuses of our study. Last but not least, the quantization conditions for momenta in section~\ref{sec:p-dis}
take a particularly simple form for non-relativistic systems with energy-independent potential and allow one
to compute the FV energies by simply solving the eigenvalue problem. 

In the past decades, nuclear forces have been extensively studied in the framework of chiral EFT. See refs.~\cite{Epelbaum:2008ga,Machleidt:2011zz,Epelbaum:2019kcf} for review articles.
State-of-the-art NN interactions at fifth order (N$^4$LO) in chiral EFT have been shown to provide an excellent
description of the available neutron-proton and proton-proton scattering data up to pion production threshold
\cite{Epelbaum:2014sza,Reinert:2017usi,Entem:2017gor}. For the purpose of this proof-of-principle study, we restrict ourselves to
next-to-next-to-leading order (NNLO) and employ the nonlocally
regularized potentials of ref.~\cite{Epelbaum:2003xx}.
The NNLO potential reads
\begin{equation}
V=V_{\text{cont}}^{(0)}+V_{1\pi}^{(0)}+V_{\text{cont}}^{(2)}+V_{2\pi}^{(2)}+V_{1\pi}^{(2)}+V_{2\pi}^{(3)},
\end{equation}
where $V_{\text{cont}}^{(0)}$ and $V_{\text{cont}}^{(2)}$ are the leading-order (LO) and next-to-leading order (NLO) contact interactions, respectively. $V_{1\pi}^{(0)}$ and $V_{1\pi}^{(2)}$ denote the LO one-pion exchange interaction and its correction at NLO. The NLO correction to the OPE is included by taking a larger value
of the (effective) nucleon axial vector coupling 
$g_A=1.29$ to account for the Goldberger-Treiman discrepancy \cite{Baru:2011bw,Reinert:2020mcu}. The two-pion-exchange interactions start contributing at NLO. For spin-singlet NN channels, we can replace the spin operators in the potentials in Ref.~\cite{Epelbaum:2003xx,Epelbaum:2003gr} as
\begin{equation}
	\sigma_{1}^{i}\sigma_{2}^{j}\to-\delta^{ij},\quad\sigma_{1}^{i}+\sigma_{2}^{i}\to0.\label{eq:spin-oprt}
\end{equation}
For example, the OPE potential $V_{1\pi}^{(0)}$ reads
\begin{equation}
V_{1\pi}^{(0)}(\bm{p},\bm{p}')=-\left(\frac{g_{A}}{2F_{\pi}}\right)^{2}\frac{(\bm{\sigma}_{1}\cdot\bm{q})(\bm{\sigma}_{2}\cdot\bm{q})}{\bm{q}^{2}+M_\pi^{2}}\bm{\tau}_{1}\cdot\bm{\tau}_{2},
\end{equation}
where $\bm{q}=\bm{p}'-\bm{p}$. For spin-singlet channels, it simplifies to 
\begin{equation}
V_{1\pi}^{(0)}(\bm{p},\bm{p}')\to\left(\frac{g_{A}}{2F_{\pi}}\right)^2\frac{\bm{q}^{2}}{\bm{q}^{2}+M_\pi^{2}}\bm{\tau}_{1}\cdot\bm{\tau}_{2}.
\end{equation} 
At NLO, the contact interactions read
\begin{equation}
V_{\text{cont}}^{(0)}(\bm{p},\bm{p}')=C_{S},\quad V_{\text{cont}}^{(2)}(\bm{p},\bm{p}')=C_{1}\bm{q}^{2}+C_{2}\bm{k}^{2},
\end{equation}
where $\bm{k}=(\bm{p}+\bm{p}')/2$, while $C_S$, $C_1$ and $C_2$ are low energy constants (LECs). Here, the general expression for the contact interactions including spin operators and consisting of $9$ independent terms has been reduced using eq.~\eqref{eq:spin-oprt}. The LECs $C_S$, $C_1$ and $C_2$ are thus linear combinations of the original LECs appearing in the chiral NN potential. We take the specific values of the LECs in Table 4 of ref.~\cite{Epelbaum:2003xx} and employ the same numerical values for all the other parameters. The expressions for the two-pion-exchange interactions can be found e.g.~in ref.~\cite{Epelbaum:2003xx}. Following the standard procedure, we introduce the regulator by the replacement,
\begin{equation}
V(\bm{p},\bm{p}')\to V(\bm{p},\bm{p}')e^{-\frac{p^{6}+p'^{6}}{\Lambda^{6}}},~\label{eq:cutoff}
\end{equation}
where the cutoff is taken as $\Lambda=0.55$ GeV.

\begin{figure}[tbp]
	\centering 
	\includegraphics[width=.32\textwidth]{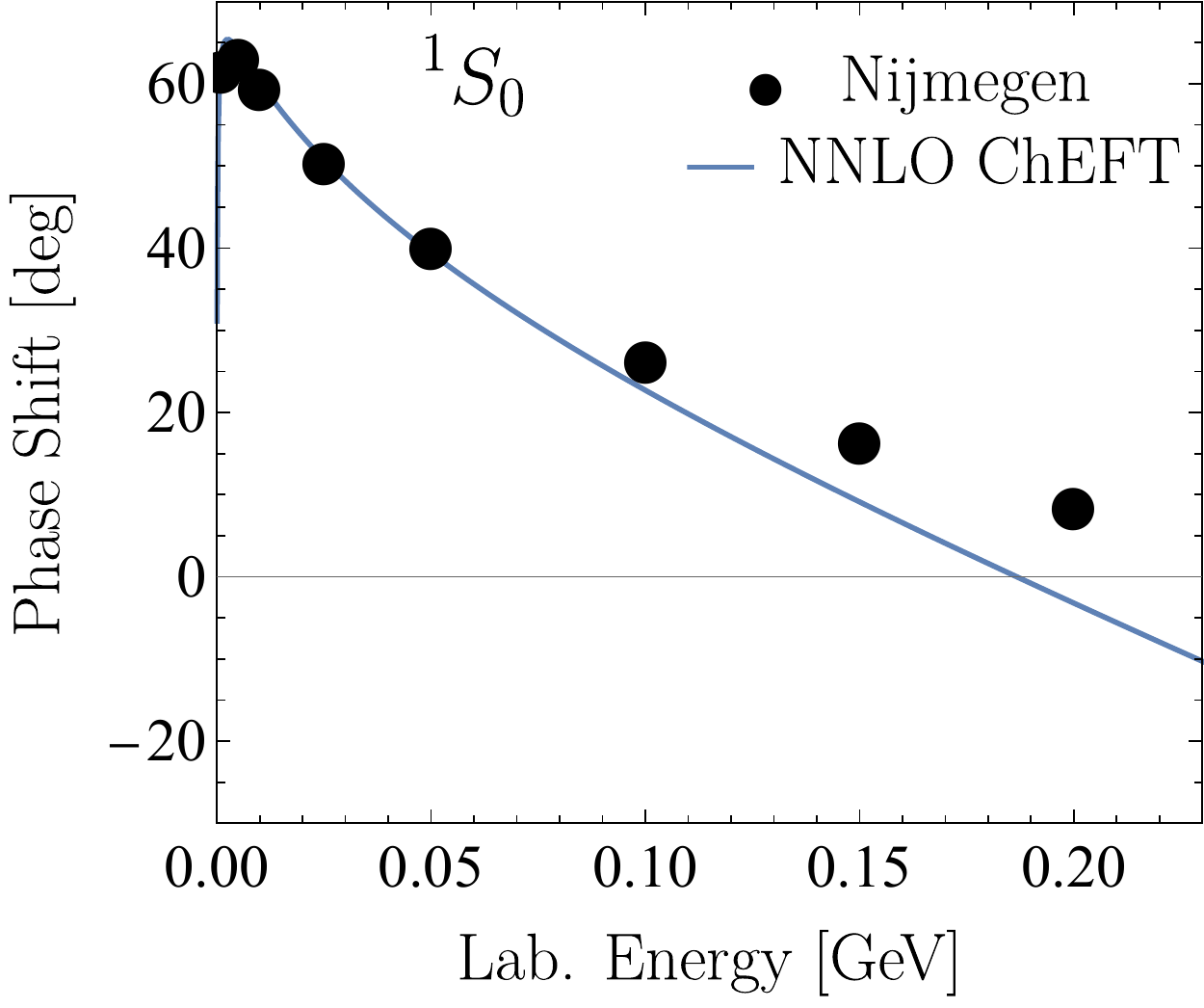}
	\includegraphics[width=.32\textwidth]{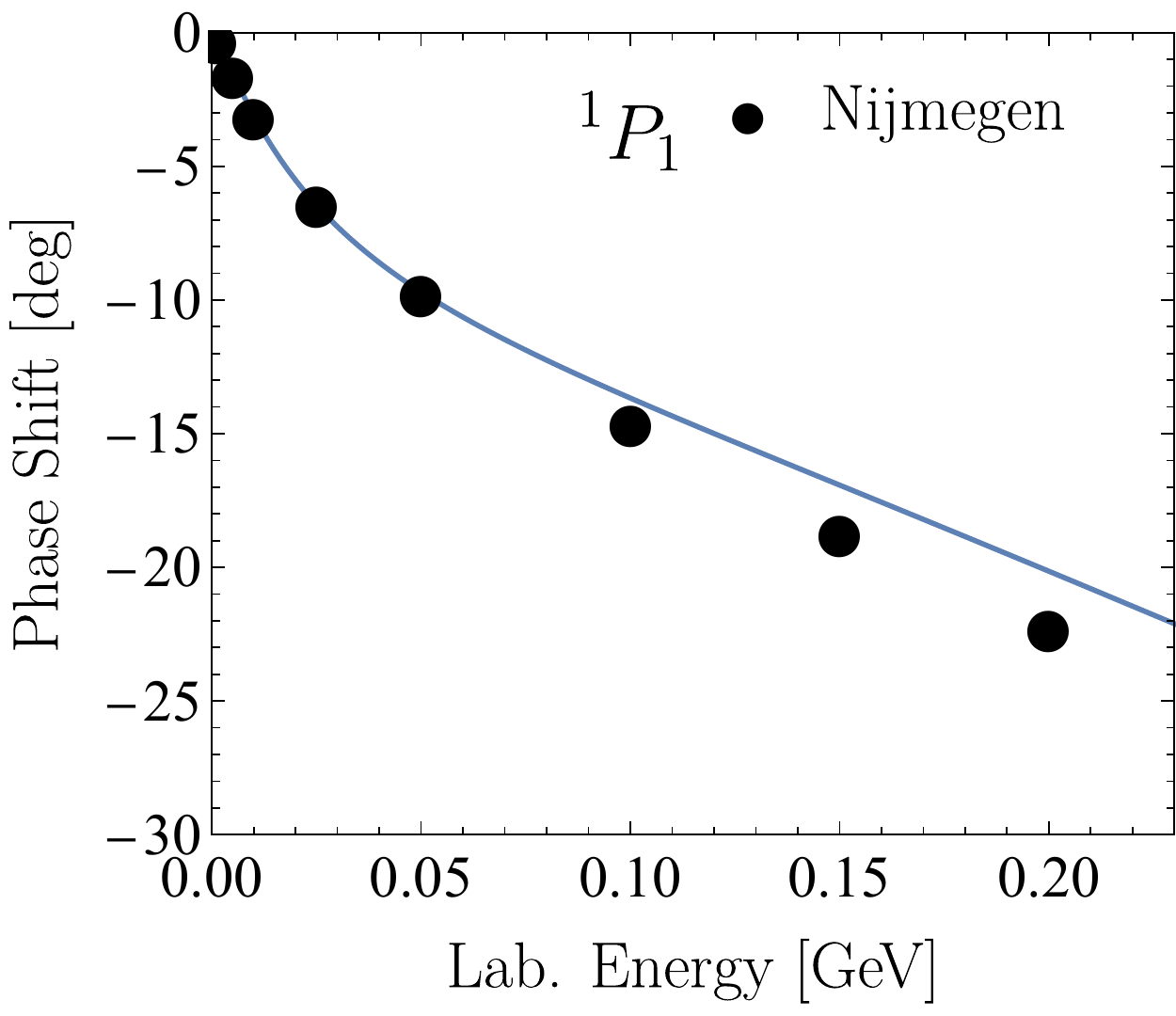}
	\includegraphics[width=.32\textwidth]{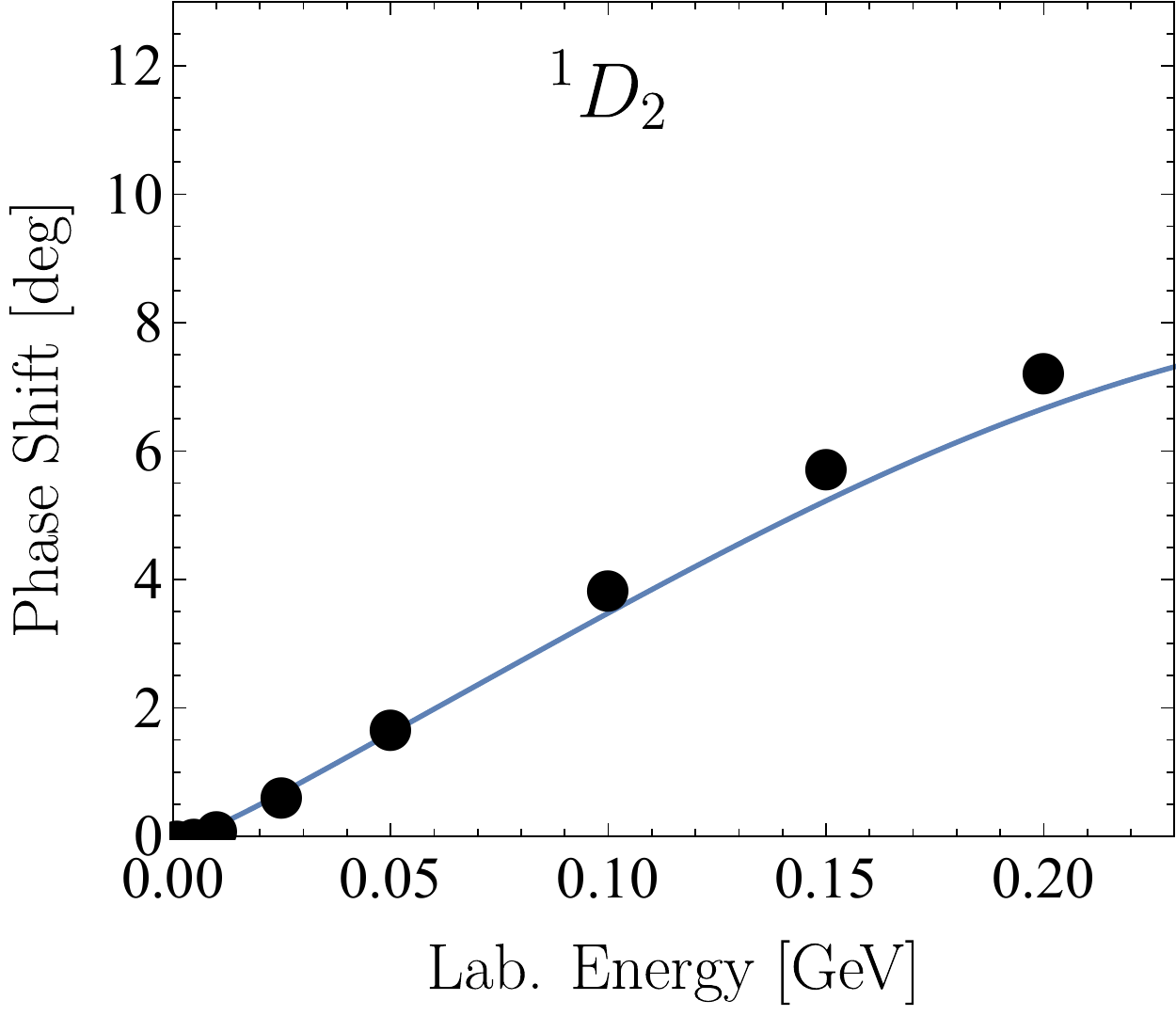}
	\caption{\label{fig:cheft}Spin-singlet NN phase shifts as functions of the laboratory energy. The results of Nijmegen PWA are shown by filled circles~\cite{Stoks:1993tb}. Solid lines correspond to the NNLO chiral EFT results~\cite{Epelbaum:2003xx,Epelbaum:2003gr}.  }
      \end{figure}
      
To calculate phase shifts in the infinite volume, we solve the LSE in the CMF 
\begin{equation}
T(\bm{p}',\bm{p};E)=V(\bm{p}',\bm{p};E)+\int\frac{d^{3}\bm{q}}{(2\pi)^{3}}\frac{V(\bm{p}',\bm{q};E) \; T(\bm{q},\bm{p};E)}{E-\frac{\bm{q}^{2}}{m_N}+i\epsilon},
\end{equation}
where $E$ is the energy of two nucleons and $m_N$ is the nucleon mass. Clearly, for infinite-volume calculations,
it is more convenient to use the standard partial wave basis instead of solving the equation in real space.  
In figure~\ref{fig:cheft}, we compare the resulting neutron-proton phase shifts at NNLO  with  the ones from the
Nijmegen partial wave analysis (PWA) \cite{Stoks:1993tb}.
This comparison clearly demonstrates that the NNLO potential already captures the most important features
of the nuclear force.

Below,  we will regard the chiral EFT potential introduced above as the underlying NN interaction
to generate the corresponding FV energy spectra. We then compare our approach to reconstruct phase shifts
from the FV energies with the single-channel L\"uscher formula and explore FV effects due to
partial wave mixing.

\subsection{Lippmann-Schwinger equation in a finite volume}
To solve the LSE in a finite volume, we employ the plane wave basis with quantized
momenta instead of expanding it in partial waves.
In the non-relativistic limit,  eq.~\eqref{eq:dis_cmf} reduces to 
\begin{equation}
	P_{d}^\text{NR}=\bigg\{\bm{n}-\frac{1}{2}\bm{d} \bigg\},\quad \bm{n}\in {Z}^{3}.
\end{equation}
We define the matrices $\mathbb{T}$ and $\mathbb{G}$ according to
\begin{eqnarray}
\mathbb{T}_{\bm{n}',\bm{n}}	=T\left(\frac{2\pi}{L}\bm{n}',\frac{2\pi}{L}\bm{n};E\right), \quad\mathbb{G}_{\bm{n},\bm{n}'}={\cal J}\frac{1}{L^{3}}\frac{1}{E-\frac{q_{\bm{n}}^{2}}{m_N}}\delta_{\bm{n}',\bm{n}},\quad \bm{n},\bm{n}'\in P_{d}^\text{NR}\,,
\end{eqnarray}
where $q_{\bm{n}}=2\pi\bm{n}/L$. Similarly, we define $\mathbb{V}_{\bm{n}',\bm{n}}$. The integration has been replaced with the summation according to eq.~\eqref{eq:tofv}. We introduce a truncation of the discretized momentum $q_{\bm{n}} < q_\text{max}$. For a non-relativistic theory, the Jacobi determinant is $\mathcal{J}=1$ and the LSE turns to a matrix equation
\begin{equation}
	\mathbb{T=\mathbb{V}+\mathbb{V}\mathbb{G}\mathbb{T}}.\label{eq:LSE_mtrx}
\end{equation}
The poles of the $T$-matrix are obtained by solving equation
\begin{equation}
\det\left(\mathbb{G}^{-1}-\mathbb{V}\right)=0.~\label{eq:pole}
\end{equation}
If the potential is energy-dependent, the root-finding algorithm should be adopted. For energy-independent potentials, eq.~\eqref{eq:pole} becomes an eigenvalue problem, 
\begin{eqnarray}
	\det\left(\mathbb{H-}E\mathbb{I}\right)=0,\quad \text{with}\quad \mathbb{H}_{\bm{m},\bm{n}}=\frac{1}{L^{3}}\mathbb{V}_{\bm{m},\bm{n}}+\frac{q_{\bm{n}}^{2}}{m_N}\delta_{\bm{m},\bm{n}},\label{eq:egvl}
\end{eqnarray}
where $\mathbb{I}$ is the identity matrix. We see that the equation becomes a Hamiltonian equation. We, however,  want to stress that this Hamiltonian equation is different from the one emerging in the approach of refs.~\cite{Hall:2013qba,Wu:2014vma,Liu:2015ktc,Li:2021mob}. In that case, partial wave expansion was made prior to discretizing the magnitudes of the momenta. In our calculation, we discretize the momenta as three-vectors with no reliance on the partial wave expansion. For a related approach see ref.~\cite{Mai:2019fba}. 

Using the projection operator technique described in section~\ref{sec:irrep}, we reduce the matrices in eq.~\eqref{eq:LSE_mtrx} to block diagonal ones. Therefore, equation (\ref{eq:egvl}) turns into a set of matrix equations according to different irreps. The FV energy levels for a given irrep $\Gamma$ are then obtained by solving the eigenvalue problem for this irrep
\begin{equation}
\det\left(\mathbb{H}_{\Gamma}-E_{\Gamma}\mathbb{I}\right)=0,\label{eq:egvlgamma}
\end{equation}
where $\mathbb{H}_\Gamma$ belongs to the block of $\mathbb{H}$ corresponding to the irrep $\Gamma$.

\subsection{The single-channel L\"uscher method}
\label{Luescher}

We consider  boxes of three different lengths: $L=3,5,8$ fm. In our  calculations, we truncate the momentum at 
\begin{equation}
	q_\text{max}^2= 100 \left(2\pi \over L\right)^2,
\end{equation}
leading to $q_\text{max}\approx 4.1,2.5,1.5$~GeV for $L=3,5,8$~fm, respectively. These momenta are
much larger than the cutoff parameter in the employed  NN potential so that the solution of the LSE is well converged. We have verified that the error 
due to the truncation is negligibly small.

\subsubsection{Benchmark calculation using contact interactions}

\begin{figure}[tbp]
	\centering 
	\includegraphics[width=.98\textwidth]{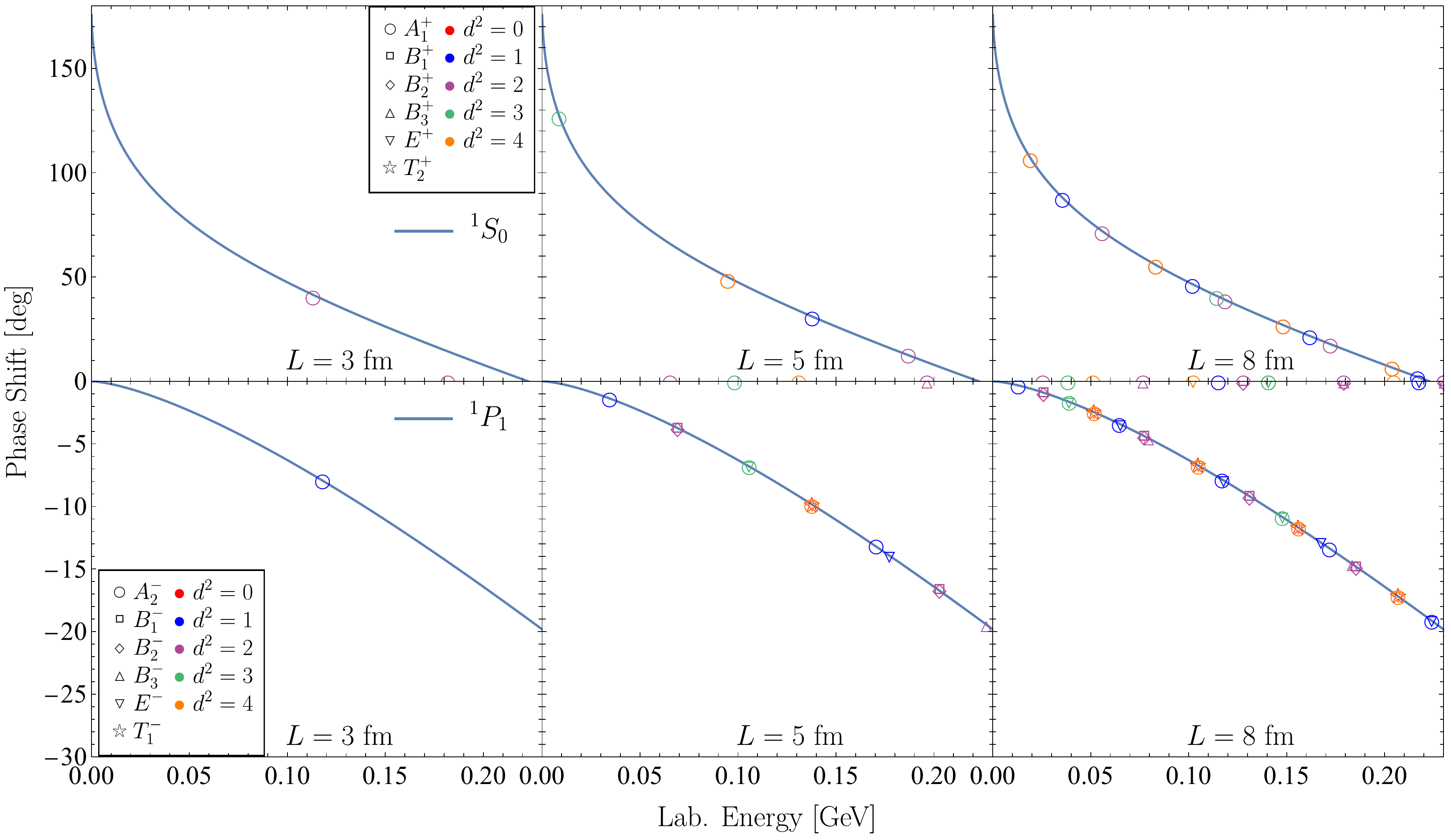}	
	\caption{\label{fig:ct} Upper (lower) row:  Various symbols show the $^1$S$_0$ ($^1$P$_1$) phase shifts calculated from the FV energy spectra using the
single-channel L\"uscher formula for the case of pionless EFT at NLO. Only those energy levels were used which belong to the irreps with the $^1$S$_0$ or $^1$P$_1$ components. Solid lines are the corresponding phase shifts calculated in the infinite volume. Left, middle and right graphs show the results obtained in the FV boxes with $L=3$, $5$ and $8$~fm, in order.}
\end{figure}

To confirm the robustness of our approach, we first calculate the FV energy levels in pionless EFT. Specifically, we include the LO and NLO contact interactions, which contribute to the $^1$S$_0$ and $^1$P$_1$ channels. The coefficients in front of the contact interactions have been chosen the same as those in chiral EFT to the NNLO in ref.~\cite{Epelbaum:2003xx}.  The regulators and cutoff parameters are introduced in eq.~\eqref{eq:cutoff}. The corresponding phase shifts, calculated by solving the LSE in the infinite volume and shown by solid lines in  figure~\ref{fig:ct},
feature a qualitatively similar behavior to the empirical 
ones in NN scattering. 
We then compute the FV energy levels by solving the LSE in the discretized plane wave basis as described in sections \ref{sec:irrep}, \ref{sec:FVft-slv}. 
We regard the calculated energy spectra as synthetic data and apply the single-channel L\"uscher formula to extract the phase shifts at these energies.
The L\"uscher quantization conditions are outlined in appendix~\ref{app:lush}. The full quantization conditions are determinant equations involving all partial waves. In this study, we will refer to the approximation involving only a leading partial wave, which sets one-to-one relations between FV energy levels and phase shifts,  as the single-channel L\"uscher approach. In figure~\ref{fig:ct}, the extracted $^1$S$_0$ and $^1$P$_1$ phase shifts
in the boxes with $L=3,5,8$ fm are compared with the infinite-volume results. For the positive parity states, we only list the $A_1^+$ states, since the other irreps have no $^1$S$_0$ components. With increased box sizes, the unit momentum $(2\pi/L)$ becomes smaller leading to denser energy spectra. In the infinite-volume limit, these discrete energy levels turn to the continuum spectrum. Meanwhile, one can see many energy levels corresponding to vanishing phase shifts (i.e.~to the non-interacting case).  In the FV, partial wave states with the same parity mix with each other due the violation of rotational symmetry. For example, the $^1$S$_0$ states mix with the $^1$D$_2$, $^1$G$_4$, etc., ones.
Thus, in general, energy levels with positive parity receive contributions from the interaction in the $^1$S$_0$, $^1$D$_2$ and even higher partial waves.
Contact interactions at NLO, regularized with an angle-independent regulator, do, however, not contribute to D- and higher partial waves.
Thus, the states in figure~\ref{fig:ct} with vanishing phase shifts correspond to the non-interacting $^1$D$_2$, $^1$F$_3$ and higher partial wave states. Apart from these points, the phase shifts calculated using the single-channel L\"uscher formula are very close to the ones obtained in the infinite volume calculation. As expected, the single-channel L\"uscher approach is found to work accurately for this case  without partial wave mixing. Notice that since for the case at hand the range of the interaction is given by the inverse cutoff, the exponentially suppressed corrections to the L\"uscher formula are negligible even for the smallest considered box size.

\subsubsection{Chiral EFT at NNLO}

\begin{figure}[tbp]
	\centering 
	\includegraphics[width=.98\textwidth]{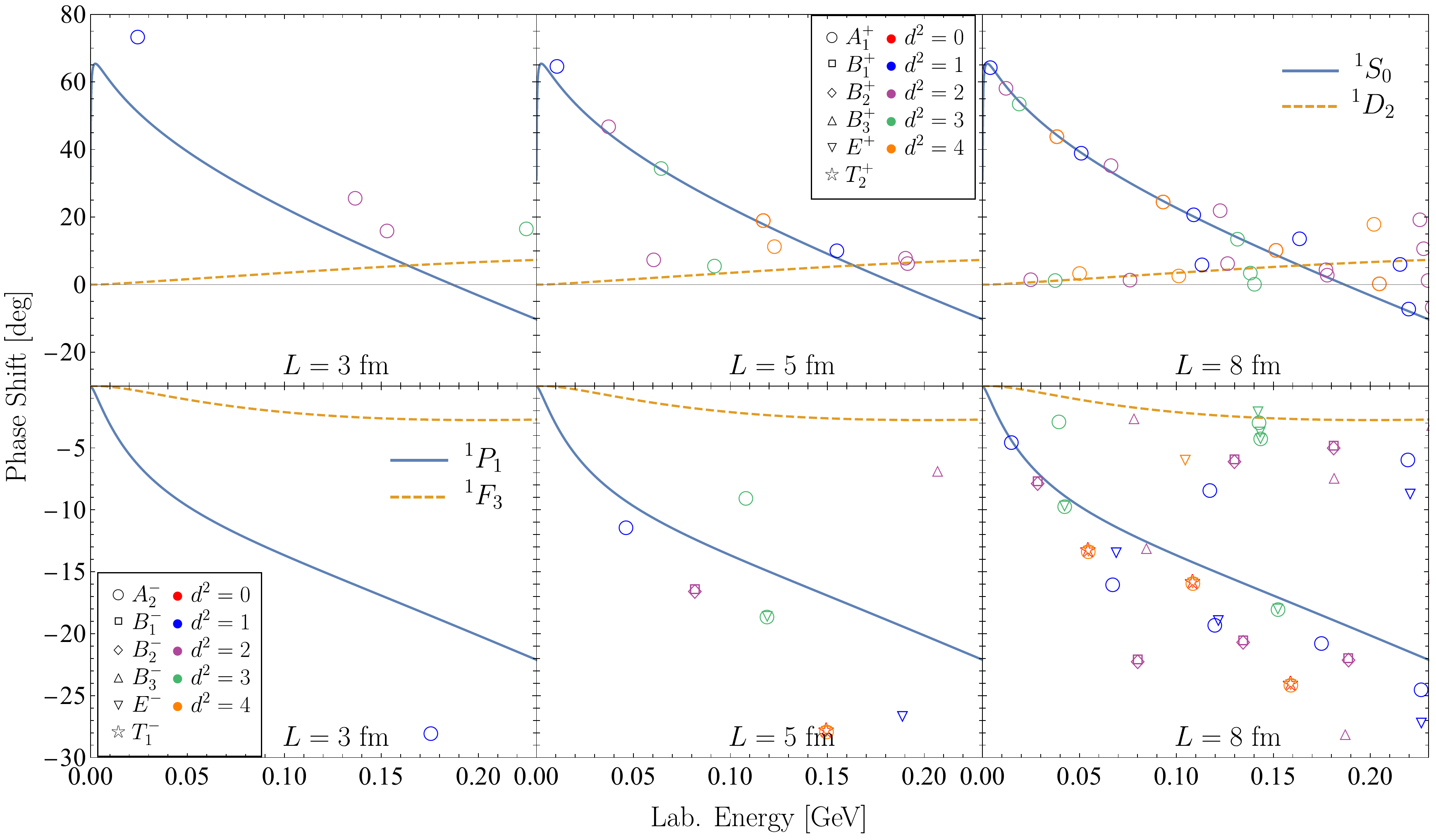}
	\caption{\label{fig:n2op}  Upper (lower) row: Various symbols show positive-parity (negative-parity) phase shifts calculated from the FV energy spectra using the S-wave (P-wave)  L\"uscher formula for the chiral EFT potential at NNLO. Solid and dashed lines in the upper row (lower row) show the $^1$S$_0$ and  $^1$D$_2$ ($^1$P$_1$ and  $^1$F$_3$) phase shifts, respectively, calculated in the infinite volume. For remaining notations see figure~\ref{fig:ct}. }
      \end{figure}

We now turn to the realistic case and repeat the analysis outlined in the previous section for the   chiral EFT NN interaction at NNLO. The results of
our FV calculations are compared with those in the infinite volume in figure~\ref{fig:n2op}. Notice that for positive parity states, we only show the results using the $A_1^+$ irrep. We use the single-channel L\"uscher formula suitable for the $^1$S$_0$ partial wave in the case of the  positive parity states and for the $^1$P$_1$ partial wave in the case of  the negative parity states.
Obviously, for the realistic NN interaction at the physical pion mass, the single-channel L\"uscher formula leads to significant deviations
from the infinite volume calculations. 

For positive-parity states shown in the upper row of figure~\ref{fig:n2op}, the deviation of single-channel L\"uscher's results from the exact ones is significant for the smallest considered box with $L=3$~fm. When the box size is increased to $5$~fm, the L\"uscher single-channel formula performs reasonably well.
Apart from the S-wave dominated states, one can identify several D-wave dominated states in figure~\ref{fig:n2op} leading to smaller-in-magnitude phase shifts.
Though we employ the L\"uscher formula for the S-wave, its leading part $w_{00}$ is the same as that for the D-wave, see appendix~\ref{app:lush}. Therefore, we obtain in figure~\ref{fig:n2op} an approximation of the D-wave phase shift from
D-wave-dominated states using the S-wave  L\"uscher formula.  

The states with negative parity in the lower row of ~figure~\ref{fig:n2op} are more interesting. Even for the largest considered box with $L=8$~fm,  one observes  large deviations from the infinite-volume results when using the single-channel L\"uscher formula. One identifies two regions where the points are concentrated, namely well above the solid line  (region-I) and below the solid line (region-II). One may expect the points in the region-I to emerge from states  dominated by the F-wave  (and possibly by even higher partial wave) interactions.  The points in the region-II are found to gradually deviate from the exact result with increasing energies. Only in the very near-threshold region, the phase shifts for several energy levels are close to those from the infinite volume calculation. Notice  that further increasing the box size\footnote{We performed calculations using the box size as large as $L = 20$~fm.} is found to yield no improvement for the negative-parity phase shifts.

\subsubsection{The one-pion-exchange potential}
\label{sec:pipi}

The results of the previous two sections suggest that the failure of the single-channel L\"uscher formula
for the NNLO chiral EFT potential is caused by partial wave mixing effects. For interactions with a finite non-zero range $R$, the
near-threshold behavior of the phase shifts $\delta_l (p_{\rm on} )$ with $E=p_{\rm on}^2/m_N$ is given by
$\delta_l (p_{\rm on} ) \sim p_{\rm on}^{2l}$. This asymptotic expression, however, only applies to on-shell momenta
well below the
lowest $t$-channel singularity, i.e.~for $p_{\rm on} < R^{-1}$. For the NN interaction, this restriction translates into
 $E_{\rm lab} \sim 2 M_\pi^2/m_N \sim 10$~MeV.  At such low energies, the single-channel L\"uscher formula
 is expected to be valid since the contributions from higher partial waves are kinematically suppressed. This expectation
 is in line with the findings of the previous section for the largest considered box size. For 
 $p_{\rm on} \gtrsim R^{-1}$, higher partial waves are still suppressed due to the centrifugal barrier, but the convergence of
 the partial wave expansion becomes slow. For example, to compute the NN differential cross section
 at $E_{\rm lab} = 300$~MeV at the $1\%$ accuracy level, it is necessary to take into account partial waves up to the total
 orbital momentum of $j_{\rm max} = 16$ \cite{Fachruddin:2001az}. For very long-range interactions such as e.g.~the
magnetic moment interaction, the scattering amplitude is not converged even for $j_{\rm max} \sim 1000$ \cite{Stoks:1990us}.   
One, therefore, may expect partial wave mixing effects in the FV calculations to be dominated by the longest-range OPE interaction.
To get further insights into this issue and to validate this conjecture, we switch off all shorter-range interactions in the chiral EFT potential
and consider the case of pure OPE.

\begin{figure}[tbp]
	\centering 
	\includegraphics[width=.98\textwidth]{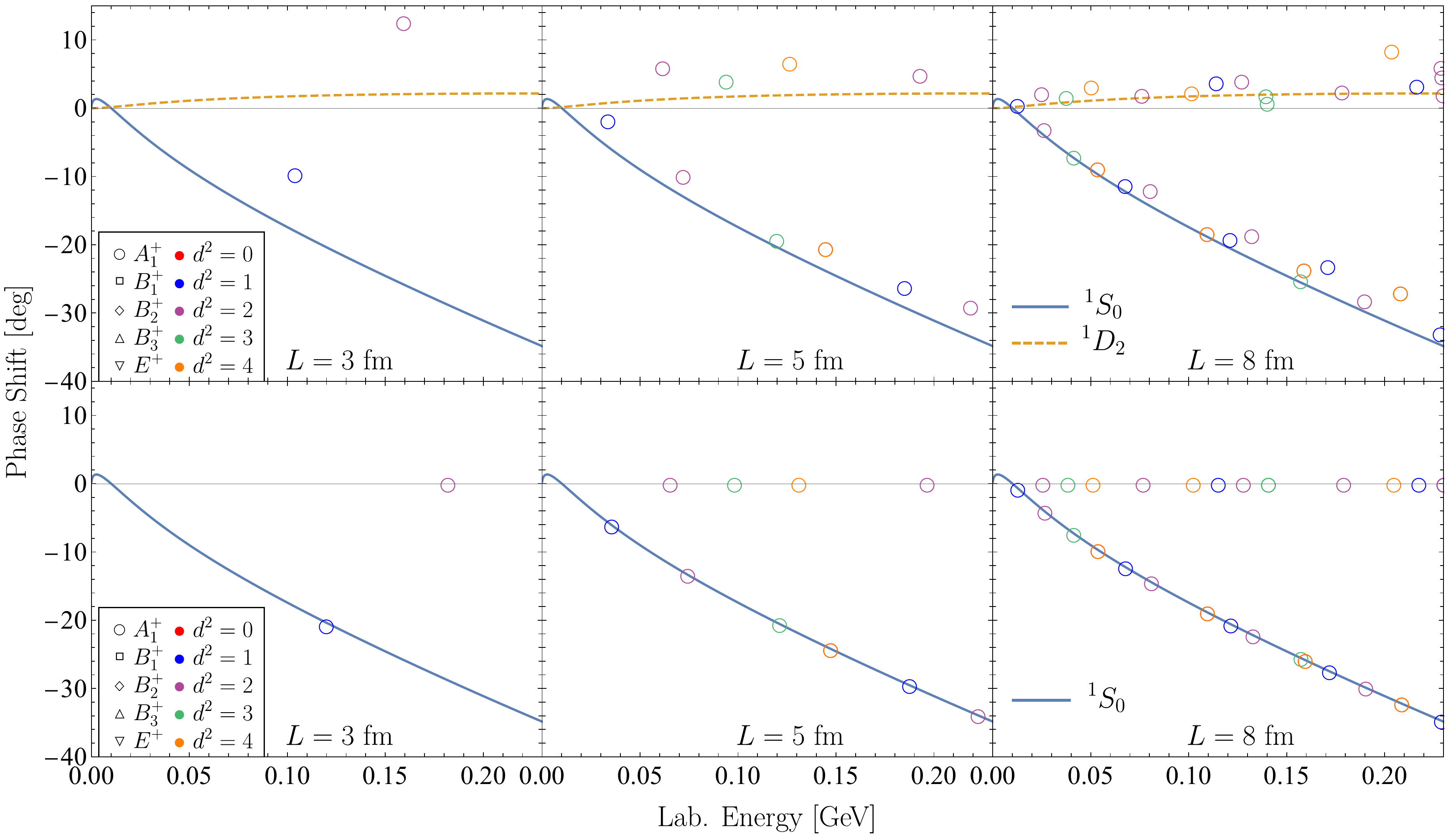}
	\caption{\label{fig:opes} Phase shifts in the positive-parity channels extracted from the FV energy spectra using the S-wave L\"uscher formula (various symbols)  in comparison with the infinite-volume results for the OPE potential (upper row) and the S-wave projected OPE potential
          (lower row). Solid and dashed lines show the  $^1$S$_0$ and  $^1$D$_2$ phase shifts calculated in the infinite volume, respectively. For remaining notations see figure~\ref{fig:ct}. }
\end{figure}

\begin{figure}[tbp]
	\centering 
	\includegraphics[width=.98\textwidth]{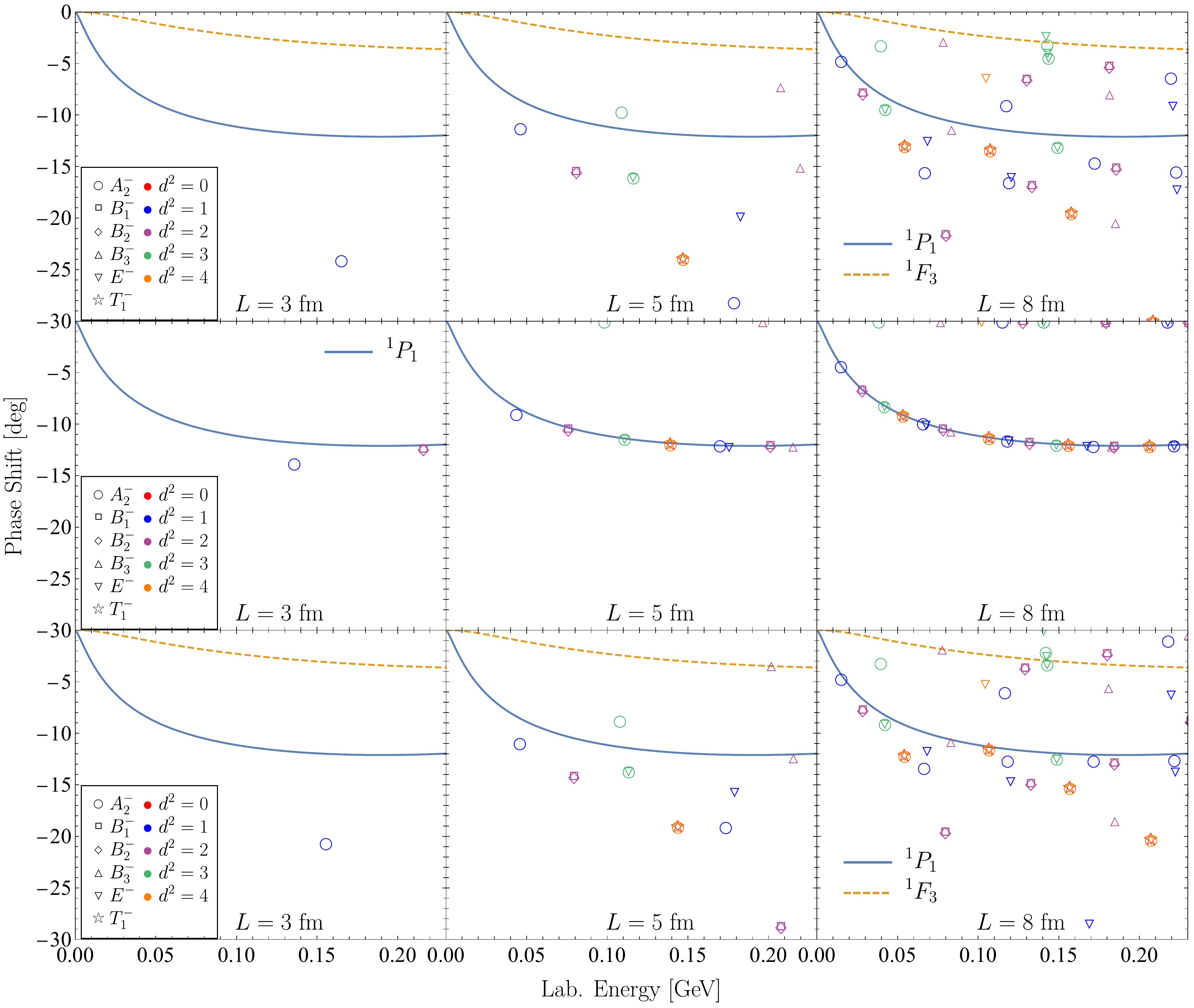}
	\caption{\label{fig:opep} Phase shifts in the negative-parity channels extracted from the FV energy spectra using the P-wave L\"uscher formula (various symbols)  in comparison with the infinite-volume results for the OPE potential (upper row),  the P-wave projected OPE potential
          (middle row) and the  P- and F-wave projected OPE potential. Solid and dashed lines show the  $^1$P$_1$ and  $^1$F$_3$ phase shifts calculated in the infinite volume. For remaining notations see figure~\ref{fig:ct}. }
\end{figure}

In the upper row of figures~\ref{fig:opes} and \ref{fig:opep}, we show the corresponding positive- and negative-parity phase shifts
extracted from the FV energies using the single-channel L\"uscher formula, along with the results in the infinite volume.
The deviations in the obtained phase shifts are qualitatively similar to those found in the calculations based on the full NNLO chiral potential.
For S-wave dominated states, the single-channel L\"uscher formula gives rise to a significant deviation for the $L=3$~fm box,
but it works reasonably well for $L\geq 5$~fm. This is in line with the findings of ref.~\cite{Sato:2007ms}.
For P-wave dominated states, we again observe large deviations which persist even for the box size of $L=8$~fm.
Meanwhile, the deviations are found to increase with energies.

We stress again that increasing the box size does not allow one to improve the results of the single-channel L\"uscher formula at higher energies.
For example, focusing on the energy levels around $0.15$~GeV in the first row of figure~\ref{fig:opep}, one sees no obvious
improvement of the results using the single-channel L\"uscher method with increasing $L$. If one, however, considers the trajectory of a single
state such as e.g.~ the ground state of the $A_2^-$ irrep in the $\bm{d}^2=1$ system as a function of the  box size,
one does observe a significant improvement when it approaches the threshold region for a sufficiently large $L$.

To unambiguously identify partial wave mixing effects as the source of failure of the single-channel L\"uscher formula,
we consider the S-wave and P-wave projected OPE interaction.
Specifically, we start with the partial wave expansion of the potential
\begin{equation}
V(\bm{p}, \bm{p}') = \sum_{l}\frac{2l+1}{4\pi}V_{l}(p,p')P_{l}(z),\label{eq:pw_dcomp}
\end{equation}
where $z=\cos \theta$ with $\theta$ being the angle between the initial and final momenta and $P_l(z)$ denotes the Legendre polynomial,
and define the potentials $V_{\rm S-wave}(\bm{p}, \bm{p}') = (4 \pi)^{-1} V_{0}(p,p')P_{0}(z) $ and
$V_{\rm P-wave}(\bm{p}, \bm{p}') = 3 (4 \pi)^{-1} V_{1}(p,p')P_{1}(z) $. Clearly, the resulting potentials
do not generate any partial wave mixing effects when used to compute the FV energy spectra, so that the
single-channel L\"uscher approach is expected to become applicable. This is indeed fully consistent with our
results, see the second rows in figures~\ref{fig:opes} and \ref{fig:opep}.
For S-wave dominated states, the small deviation disappears after switching off the interaction in higher partial waves.
For states with negative parity, the change is more pronounced. After switching off the interaction in  higher partial waves,
the P-wave phase shift is reproduced accurately with the single-channel L\"uscher formula except for the smallest box size (presumably due to 
exponentially suppressed corrections).

Finally, the lower row in figure~\ref{fig:opep} shows the effect of including the F-wave interaction,
i.e.~we consider
$V_{\rm P,F-wave}(\bm{p}, \bm{p}') = 3 (4 \pi)^{-1} V_{1}(p,p')P_{1}(z) +  7 (4 \pi)^{-1} V_{3}(p,p')P_{3}(z) $. 
While we have not explicitly investigated the impact of even higher partial waves, a comparison of the
upper and lower rows in  figure~\ref{fig:opep} suggests that their mixing effects may also be significant at higher energies.

\subsection{Phase shifts from FV energies using EFT}

In section \ref{Luescher}, we have shown in detail that partial wave mixing effects can \emph{not} be neglected when calculating FV energies
of two interacting nucleons at the physical pion masses. Such large partial wave mixing effects are caused by
the long-range (pion-exchange) interaction, and they are responsible for the failure of the single-channel L\"uscher
approach to extract phase shifts from FV energy spectra. This will pose a significant challenge for future
lattice QCD calculations in the NN sector close to the physical point. In the lattice QCD community, the issue is addressed by including several partial waves in the L\"uscher's quantization conditions, see e.g.~refs.~\cite{Dudek:2012gj,Cheung:2020mql}. When more than one partial wave is included, there is no longer a one-to-one mapping between energy levels and phase shifts, and one has to choose a theoretical framework to parameterize the $T$-matrix.  As an alternative to the L\"uscher method,
one can benefit from the known model-independent OPE interaction using an EFT-inspired approach. Specifically,
we propose to determine the short-range part of the NN interaction, parametrized in a systematic way by means
of contact interactions, via a direct matching to lattice QCD FV energy levels. The method is, to some extent, similar
to the low-energy theorems used in refs.~\cite{Baru:2015ira,Baru:2016evv} to restore the energy dependence of
the NN scattering amplitude at unphysical pion masses.

\begin{figure}[tbp]
	\centering 
	\includegraphics[width=0.98\textwidth]{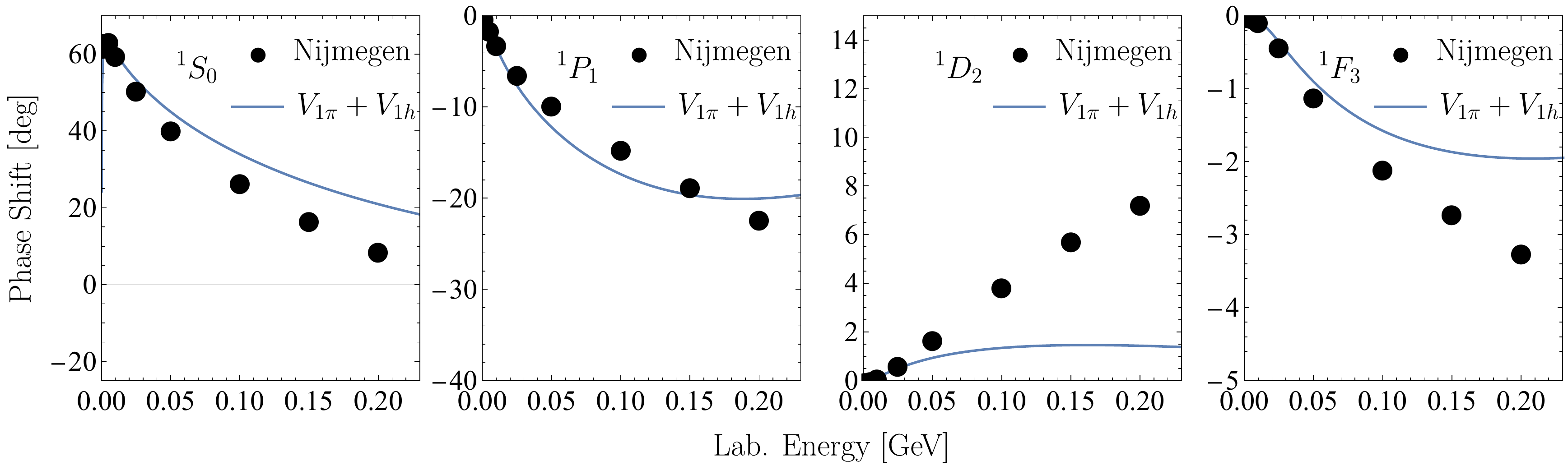}\\
	\caption{\label{fig:toyIV} $S$, $P$, $D$ and F-wave phase shifts for the toy-model potential in eq.~(\ref{eq:pw_dcomp}) in
          comparison with the corresponding spin-singlet NN phase shifts from the Nijmegen PWA \cite{Stoks:1993tb}.  }
\end{figure}

\begin{figure}[tbp]
	\centering 
	\includegraphics[width=\textwidth]{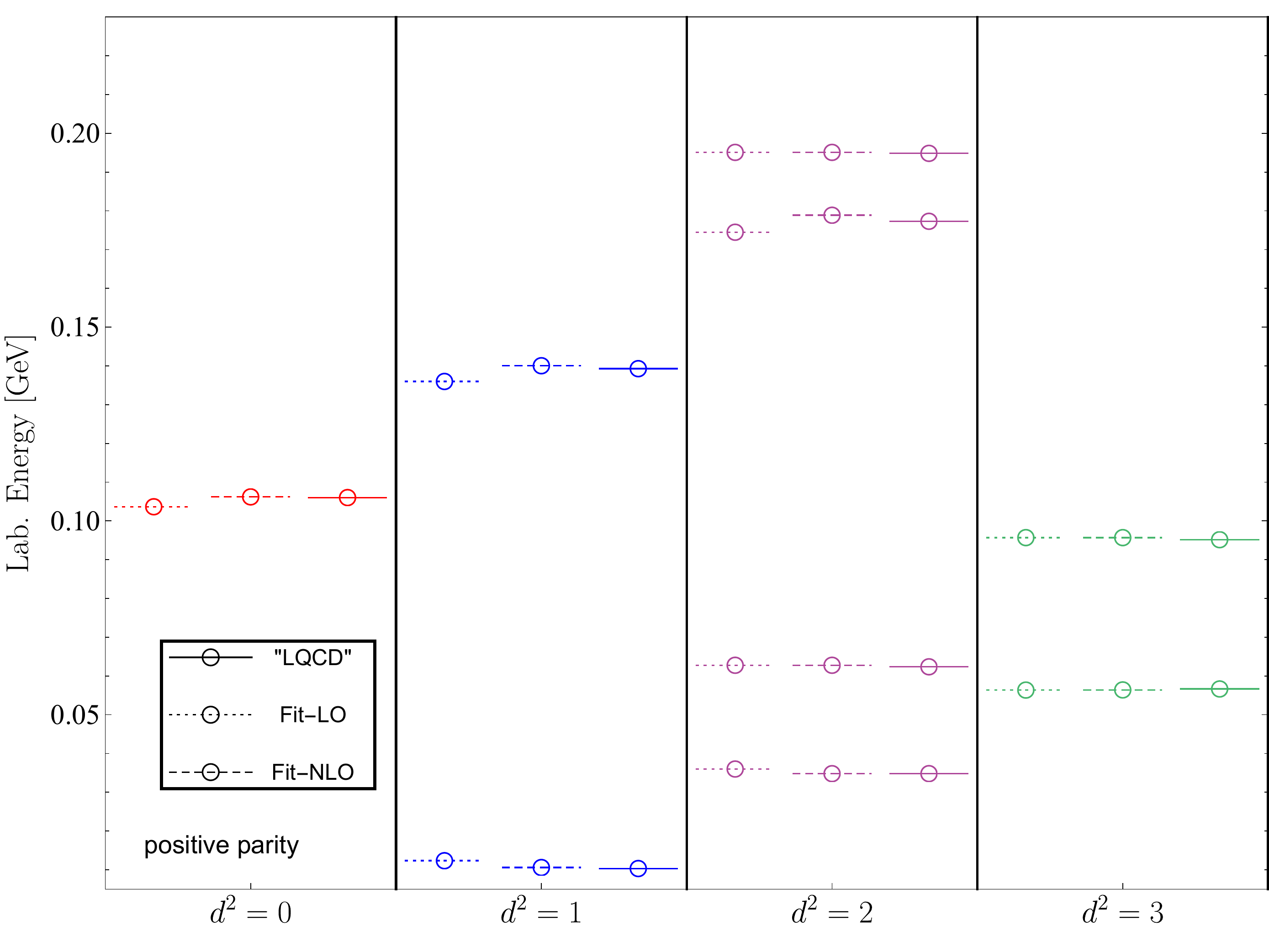}
	\caption{\label{fig:toyelp} Comparisons of the synthetic lattice energy levels with positive parity in the box with $L=5$ fm and those from the EFT determined by fitting. The marker shapes represent the irreps, which are the same as those in figure~\ref{fig:ct}.   }
\end{figure}

\begin{figure}[tbp]
	\centering 
	\includegraphics[width=\textwidth]{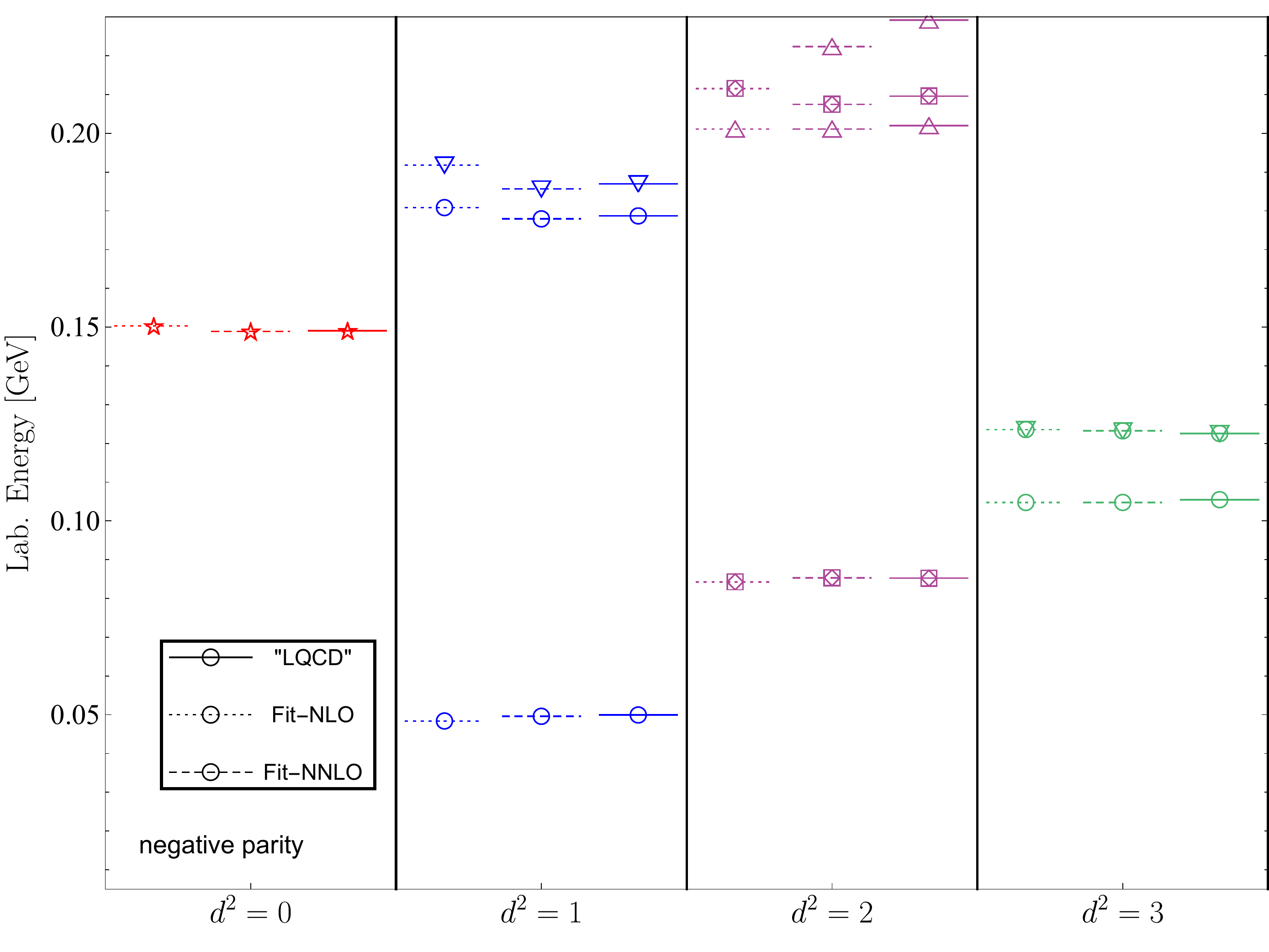}
	\caption{\label{fig:toyelm} Comparison of the synthetic lattice energy levels with negative parity in the box with $L=5$ fm and those from the EFT determined by fitting. The marker shapes represent the irreps, which are the same as those in figure~\ref{fig:ct}.   }
\end{figure}

To illustrate the method, we first define  a toy model comprising the OPE  and the heavy-meson-exchange potentials
to generate synthetic data for the FV energies. Specifically, we consider 
\begin{equation}	V_\text{toy}=V_{1\pi}+V_{1h}=-\left(\frac{g_{A}}{2F_{\pi}}\right)^{2}\frac{M_\pi^{2}}{\bm{q}^{2}+M_\pi^{2}}\bm{\tau}_{1}\cdot\bm{\tau}_{2}+\left(c_{h1}+c_{h2}\bm{\tau}_{1}\cdot\bm{\tau}_{2}\right){1\over \bm{q^2}+m_h^2},\label{eq:vtoy}
	\end{equation}
        where $V_{1\pi}$ is the OPE potential considered in the previous sections (up to an S-wave contact interaction).  For the various parameters, we choose the numerical values of $M_\pi=139$~MeV, $F_{\pi}=92.4$~MeV and $g_{A}=1.26$.  For the heavy-meson-exchange interaction, we introduce both the isospin-triplet and isospin-singlet potentials with the same meson mass $m_{h}=0.5$~GeV. Further, 
        $c_{h1}$ and $c_{h2}$ denote the corresponding dimensionless  coupling constants. In order to regularize the UV divergences, we introduce a nonlocal Gaussian cutoff
\begin{equation}
	V_\text{toy}\; \to \;  V_\text{toy}\; e^{-\frac{p^{2}+p'^{2}}{\Lambda^{2}}} , \label{eq:rgl}
\end{equation}
and choose $\Lambda=0.45$~GeV. The couplings $c_{h1}$ and $c_{h2}$ are adjusted in such a way that the toy-model interaction mimics
the behavior of the NN $^1$S$_0$ and $^1$P$_1$ phase shifts as shown in figure~\ref{fig:toyIV}.
Using the toy model introduced above, we compute the corresponding FV energy levels, which are  regarded as synthetic lattice data,
see the right-most symbols  in figures~\ref{fig:toyelp} and \ref{fig:toyelm} for a given $\bm{d}^2$-value, where the results are only shown for the box
with $L=5$~fm.

Having generated the synthetic data as described before, we are now in the position to describe our
approach for extracting the corresponding phase shifts. To this aim,  we exploit the knowledge of the long-range
interaction in the underlying model to construct the EFT interaction
\begin{equation}
V_{\text{EFT}}	=V_{\text{OPE}}^{(0)}+V_{\text{cont}}^{(0)}+V_{\text{cont}}^{(2)}+V_{\text{cont}}^{(4)}+...\label{eq:veft}
\end{equation}
We use the same regulator as in eq.~\eqref{eq:rgl} and employ the same value for the cutoff parameter $\Lambda=0.45$ GeV. In order to fit the FV energy levels with positive parity, we introduce the interaction at NLO,
\begin{equation}
	V_{\text{cont}}^{(0)}=\frac{1}{4\pi}\tilde{C}_{^{1}S_{0}},\quad V_{\text{cont}}^{(2)}=\frac{1}{4\pi}C_{^{1}S_{0}}(p^{2}+p'^{2})~\label{eq:vctcs}.
\end{equation}
where $\tilde{C}_{^{1}S_{0}}$ and $C_{^{1}S_{0}}$ refer to the corresponding LECs. For the states with negative parity, we include the contact interactions
up to NNLO, 
\begin{equation}
V_{\text{cont}}^{(2)}(p,p',z)=\frac{3}{4\pi}C_{^{1}P_{1}}pp'z,\quad V_{\text{cont}}^{(4)}(p,p',z)=\frac{3}{4\pi}D_{^{1}P_{1}}pp'(p^{2}+p'^{2})z~\label{eq:vctcp},
\end{equation}
where  $C_{^{1}P_{1}}$ and $D_{^{1}P_{1}}$ are the LECs. Since the contact interactions introduced above only contribute to the S- and P-wave channels,
 the FV partial wave mixing effects at the considered EFT order only arise from the OPE interaction. 

 To fix the LECs  $\tilde{C}_{^{1}S_{0}}$, $C_{^{1}S_{0}}$, $C_{^{1}P_{1}}$ and $D_{^{1}P_{1}}$, we employ the determinant residual method.
 For the considered toy-model example, we neglect the uncertainties of the synthetic  data.
 First, we perform single-parameter fits by including only the dominant contact interaction in the corresponding parity channel, i.e.~at
 LO (NLO) for positive- (negative-) parity states.
 In the second step,  we also take into account  the corresponding subdominant contact terms and perform two-parameter fits to the FV energies.
 As for the synthetic data, we only include  the ground state energy of each irrep as input. Meanwhile, we ignore the energy levels of the
 $\bm{d}=(0,0,2)$ systems because they are identical to those of the $\bm{d}=(0,0,0)$ system in the non-relativistic case. For positive-parity
 channels, this leaves us with three and four energy levels for the boxes with $L=3$ and $5$~fm, respectively, up to the maximal energy of 0.23 GeV. For the negative-parity case, we have one and eight\footnote{This number includes degenerate energy levels due to the non-relativistic nature of the considered system as explained in section \ref{sec:p-dis}.} energy levels for the boxes with $L=3$ and $5$~fm, respectively.
 Thus, for the smallest considered box size, our calculations for the negative-parity states are restricted to
 NLO in the EFT expansion due to the lack of input information.

 In figures~\ref{fig:toyelp} and \ref{fig:toyelm}, we show the quality of the reproduction of the FV energy levels.
 The results from a single-parameter fit using the dominant contact interactions are shown by the dotted lines
 and the leftmost symbols for the considered $\bm{d}^2$ value, while the dashed lines with symbols in the middle  
 correspond to two-parameter fits including the contributions of the subdominant contact terms. As expected, one observes 
 a clear improvement in the description of the energy levels by including the subdominant short-range interactions. The
 largest deviations from the synthetic data are observed for higher-energy states, namely for the third $A_1^+$ state and
 the second $B_3^-$ state in the $\bm{d}^2=2$ box (which have not been used in the fits). This pattern is expected and
 reflects a slower convergence of the EFT at higher energies. 
 
\begin{figure}[tbp]
	\centering 
	\includegraphics[width=.98\textwidth]{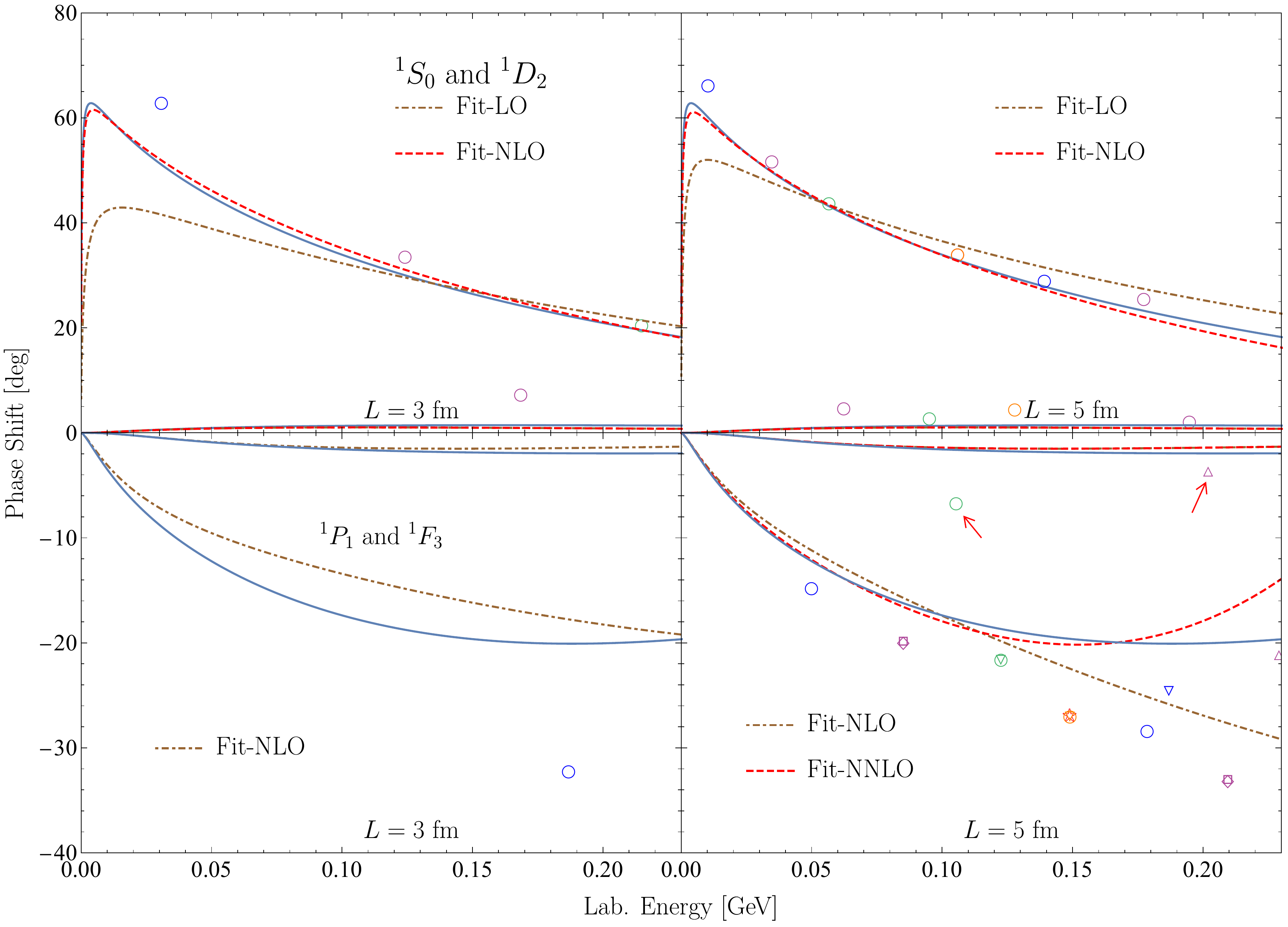}
	\caption{\label{fig:toy} $^1$S$_0$ (upper row) and $^1$P$_1$ (lower row) phase shifts extracted by matching the EFT to the finite-volume spectra for the toy-model example in comparison with the underlying phase shifts shown by the solid blue lines. Brown dashed-dotted (red dashed) lines show the results from single-parameter  (two-parameter)
          fits to the FV energies as visualized in figures ~\ref{fig:toyelp} and \ref{fig:toyelm}. Left and right panels correspond to the boxes of the size $L=3$~fm and  $L=5$~fm, respectively.
          Also shown by various symbols are the phase shifts extracted from the FV energies using the single-channel L\"uscher formula. The lines near $\delta_l \sim 0$ show the phase shifts in the next-higher partial wave for a given parity quantum number (i.e.~in the $^1$D$_2$ and $^1$F$_3$ channels in the upper and lower row, respectively). For remaining notations see  figure~\ref{fig:ct}. }
\end{figure}

Having determined the values of the LECs $\tilde{C}_{^{1}S_{0}}$, $C_{^{1}S_{0}}$, $C_{^{1}P_{1}}$ and $D_{^{1}P_{1}}$ from
the FV spectra as discussed below, we are now in the position to compute the phase shifts. This is achieved by solving
the partial wave LSE in the infinite volume using the standard methods. In figure~\ref{fig:toy}, we compare the phase shifts
resulting from matching the EFT in the finite volume with the ones from the underlying model. We also show in this
figure the phase shifts extracted using the single-channel L\"uscher formula.\footnote{As seen in figure~\ref{fig:toy},
  the phase shifts obtained using the single-channel L\"uscher approach for the considered toy model show a rather similar behavior to the
  more realistic case discussed in section \ref{Luescher}. This is to be expected since partial wave mixing effects are dominated by the
  longest-range interaction, which is the same in all considered cases.} For the $^1$S$_0$ channel, the one-parameter fit at LO only qualitatively captures
the behavior of the underlying phase shift.  This can be attributed to the strongly fine-tuned nature of the interaction in this channel as reflected by
the very large absolute value of the scattering length and to the well-known important role played by the range corrections. The fit results at NLO  are
strongly improved, and the underlying phase shifts are correctly reproduced even using the smallest considered box size of $L=3$~fm.
For the $^1$P$_1$ channel, we only have a single data point for the smallest box at our disposal. Therefore, we can only perform
a single-parameter fit for the $L=3$~fm box. Using the single channel L\"uscher formula, the resulting phase shift deviates strongly from
the underlying one, thus pointing towards a significant contribution of higher partial waves to this energy level.
Nevertheless,  in the approach we propose, the one-parameter fit to this energy level already allows one to capture
the qualitative behavior of the phase shift in this channel. In the larger box with $L=5$~fm, we have more synthetic data
at our disposal. The lower interacting energies as compared with the smaller box size allow for a more accurate determination
of the LEC $C_{^1P_1}$ in the NLO fit, which results in an improved description of the phase shifts at low energies as
compared with the one using the smaller box of $L=3$~fm. Including the subdominant contact interactions
$\propto C_{^1S_0},  \, D_{^1D_2}$, the description of both the $^1$S$_0$ and $^1$P$_1$ phase shifts improves considerably
in a wide energy range. It should be emphasized that our input includes several energy levels dominated by higher partial wave components
such as the ground state in the  $A_2^-$ irrep  with $\bm{d}^2=3$ and the ground state in the  $B_3^-$ irrep  with $\bm{d}^2=2$.
Contrary to the single-channel L\"uscher approach, which in these cases leads to large deviations, our method is insensitive
to such FV partial wave mixing artifacts.

\section{Application II: P-wave pion-pion scattering}
\label{sec:pipiSystem}

We now turn to our second application and consider $\pi \pi$ scattering as an example of a relativistic system.
In this exploratory example, we employ a simple phenomenological model for $\pi \pi$ interaction instead of using
EFT. For a formulation of chiral perturbation theory with resonances see e.g.~refs.~\cite{Bernard:1991zc,Bruns:2004tj}. 

\subsection{Reduced Bethe-Salpeter equation in the finite volume}\label{subsec:bseFV}
The relativistic Bethe-Salpeter equation is a four-dimensional integral equation. One has to reduce it into three-dimensional one to perform the plane wave expansion. In the literature, there are various approaches to achieve it~\cite{Woloshyn:1973mce,Kim:2005gf,Albaladejo:2012jr}. In this work, we choose the Blankenbecler-Sugar equation as discussed
e.g.~in ref.~\cite{Woloshyn:1973mce} as our starting point. One can analytically perform the integration over the $0$-th components of the relativistic two-body propagator as follows,
\begin{equation}
	\label{temp1}
	{\cal G}	=i\int\frac{d^{4}q}{(2\pi)^{4}}\frac{1}{(P-q)^{2}-M_\pi^{2}+i\epsilon}\frac{1}{q^{2}-M_\pi^{2}+i\epsilon}=i\int\frac{d^{3}\bm{q}}{(2\pi)^{3}}G(q;E),
\end{equation}
with 
\begin{equation}
	G(q,E)=\frac{1}{2\omega_{1}(q)\omega_{2}(q)}\frac{\omega_{1}(q)+\omega_{2}(q)}{E^{2}-[\omega_{1}(q)+\omega_{2}(q)]^{2}+i\epsilon},
\end{equation}
where $\omega_i(q)\equiv \sqrt{m_i^2+q^2}$ and $m_{1}=m_2=M_\pi$. 
The reduced BSE for the $\pi\pi$ scattering reads 
\begin{equation}
T(\bm{p},\bm{p}';E)=V(\bm{p},\bm{p}';E)+i\int_{q<q_\text{max}}\frac{d^{3}\bm{q}}{(2\pi)^{3}}V(\bm{p},\bm{q};E)G(q;E)T(\bm{q},\bm{p}';E).~\label{eq:bse}
\end{equation}
Notice that one can choose different reduced BSEs, but the procedure of performing the plane wave expansion is the same.

We expand the reduced BSE in the plane wave basis with discrete  momenta to obtain the matrix equation
\begin{equation}
	\mathbb{T}(E)=\mathbb{V}(E)+\mathbb{V}(E)\mathbb{G}(E)\mathbb{T}(E),
\end{equation}
where the discretized propagator $\mathbb{G}$ is defined as
\begin{equation}
	\mathbb{G}_{\bm{n},\bm{n}'}={\cal J}\frac{1}{L^{3}}G(q_{\bm{n}},E)\delta_{\bm{n}',\bm{n}}.
\end{equation}
Further, $\mathcal{J}$ is the Jacobi determinant arising from the transformation between the BF and CMF as shown in eq.~\eqref{eq:tofv}, which is given explicitly in eq.~\eqref{eq:jcob} for two different schemes. The FV energy levels are obtained by solving
the determinant equation
\begin{equation}
	\det\left[\mathbb{G}^{-1}(E)-\mathbb{V}(E)\right]=0.\label{eq:rel_det0}
\end{equation}

The procedure for dealing with relativistic systems is more demanding than that for non-relativistic ones since the effective interaction $V$ in the three-dimensionally reduced eq.~\eqref{eq:bse} is, in general, energy-dependent.
Meanwhile, the Jacobi determinant can introduce additional energy-dependence.
Such energy-dependence prevents one from reducing eq.~\eqref{eq:rel_det0} to the eigenvalue problem like in the non-relativistic case.
In a special case when the interaction can be assumed to be energy-independent and the scheme-II is employed to
relate the momenta in the BF and CMF as discussed in section \ref{subsec:intac}, leading to the  Jacobi determinant given in the second line of eq.~(\ref{eq:jcob}), 
equation~\eqref{eq:rel_det0} reduces to the eigenvalue problem. This case is considered e.g.~in refs.~\cite{Hall:2013qba,Wu:2014vma,Liu:2015ktc,Li:2021mob}. 

In this paper we do not make use of the above-mentioned simplifications and keep the energy dependence
of the interaction and the Jacobi determinant. We decompose eq.~\eqref{eq:rel_det0} into decoupled matrix equations corresponding
to different irreps. The FV energy levels are obtained by finding roots of the determinant equations for a given irrep $\Gamma$:
\begin{equation}
	\det\left[\mathbb{G}^{-1}_\Gamma(E)-\mathbb{V}_\Gamma (E)\right]=0.\label{eq:rel_det}
\end{equation}

\subsection{FV energy levels using a phenomenological model for P-wave $\pi\pi$ interaction}

We employ a phenomenological interaction model for P-wave $\pi\pi$ scattering rather than chiral perturbation theory.
Specifically, we parametrize the interaction via an energy-dependent potential 
\begin{equation}
	V(\bm{p},\bm{p}';E)=-\frac{2\bm{p}\cdot\bm{p}'}{f^{2}}\left(1+\frac{2G_{V}^{2}}{f^{2}}\frac{E^{2}}{M_{0}^2-E^{2}}\right),\label{eq:vpipi}
\end{equation}
where $f$, $M_0$ and $G_V$ are free parameters. We introduce the CDD (Castillejo, Dalitz, Dyson) pole in the interaction~\cite{Castillejo:1955ed}, which naturally appears in the description of the vector form factor of the $\pi\pi$ system~\cite{Oller:2000ug} based on the Lagrangians from  refs.~\cite{Gasser:1984gg,Ecker:1988te}. In this formalism, we neglect possible coupled-channel effects from the $K\bar{K}$ system. The interaction in eq.~\eqref{eq:vpipi} corresponds to a short-range contact potential. The angular-dependent term $\bm{p}\cdot \bm{p}'$ ensures that the interaction can only contribute to the P-wave, which switches off possible partial wave mixing effects in the finite volume. Thus, for such a short-range potential 
contributing to the P-wave alone, the single-channel L\"uscher formula is perfectly applicable. The potential in the partial wave basis reads
\begin{equation}
	V^{l=1}(p,p';E)=-4\pi\frac{2pp'}{3f^{2}}\left(1+\frac{2G_{V}^{2}}{f^{2}}\frac{E^{2}}{M_{0}^2-E^{2}}\right).
\end{equation}  
Notice that if we would make an on-shell approximation, we would obtain the interaction of ref.~\cite{Chen:2012rp} (up to the $4\pi$-factor), which was used to investigate finite volume effects in the same channel within the chiral unitary approach. In our normalization, the phase shift can be extracted from the partial wave $T$-matrix as
\begin{equation}
	T^{l}(E)=\frac{-32\pi^{2}E}{p\cot\delta^l-ip}.
\end{equation}

\begin{figure}[tbp]
	\centering 
	\includegraphics[width=.53\textwidth]{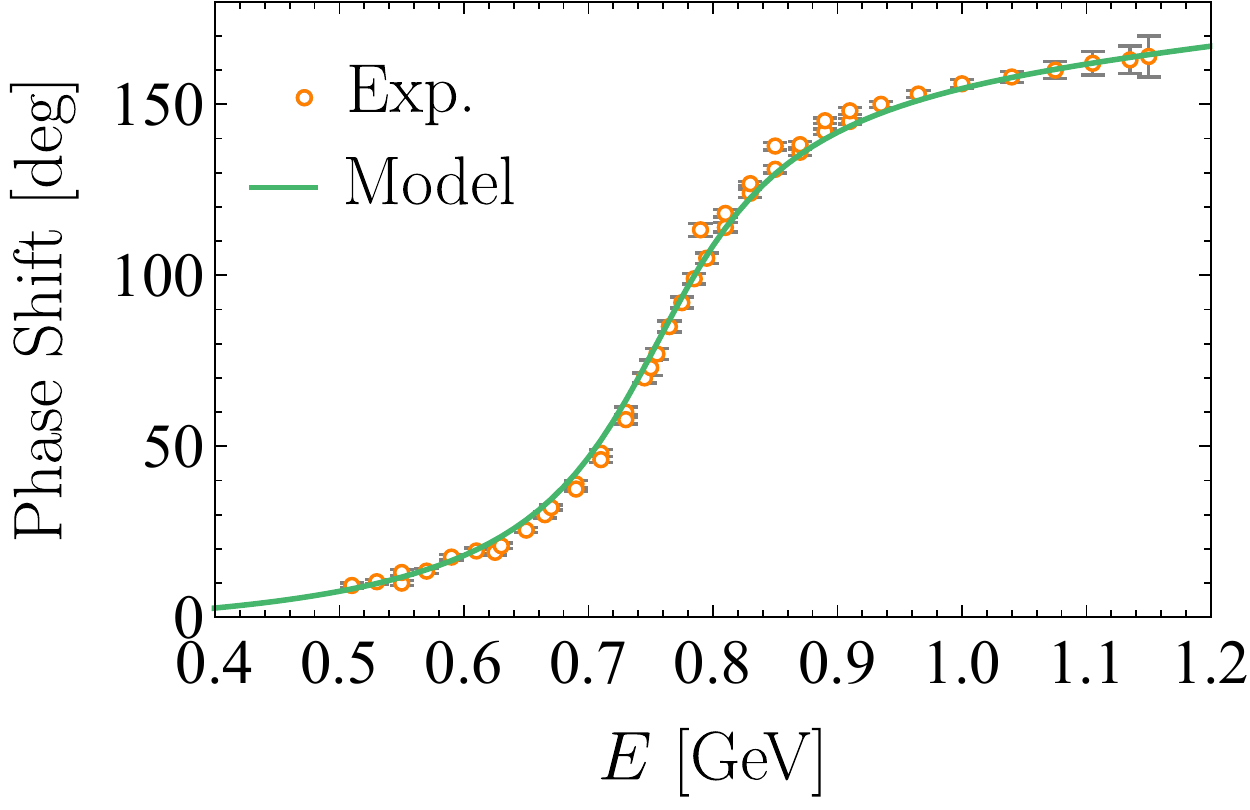}
	\caption{\label{fig:exmod} P-wave $\pi\pi$ phase shifts of the phenomenological model defined in eq.~\eqref{eq:vpipi}  (solid line) in comparison with the experimental data from refs.~\cite{Estabrooks:1974vu,Protopopescu:1973sh} (open dots).  }
\end{figure}

Recently, lattice simulations of $\pi\pi$ scattering in the $\rho$-channel at the physical value of the pion mass were performed~\cite{Fischer:2020fvl}. In our calculation, we choose the same box size of $L=4.3872$~fm and pion mass of $M_\pi=132$~MeV as used
in that work. Since the root-finding algorithm is very time-consuming,  we choose a smaller cutoff $q_\text{max}=1.5$~GeV as compared to
the one employed in ref.~\cite{Chen:2012rp}. The three parameters $f$, $M_0$ and $G_V$ are determined by fitting the experimental $\pi\pi$ scattering phase shift~\cite{Estabrooks:1974vu,Protopopescu:1973sh} leading to the values of 
\begin{equation}
f=0.124\text{ GeV},\quad \quad G_V=0.059\text{ GeV},\quad \quad M_0=1.110 \text{ GeV}.
\end{equation}
In figure~\ref{fig:exmod}, we show that the resulting phase shifts from our model provide a very good description of the experimental ones. 

\begin{figure}[tbp]
	\centering 
	\includegraphics[width=.43\textwidth]{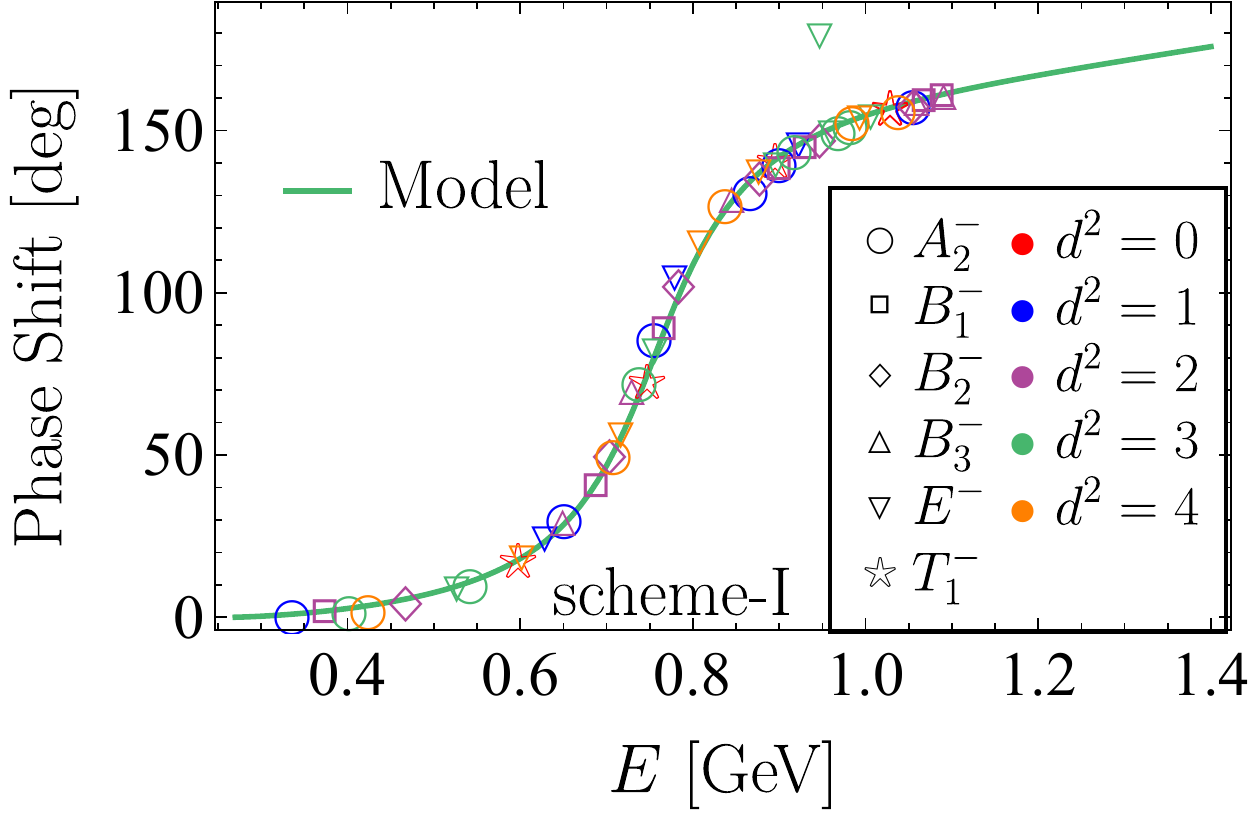}
	\includegraphics[width=.43\textwidth]{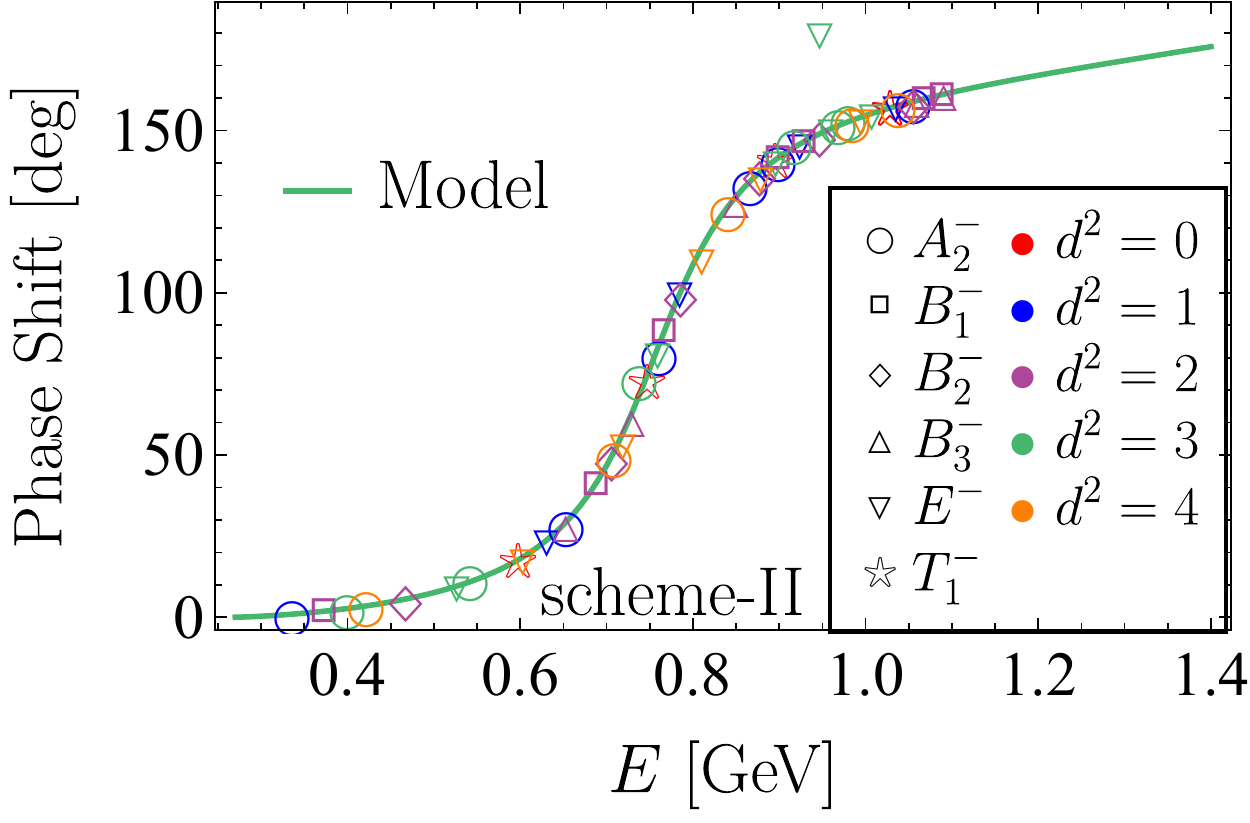}
	\caption{\label{fig:ellsch} P-wave $\pi\pi$ phase shifts of the phenomenological model defined in eq.~\eqref{eq:vpipi}  (solid lines) in comparison with the phase shifts extracted from the FV energies using the single-channel L\"uscher formula (various symbols). Left and right panels correspond to two different schemes for discretizing the momenta in the moving frames as explained in section \ref{subsec:intac}. }
\end{figure}

\begin{figure}[tbp]
	\centering 
	\includegraphics[width=\textwidth]{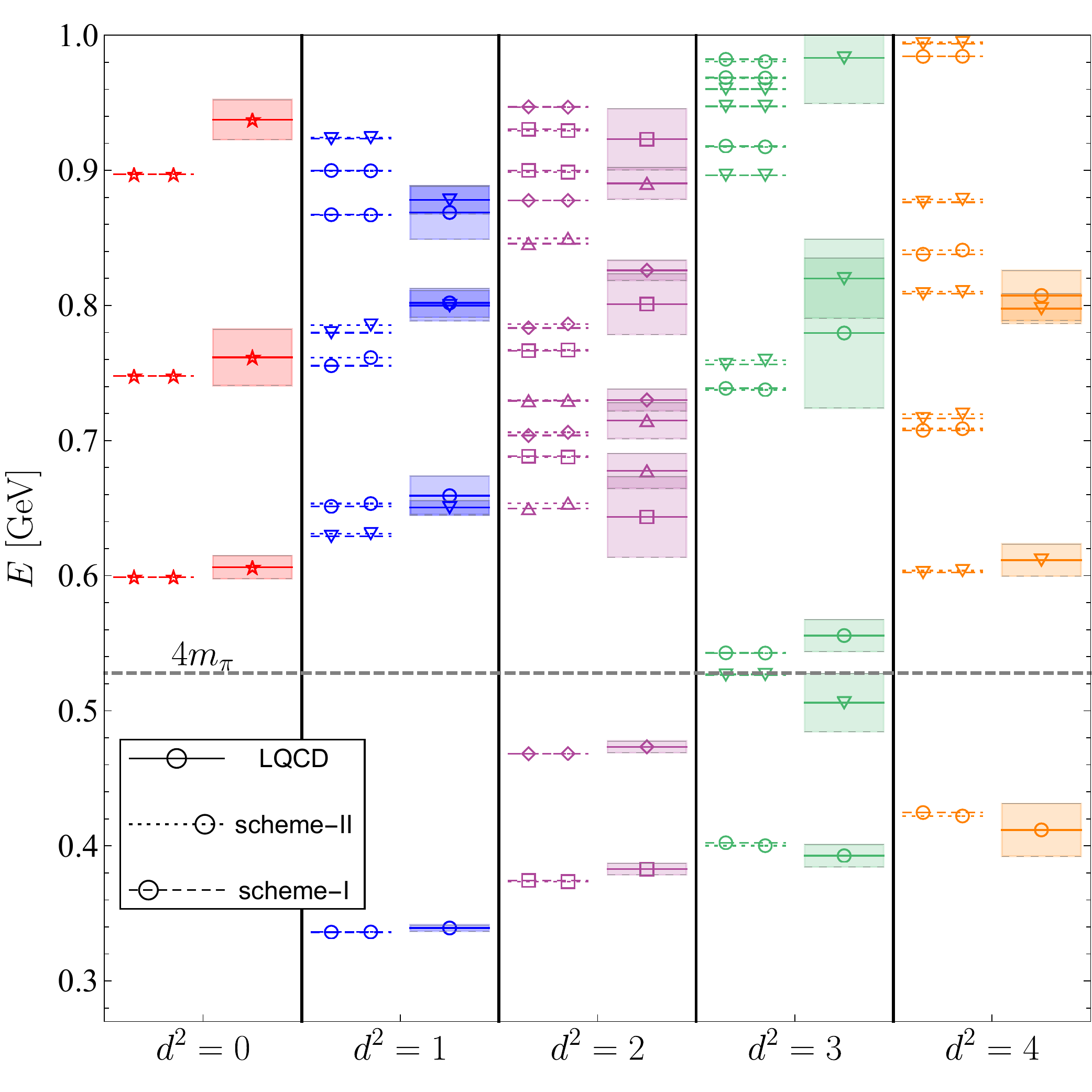}
	\caption{\label{fig:fve} Comparison of the FV energy levels from lattice QCD  and our calculations using the two different schemes for discretizing the momenta in the moving frames as explained in section \ref{subsec:intac}. The notation for various symbols is the same as those in figure~\ref{fig:ellsch}. The uncertainties of the lattice QCD data are taken from ref.~\cite{data}. }
\end{figure}

Using the constructed model, we compute the FV energy levels of the $\pi\pi$ system in two different schemes to discretize the momenta as discussed in subsection~\ref{subsec:intac}. We then extract the phase shifts corresponding to these FV energy levels using the single-channel L\"uscher formula as shown in figure~\ref{fig:ellsch}.  As expected, for the considered short-range interaction without partial waving mixing,  the single-channel
L\"uscher formula leads to accurate results, which holds true for both considered discretization schemes. The outlier points with phase shifts about 180 degrees correspond to the non-interacting higher partial wave components. Further,  in figure~\ref{fig:fve} we compare our calculated FV energy levels with those from lattice QCD simulations~\cite{Fischer:2020fvl,data}. The differences of the FV energy levels from two discretization schemes appear to be tiny. Meanwhile, one can see the clear correspondence between our calculation and lattice QCD results for the low-lying energy levels in each irrep.
For states below the inelastic threshold $4M_\pi$, our results are close to the lattice QCD ones. 

\subsection{P-wave $\pi\pi$ scattering from lattice QCD data for FV energies}

We are now in the position to apply our approach for extracting the phase shifts from the FV spectra by
fixing the model parameters $f$, $G_V$ and $M_0$ directly from the lattice QCD
energy levels. To this end, we adopt the determinant residual method~\cite{Morningstar:2017spu} and  use 
the six energy levels below the $4M_\pi$ threshold from the lattice QCD simulations of ref.~\cite{Fischer:2020fvl}
shown in figure~\ref{fig:fve}.  

We define the $\chi^2$-function as in eq.~\eqref{eq:chi2}. To calculate the uncertainties $\sigma[\Omega_{\Gamma}(E_{\Gamma,i})]$,
one should propagate the errors of the energy levels from lattice QCD to the residual function $\Omega_{\Gamma}(E_{\Gamma,i})$, which is a tedious and
time-consuming task. In our exploratory calculation we neglect possible correlations among the lattice data, which will have to be taken into account for more precise determinations of the phase shifts in future studies,
and make an approximation
\begin{equation}
\sigma[\Omega_{\Gamma}(E_{\Gamma,i})]\approx\frac{1}{2}\left|\Omega(E_{\Gamma,i}^{\text{upper}})-\Omega(E_{\Gamma,i}^{\text{lower}})\right|,
\end{equation}
where $E_{\Gamma,i}^{\text{upper}}$ and $E_{\Gamma,i}^{\text{lower}}$ are the upper and lower limits of the energy levels from the lattice data. We use the discretization scheme-I to fit the lattice QCD energy levels.
We minimize the $\chi^2$-function using the package MINUIT~\cite{James:1994vla} and obtain
\begin{equation}
 f=118.6\pm8.3\text{ MeV}, \quad \quad M_0=1230.8\pm94.0 \text{ MeV}, \quad \quad  G_V=53.3\pm5.4 \text{ MeV}.
  \label{ModelParam}
\end{equation}

\begin{figure}[tbp]
	\centering 
	\includegraphics[width=.53\textwidth]{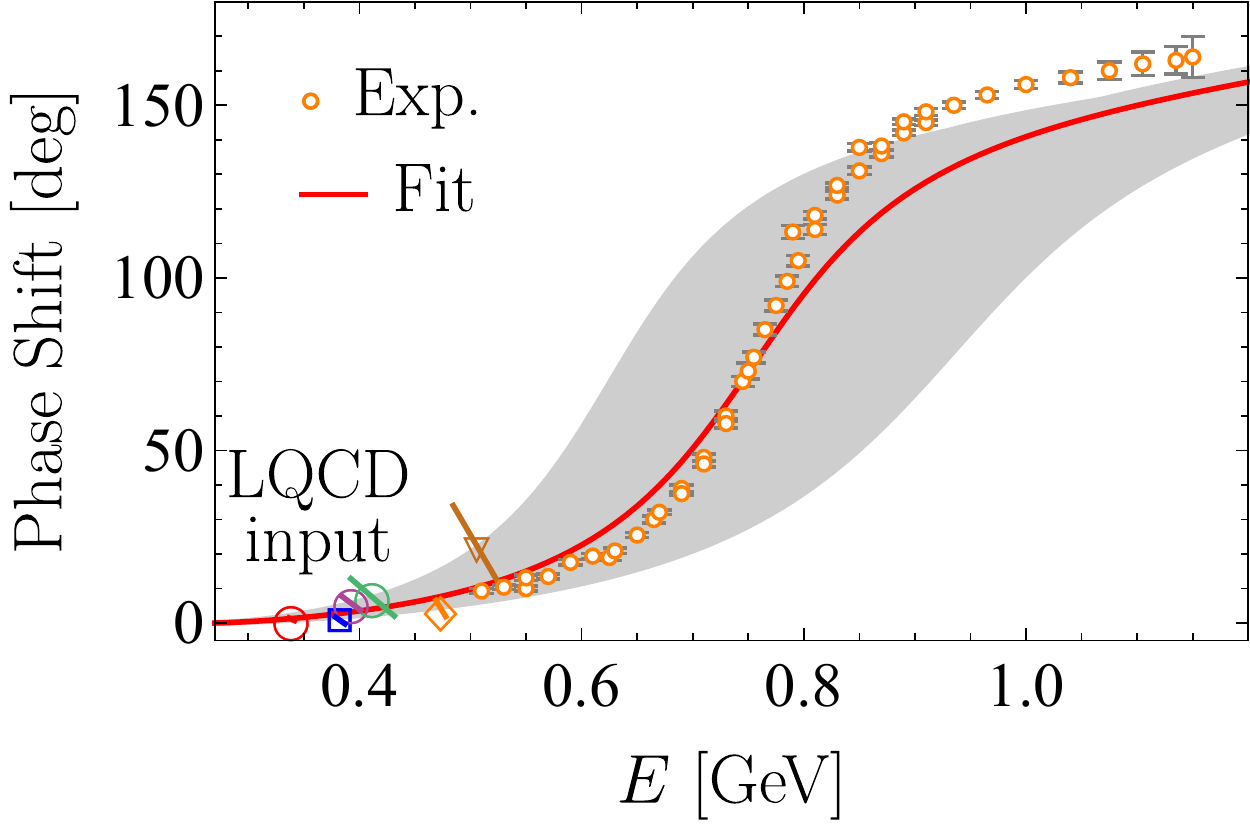}
	\caption{\label{fig:fit} P-wave $\pi\pi$ phase shifts extracted by matching the phenomenological model to the 
          FV energy levels  below the first inelastic channel computed in lattice QCD \cite{Fischer:2020fvl}. Experimental data are taken from refs.~\cite{Estabrooks:1974vu,Protopopescu:1973sh}. The red line shows the central result of the fit with the shaded area representing
          the fit uncertainty. The six lattice QCD energy levels with uncertainties are shown by various open symbols using the same notation
          as in figure~\ref{fig:ellsch}.}
\end{figure}

Having extracted the parameters of the model from the lattice QCD FV energy levels, we calculate the $\pi \pi$ P-wave
phase shifts by solving the reduced BSE for the $T$-matrix in the infinite volume. In figure ~\ref{fig:fit}, the resulting phase shifts
are shown, along with the uncertainty stemming from the errors in the extracted values of the model parameters in eq.~\eqref{ModelParam}. The pole mass and width extracted from our fitting results read,
	\begin{eqnarray}
		M=754^{+189}_{-195}\text{ MeV},\quad \Gamma=247^{+114}_{-107}\text{ MeV}.
	\end{eqnarray}
As a comparison, the mass and width obtained within the inverse amplitude method in ref.~\cite{Fischer:2020fvl} are 786(20) MeV and 180(6) MeV, respectively. Though the uncertainties of the extracted phase shifts, mass and widths appear to be large, owing to the employed simplistic model and the restriction
of the used lattice QCD data to energies below the first inelastic channel, these results confirm that our  method can
also be successfully applied to relativistic systems. 

\section{Summary and outlook}\label{sec:sum}

In this work, we propose an alternative approach to L\"uscher's method for extracting two-body scattering information from FV energy levels. We employ the plane wave basis instead of relying on the commonly used partial wave expansion.  Using the projection operator technique,  we reduce the discrete plane wave basis into a direct sum of irreducible representations of the corresponding discrete groups. The FV energy levels are computed by solving the LSE or BSE for nonrelativistic or relativistic systems, respectively, for a given irrep and finding the poles of the resulting amplitude. The formalism is applied to 
both static and moving two-particle systems in finite periodic boxes. Since we do not use the partial wave expansion, all partial wave mixing effects due to rotational symmetry breaking in a cubic box are naturally embedded.

We have used the above method to study spin-singlet NN scattering below pion production threshold as an example of a nonrelativistic system. Specifically, our goal was to study the impact of the long-range interaction due to the one-pion exchange on the severity of partial wave mixing effects when using the L\"uscher formula to relate scattering phase shifts and FV energy levels.
Throughout this study, we have restricted ourselves to the physical value of the pion mass. For S-wave dominated states, we found the single-channel L\"uscher formula
to lead to significant deviations for the smallest considered box size of $L=3$~fm. For larger boxes with $L \gtrsim 5$~fm,
the $^1$S$_0$ phase shift is, however, well reproduced for energies below $E_{\rm lab} \lesssim 150$~MeV, indicating that the contributions of the $^1$D$_2$ and higher partial waves to the corresponding FV energy levels are negligible.  On the other hand, we found severe partial wave mixing effects
for P-wave dominated states, which originate from the longest-range interaction due to the OPE and make the single-channel L\"uscher method inapplicable regardless of the considered box size (except for the near-threshold kinematics). While present-day lattice QCD calculations of the NN system are limited to unphysically large pion masses, this issue will pose a challenge once the lattice results will get closer to the physical point.   

A natural and efficient solution to this problem can be achieved using the EFT framework, as it allows one to take into account the
implications of long-range interactions in a systematic and model-independent way. To illustrate the method, we have considered a toy model
of the NN interaction comprising the long-range OPE accompanied by the heavy-meson exchange potentials, which reproduces the qualitative
behavior of the $^1$S$_0$ and $^1$P$_1$ neutron-proton phase shifts.  In the absence of lattice QCD results for the NN system near the physical point,
we used this model to generate synthetic data for FV energy levels. We then constructed the corresponding EFT featuring the same long-range interaction due to the OPE and involving contact terms with an increasing number of derivatives. Having determined the corresponding LECs
by fitting them to the  synthetic lattice QCD data, we have succeeded to reproduce the phase shifts of the underlying model using even the
smallest box of $L=3$~fm. All FV calculations are carried out using the discretized plane wave basis. The accuracy of the resulting phase shifts is systematically improvable by going to higher orders in the EFT expansion,
and the method is completely insensitive to partial wave mixing effects that plague the applications of the L\"uscher formula.
Our results suggest that a precise determination of spin-singlet NN phase shifts from lattice QCD
simulations at the physical point should be possible with box sizes no larger than $L \sim 4\ldots 5$~fm. The lower bound on the
box size in our approach is set by the requirement to have a sufficient number of resolved energy levels within the EFT applicability domain rather than
by the interaction range as it is the case for the L\"uscher approach. 

As a second application, we considered P-wave $\pi\pi$ scattering. Partial wave mixing effects are not expected to be an issue for the $\pi\pi$ system in the FV, and our main motivation here was to demonstrate the applicability of our method in a relativistic setting. 
We employed a phenomenological model to parametrize the $\pi\pi$ interaction instead of relying on EFT. The three adjustable parameters of the model were tuned to reproduce the experimental behavior of the  P-wave phase shift. We then computed the FV energy levels using the plane wave basis and
compared them with those from lattice QCD simulations for the same pion mass and box size, finding a good agreement for the low-lying states in every irrep.
We also applied our method to extract the phase shift by tuning the parameters of the model directly to the lattice QCD energy levels below the four-pion threshold. Though the uncertainties in the resulting determination are large, our central results agree well with the experimental $\pi\pi$ scattering phase shifts.

The considered examples provide a proof-of-principle that our approach can be successfully employed to extract the infinite-volume scattering information from FV energy levels accessible in lattice QCD. The essential ingredients of our method are (i) EFT or, possibly, phenomenological approaches to parametrize the interaction between the considered particles in a systematic way, (ii)  the plane-wave basis in momentum space to avoid complications caused by partial wave mixing in the FV and (iii) matching to the FV energy spectra computable in lattice QCD.  Compared with the conventional L\"uscher approach, our method offers the advantage of being insensitive to partial wave mixing artifacts in the FV energy levels, and it also allows one to considerably reduce the box size by taking into account exponentially suppressed corrections to the L\"uscher formula in the cases where the long-range interaction is known. Compared to alternative schemes discussed in literature such as the FV unitarized chiral perturbation theory for Goldstone boson scattering  \cite{Doring:2011vk,Doring:2012eu} and the effective Hamiltonian approach of refs.~\cite{Hall:2013qba,Wu:2014vma,Liu:2015ktc,Li:2021mob},
our method allows for a clean decomposition of two-particle states into irreps of the FV symmetry groups and does not rely on the partial wave expansion known to converge slowly in the presence of long-range interactions. 
We have also addressed the complication due to a possible energy dependence of the effective interactions. In that case,  one has to resort to the root-finding algorithm to obtain the poles of the $T$-matrix, which is computationally time consuming. We have explored an alternative technique to speed up  the calculations by using the determinant residual method when tuning the parameters of the interaction to the FV energies. 

The method contains two different ingredients, the EFT as the dynamics framework and plane wave expansion as the tool for numerical solution of FV energy levels. One can choose to combine one of them with other frameworks. For example, the plane wave expansion method in this work can be used to obtain the FV energy levels in the interaction of refs.~\cite{Wu:2014vma,Liu:2015ktc}. The EFT framework in this work can be used to parameterize the multi-channel L\"uscher quantization conditions. At the cost of more numerical effect, one might expect similar results to those in this work if the partial wave expansion is truncated such that the expansion of the amplitudes converges below a given energy. In the partial wave expansion, one should use the infinite-volume EFT framework to calculate NN phase shifts in different partial waves and then write down the
multi-channel L\"uscher equation containing all these partial waves. Using plane wave expansion, one can write down the FV quantization conditions directly without considering the scattering in infinite volume, which saves considerable effort. 

As already pointed out above, we expect our method to show largest benefits for lattice QCD studies of systems featuring long-range interactions
such as e.g.~nucleon-nucleon scattering with the pion mass near or even smaller than the physical value. In the past few years, chiral EFT for the two-nucleon
system has been developed into a precision tool, see e.g.~\cite{Reinert:2017usi,Reinert:2020mcu,Filin:2019eoe,Filin:2020tcs} for recent precision studies. In ref.~\cite{Reinert:2020mcu}, a full fledged partial wave analysis of the
available neutron-proton and proton-proton scattering data up to the pion production threshold has, for the first time, been performed
within the framework of chiral EFT, thereby achieving a statistically perfect description of the experimental data. However, away from the physical quark masses,
no experimental data are available, and one will have to rely on lattice QCD simulations, see \cite{NPLQCD:2011naw,Inoue:2011ai,Yamazaki:2012hi,Yamazaki:2015asa,Orginos:2015aya,Horz:2020zvv} for examples of lattice QCD studies in the NN sector and 
ref.~\cite{Davoudi:2020ngi} for a recent review article.
For heavy pion masses \cite{NPLQCD:2012mex,Yamazaki:2012hi}, lattice QCD results have already been matched to pionless EFT \cite{Barnea:2013uqa}, see also ref.~\cite{Drischler:2019xuo,Davoudi:2020ngi}. Our study provides a first step towards establishing an efficient and robust interface between lattice QCD and chiral EFT that will become necessary once the simulations at lower pion masses will become available. While we have restricted ourselves to spinless systems of two particles with equal masses in the present work, a generalization to nonzero spin and to particles with different masses in
elongated boxes is straightforward. Work along these lines is in progress.

\appendix
\section{Discrete groups}\label{app:pgrp}
In this appendix, we briefly describe the elements, conjugacy classes and characters of the $O_h$, $D_{4h}$, $D_{2h}$ and $D_{3d}$ groups needed in our analysis. For more details see textbooks about point groups, such as e.g.~ref.~\cite{dresselhaus2007group}. 
In table~\ref{tab:ele_o} the elements and conjugacy classes of the point group $O$ are presented. We label the $24$ rotation elements as $R_1,...,R_{24}$. The elements of $O_h$ can be obtained by using $O_h=O\otimes \{E,I\}$. The character tables of $O_h$ and $O$ are presented in tables~\ref{tab:chr_o} and ~\ref{tab:chr_oh}, respectively. The elements and characters of $D_4$, $D_2$ and $D_3$ groups are given in tables~\ref{tab:chr_d4}, \ref{tab:chr_d2} and \ref{tab:chr_d3}, respectively. The group elements and characters of  $D_{4h}$, $D_{2h}$ and $D_{3d}$ can be obtained easily.
\begin{table}[tbp]
	\centering
\begin{tabular}{|r|c|rr|rrr|r|}
	\hline 
	Class & $R_{i}$ & ${\bm n}$ & $\omega$ & $\alpha$ & $\beta$ & $\gamma$ & $|R_{i}\bm{p}\rangle$\tabularnewline
	\hline 
	$E$ & $R_{1}$ & any & $0$ & $0$ & $0$ & $0$ & $|p_{x},p_{y},p_{z}\rangle$\tabularnewline
	\hline 
	$8C_{3}$ & $R_{2}$ & $(1,1,1)$ & $-2\pi/3$ & $-\pi/2$ & $-\pi/2$ & $0$ & $|p_{y},p_{z},p_{x}\rangle$\tabularnewline
	& $R_{3}$ & $(1,1,1)$ & $2\pi/3$ & $0$ & $\pi/2$ & $\pi/2$ & $|p_{z},p_{x},p_{y}\rangle$\tabularnewline
	& $R_{4}$ & $(-1,1,1)$ & $-2\pi/3$ & $0$ & $-\pi/2$ & $-\pi/2$ & $|-p_{z},-p_{x},p_{y}\rangle$\tabularnewline
	& $R_{5}$ & $(-1,1,1)$ & $2\pi/3$ & $\pi/2$ & $\pi/2$ & $0$ & $|-p_{y},p_{z},-p_{x}\rangle$\tabularnewline
	& $R_{6}$ & $(-1,-1,1)$ & $-2\pi/3$ & $-\pi/2$ & $\pi/2$ & $0$ & $|p_{y},-p_{z},-p_{x}\rangle$\tabularnewline
	& $R_{7}$ & $(-1,-1,1)$ & $2\pi/3$ & $0$ & $-\pi/2$ & $\pi/2$ & $|-p_{z},p_{x},-p_{y}\rangle$\tabularnewline
	& $R_{8}$ & $(1,-1,1)$ & $-2\pi/3$ & $0$ & $\pi/2$ & $-\pi/2$ & $|p_{z},-p_{x},-p_{y}\rangle$\tabularnewline
	& $R_{9}$ & $(1,-1,1)$ & $2\pi/3$ & $\pi/2$ & $-\pi/2$ & $0$ & $|-p_{y},-p_{z},p_{x}\rangle$\tabularnewline
	\hline 
	$6C_{4}$ & $R_{10}$ & $(1,0,0)$ & $-\pi/2$ & $-\pi/2$ & $-\pi/2$ & $\pi/2$ & $|p_{x},p_{z},-p_{y}\rangle$\tabularnewline
	& $R_{11}$ & $(1,0,0)$ & $\pi/2$ & $\pi/2$ & $-\pi/2$ & $-\pi/2$ & $|p_{x},-p_{z},p_{y}\rangle$\tabularnewline
	& $R_{12}$ & $(0,1,0)$ & $-\pi/2$ & $0$ & $-\pi/2$ & $0$ & $|-p_{z},p_{y},p_{x}\rangle$\tabularnewline
	& $R_{13}$ & $(0,1,0)$ & $\pi/2$ & $0$ & $\pi/2$ & $0$ & $|p_{z},p_{y},-p_{x}\rangle$\tabularnewline
	& $R_{14}$ & $(0,0,1)$ & $-\pi/2$ & $-\pi/2$ & $0$ & $0$ & $|p_{y},-p_{x},p_{z}\rangle$\tabularnewline
	& $R_{15}$ & $(0,0,1)$ & $\pi/2$ & $\pi/2$ & $0$ & $0$ & $|-p_{y},p_{x},p_{z}\rangle$\tabularnewline
	\hline 
	$6C'_{2}$ & $R_{16}$ & $(0,1,1)$ & $-\pi$ & $-\pi/2$ & $-\pi/2$ & $-\pi/2$ & $|-p_{x},p_{z},p_{y}\rangle$\tabularnewline
	& $R_{17}$ & $(0,-1,1)$ & $-\pi$ & $-\pi/2$ & $\pi/2$ & $-\pi/2$ & $|-p_{x},-p_{z},-p_{y}\rangle$\tabularnewline
	& $R_{18}$ & $(1,1,0)$ & $-\pi$ & $-\pi/2$ & $-\pi$ & $0$ & $|p_{y},p_{x},-p_{z}\rangle$\tabularnewline
	& $R_{19}$ & $(1,-1,0)$ & $-\pi$ & $0$ & $\pi$ & $-\pi/2$ & $|-p_{y},-p_{x},-p_{z}\rangle$\tabularnewline
	& $R_{20}$ & $(1,0,1)$ & $-\pi$ & $0$ & $\pi/2$ & $-\pi$ & $|p_{z},-p_{y},p_{x}\rangle$\tabularnewline
	& $R_{21}$ & $(-1,0,1)$ & $-\pi$ & $0$ & $-\pi/2$ & $-\pi$ & $|-p_{z},-p_{y},-p_{x}\rangle$\tabularnewline
	\hline 
	$3C_{2}=3C_{4}^{2}$ & $R_{22}$ & $(1,0,0)$ & $-\pi$ & $\pi$ & $\pi$ & $0$ & $|p_{x},-p_{y},-p_{z}\rangle$\tabularnewline
	& $R_{23}$ & $(0,1,0)$ & $-\pi$ & $0$ & $-\pi$ & $0$ & $|-p_{x},p_{y},-p_{z}\rangle$\tabularnewline
	& $R_{24}$ & $(0,0,1)$ & $-\pi$ & $0$ & $0$ & $-\pi$ & $|-p_{x},-p_{y},p_{z}\rangle$\tabularnewline
	\hline 
\end{tabular}
	\caption{\label{tab:ele_o}Elements and conjugacy classes of the O group. The rotations are parametrized by the rotation axis $\bm{n}$ and angle $\omega$ or by the Euler angles $\alpha,\beta,\gamma$.}
\end{table}

\begin{table}[tbp]
	\centering
	\begin{tabular}{|c|c|c|c|c|c|}
		\hline 
	\diagbox{$\Gamma$}{$O$ group}{Class}	 & $E$ & $8C_{3}$ & $6C_{4}$ & $6C'_{2}$ & $3C_{2}=3C_{4}^{2}$\tabularnewline
		\hline 
		$A_{1}$ & 1 & 1 & 1 & 1 & 1\tabularnewline
		$A_{2}$ & 1 & 1 & -1 & -1 & 1\tabularnewline
		$E$ & 2 & -1 & 0 & 0 & 2\tabularnewline
		$T_{1}$ & 3 & 0 & 1 & -1 & -1\tabularnewline
		$T_{2}$ & 3 & 0 & -1 & 1 & -1\tabularnewline
		\hline 
	\end{tabular}
	\caption{\label{tab:chr_o}Character table for the $O$ group.}
\end{table}

\begin{table}[tbp]
	\centering
\begin{tabular}{|c|ccc|ccc|}
	\hline 
	\diagbox{$\Gamma$}{$O_h$ group}{Class} & $E$ & $\cdots$ & $3C_{2}$ & $I$ & $\cdots$ & $3IC_{2}$\tabularnewline
	\hline 
	$A_{1}^{+}$ & \multicolumn{1}{c}{} &  &  &  &  & \tabularnewline
	$\vdots$ &  & $\chi(O)$ &  &  & $\mathbb{\chi}(O)$ & \tabularnewline
	$T_{2}^{+}$ &  &  &  &  &  & \tabularnewline
	\hline 
	$A_{1}^{-}$ &  &  &  &  &  & \tabularnewline
	$\vdots$ &  & $\mathbb{\chi}(O)$ &  &  & $-\mathbb{\chi}(O)$ & \tabularnewline
	$T_{2}^{-}$ &  &  &  &  &  & \tabularnewline
	\hline 
\end{tabular}	
	\caption{\label{tab:chr_oh}Character table for the $O_h$ group. $\chi(O)$ stands for the character table of the $O$ group given in table~\ref{tab:chr_o}.}
\end{table}

\begin{table}[tbp]
	\centering
\begin{tabular}{|c|c|c|c|c|c|}
	\hline 
	$D_{4}$ group & $R_{1}$ & $R_{24}$ & $R_{14},R_{15}$ & $R_{22},R_{23}$ & $R_{18},R_{19}$\tabularnewline
	\hline 
\diagbox{$\Gamma$}{Class}	& $E$ & $C_{2}=C_{4}^{2}$ & $2C_{4}$ & $2C_{2}'$ & $2C_{2}''$\tabularnewline
	\hline 
	$A_{1}$ & 1 & 1 & 1 & 1 & 1\tabularnewline
	$A_{2}$ & 1 & 1 & 1 & -1 & -1\tabularnewline
	$B_{1}$ & 1 & 1 & -1 & 1 & -1\tabularnewline
	$B_{2}$ & 1 & 1 & -1 & -1 & 1\tabularnewline
	$E$ & 2 & -2 & 0 & 0 & 0\tabularnewline
	\hline 
\end{tabular}
	\caption{\label{tab:chr_d4}Character table for the $D_4$ group. The group elements $R_i$ are listed in table~\ref{tab:ele_o}.}
\end{table}

\begin{table}[tbp]
	\centering
\begin{tabular}{|c|c|c|c|c|}
	\hline 
	$D_{2}$ group & $R_{1}$ & $R_{18}$ & $R_{19}$ & $R_{24}$\tabularnewline
	\hline 
\diagbox{$\Gamma$}{Class}	& $E$ & $C_{2}(e_{x}+e_{y})$ & $C_{2}(e_{x}-e_{y})$ & $C_{2}(e_{z})$\tabularnewline
	\hline 
	$A_{1}$ & 1 & 1 & 1 & 1\tabularnewline
	$B_{1}$ & 1 & 1 & -1 & -1\tabularnewline
	$B_{2}$ & 1 & -1 & 1 & -1\tabularnewline
	$B_{3}$ & 1 & -1 & -1 & 1\tabularnewline
	\hline 
\end{tabular}	
	\caption{\label{tab:chr_d2}Character table for the $D_2$ group.  The group elements $R_i$  are listed in table~\ref{tab:ele_o}.}
\end{table}

\begin{table}[tbp]
	\centering
\begin{tabular}{|c|c|c|c|}
	\hline 
	$D_{3}$ group & $R_{1}$ & $R_{2},R_{3}$ & $R_{17},R_{19},R_{21}$\tabularnewline
	\hline 
\diagbox{$\Gamma$}{Class}	& $E$ & $2C_{3}$ & $3C_{2}'$\tabularnewline
	\hline 
	$A_{1}$ & 1 & 1 & 1\tabularnewline
	$A_{2}$ & 1 & 1 & -1\tabularnewline
	$E$ & 2 & -1 & 0\tabularnewline
	\hline 
\end{tabular}	
	\caption{\label{tab:chr_d3}Character table for the $D_3$ group.  The group elements $R_i$  are listed in table~\ref{tab:ele_o}.}
\end{table}

\section{L\"uscher quantization conditions}~\label{app:lush}
For the spinless systems, the L\"uscher quantization conditions read~\cite{Gockeler:2012yj},
\begin{equation}
	\det\left[M_{ln,l'n'}^{(\Gamma,\bm{P})}-\delta_{ll'}\delta_{nn'}\cot\delta_{l}\right]=0,
\end{equation}
where, $l$ and $l'$ are the angular momentum quantum numbers, $n$ and $n'$ are used to label the multiple occurrences of $\Gamma$ in the representation spaces with the angular momentum $l$ and $l'$, while $M_{ln,l'n'}^{(\Gamma,\bm{P})}$ is the interaction-independent matrix. For two-body systems of particles with equal masses, we list the matrix elements $M_{ln,l'n'}^{(\Gamma,\bm{P})}$ with positive and negative parities in tables~\ref{tab:qt_sd} and \ref{tab:qt_p}, respectively. In principle, $M_{ln,l'n'}^{(\Gamma,\bm{P})}$ is an infinite-dimensional matrix. We truncate the angular momenta to $l\leq 2$. To make the expression concise, we adopt the short-hand notation,
\begin{equation}
	w_{lm}=\frac{1}{\pi^{3/2}\sqrt{2l+1}}\gamma^{-1}q^{-l-1}{\cal Z}_{lm}^{\bm{P}}(1,q^{2}),
\end{equation}
where ${\cal Z}_{lm}^{\bm{P}}(1,q^{2})$ are the conventional L\"uscher zeta functions, which can be evaluated numerically~\cite{Leskovec:2012gb}.  

\begin{table}
	\centering
	 \renewcommand\arraystretch{1.5}
	  \setlength{\tabcolsep}{1.3mm}
\begin{tabular}{|c|c|c|c|c|}
	\hline 
	$\bm{d}$ & $\Gamma$-I & $\Gamma$-II & $M^{(\Gamma_{i},\bm{P})}$ & Partial wave\tabularnewline
	\hline 
	$(0,0,0)$ & $A_{1}^{+}$ & $A_{1}^{+}$ & $\left[w_{00}\right]$ & $\{S\}$\tabularnewline
	\cline{2-5} \cline{3-5} \cline{4-5} \cline{5-5} 
	& $E^{+}$ & $E^{+}$ & $\left[w_{00}+\frac{18}{7}w_{40}\right]$ & $\{D\}$\tabularnewline
	\cline{2-5} \cline{3-5} \cline{4-5} \cline{5-5} 
	& $T_{2}^{+}$ & $T_{2}^{+}$ & $\left[w_{00}-\frac{12}{7}w_{40}\right]$ & $\{D\}$\tabularnewline
	\hline 
	$(0,0,a)$ & $A_{1}^{+}$ & $A_{1}^{+}$ & $\left[\begin{array}{cc}
		w_{00} & -\sqrt{5}w_{20}\\
		-\sqrt{5}w_{20} & w_{00}+\frac{10}{7}w_{20}+\frac{18}{7}w_{40}
	\end{array}\right]$ & $\{S,D\}$\tabularnewline
	\cline{2-5} \cline{3-5} \cline{4-5} \cline{5-5} 
	& $B_{1}^{+}$ & $B_{1}^{+}$ & $\left[w_{00}-\frac{10}{7}w_{20}+\frac{3}{7}w_{40}+\frac{3\sqrt{70}}{7}w_{44}\right]$ & $\{D\}$\tabularnewline
	\cline{2-5} \cline{3-5} \cline{4-5} \cline{5-5} 
	& $B_{2}^{+}$ & $B_{2}^{+}$ & $\left[w_{00}-\frac{10}{7}w_{20}+\frac{3}{7}w_{40}-\frac{3\sqrt{70}}{7}w_{44}\right]$ & $\{D\}$\tabularnewline
	\cline{2-5} \cline{3-5} \cline{4-5} \cline{5-5} 
	& $E^{+}$ & $E^{+}$ & $\left[w_{00}+\frac{5}{7}w_{20}-\frac{12}{7}w_{40}\right]$ & $\{D\}$\tabularnewline
	\hline 
	$(a,a,0)$ & \multirow{2}{*}{$A_{1}^{+}$} & \multirow{2}{*}{$A_{1}^{+}$} & $\left[\begin{array}{ccc}
		w_{00} & \sqrt{10}w_{22} & -\sqrt{5}w_{20}\\
		-\sqrt{10}w_{22} & A_{22} & A_{23}\\
		-\sqrt{5}w_{20} & -A_{23} & A_{33}
	\end{array}\right]$ & \multirow{2}{*}{$\{S,D_{a},D_{b}\}$}\tabularnewline
	&  &  & $\begin{array}{cl}
		A_{22} & =w_{00}-\frac{10}{7}w_{20}+\frac{3}{7}w_{40}-\frac{3\sqrt{70}}{7}w_{44}\\
		A_{33} & =w_{00}+\frac{10}{7}w_{20}+\frac{18}{7}w_{40}\\
		A_{23} & =-\frac{10\sqrt{2}}{7}w_{22}+\frac{3\sqrt{30}}{7}w_{42}
	\end{array}$ & \tabularnewline
	\cline{2-5} \cline{3-5} \cline{4-5} \cline{5-5} 
	& $B_{1}^{+}$ & $A_{2}^{+}$ & $\left[w_{00}+\frac{5}{7}w_{20}-\frac{12}{7}w_{40}+i\frac{5\sqrt{6}}{7}w_{22}+i\frac{6\sqrt{10}}{7}w_{42}\right]$ & $\{D\}$\tabularnewline
	\cline{2-5} \cline{3-5} \cline{4-5} \cline{5-5} 
	& $B_{2}^{+}$ & $B_{1}^{+}$ & $\left[w_{00}+\frac{5}{7}w_{20}-\frac{12}{7}w_{40}-i\frac{5\sqrt{6}}{7}w_{22}-i\frac{6\sqrt{10}}{7}w_{42}\right]$ & $\{D\}$\tabularnewline
	\cline{2-5} \cline{3-5} \cline{4-5} \cline{5-5} 
	& $B_{3}^{+}$ & $B_{2}^{+}$ & $\left[w_{00}-\frac{10}{7}w_{20}+\frac{3}{7}w_{40}+\frac{3\sqrt{70}}{4}w_{44}\right]$ & $\{D\}$\tabularnewline
	\hline 
	($a,a,a)$ & \multirow{2}{*}{$A_{1}^{+}$} & \multirow{2}{*}{$A_{1}^{+}$} & $\left[\begin{array}{cc}
		w_{00} & \sqrt{30}w_{22}\\
		-\sqrt{30}w_{22} & A_{22}
	\end{array}\right]$ & \multirow{2}{*}{$\{S,D\}$}\tabularnewline
	&  &  & $A_{22}=w_{00}-\frac{12}{7}w_{40}-i\frac{10\sqrt{6}}{7}w_{22}-i\frac{12\sqrt{10}}{7}w_{42}$ & \tabularnewline
	\cline{2-5} \cline{3-5} \cline{4-5} \cline{5-5} 
	& \multirow{2}{*}{$E^{+}$} & \multirow{2}{*}{$E^{+}$} & $\left[\begin{array}{cc}
		w_{00}+\frac{18}{7}w_{40} & A_{12}^{-}\\
		-A_{12}^{+} & A_{22}
	\end{array}\right]$ & \multirow{2}{*}{$\{D_{a},D_{b}\}$}\tabularnewline
	&  &  & $\begin{array}{cl}
		A_{12}^{\pm} & =\frac{5\sqrt{6}}{7}(1\pm i)w_{22}-\frac{9\sqrt{10}}{14}(1\pm i)w_{42}\\
		A_{22} & =w_{00}-\frac{12}{7}w_{40}+i\frac{5\sqrt{6}}{7}w_{22}+i\frac{6\sqrt{10}}{7}w_{42}
	\end{array}$ & \tabularnewline
	\hline 
\end{tabular}
\caption{\label{tab:qt_sd}L\"uscher's quantization conditions for systems with even parity~\cite{Gockeler:2012yj}. We label the multiple occurrences of D-wave in some irreps with $D_a$ and $D_b$. }
\end{table}

\begin{table}
	\centering
	\renewcommand\arraystretch{1.5}
	\setlength{\tabcolsep}{1.3mm}
\begin{tabular}{|c|c|c|c|c|}
	\hline 
	$\bm{d}$ & $\Gamma$-I & $\Gamma$-II & $M^{(\Gamma_{i},\bm{P})}$ & Partial wave\tabularnewline
	\hline 
	$(0,0,0)$ & $T_{1}^{-}$ & $T_{1}^{-}$ & $w_{0,0}$ & $\{P\}$\tabularnewline
	\hline 
	$(0,0,a)$ & $A_{2}^{-}$ & $A_{1}^{-}$ & $\ensuremath{w_{00}+2w_{20}}$ & $\{P\}$\tabularnewline
	\cline{2-5} \cline{3-5} \cline{4-5} \cline{5-5} 
	& $E^{-}$ & $E^{-}$ & $w_{00}-w_{20}$ & $\{P\}$\tabularnewline
	\hline 
	$(a,a,0)$ & $B_{1}^{-}$ & $A_{1}^{-}$ & $\ensuremath{w_{00}-w_{20}-i\sqrt{6}w_{22}}$ & $\{P\}$\tabularnewline
	\cline{2-5} \cline{3-5} \cline{4-5} \cline{5-5} 
	& $B_{2}^{-}$ & $B_{2}^{-}$ & $w_{00}-w_{20}+i\sqrt{6}w_{22}$ & $\{P\}$\tabularnewline
	\cline{2-5} \cline{3-5} \cline{4-5} \cline{5-5} 
	& $B_{3}^{-}$ & $B_{1}^{-}$ & $w_{00}+2w_{20}$ & $\{P\}$\tabularnewline
	\hline 
	$(a,a,a)$ & $A_{2}^{-}$ & $A_{1}^{-}$ & $w_{00}-i2\sqrt{6}w_{22}$ & $\{P\}$\tabularnewline
	\cline{2-5} \cline{3-5} \cline{4-5} \cline{5-5} 
	& $E^{-}$ & $E^{-}$ & $w_{00}+i\sqrt{6}w_{22}$ & $\{P\}$\tabularnewline
	\hline 
\end{tabular}
	\caption{\label{tab:qt_p}L\"uscher's quantization conditions for systems with odd parity~\cite{Gockeler:2012yj}.}
	\end{table}

\acknowledgments

We are grateful to Jambul Gegelia for sharing his insights into the considered topics, numerous discussions and
comments on the manuscript. We also thank Zhan-Wei Liu for helpful discussions as well as Ra\'{u}l Brice\~{n}o, Michael D\"oring, Maxim Mai, Ulf-G.~Mei{\ss}ner, Akaki Rusetsky and Andr\'{e} Walker-Loud for useful comments 
on the manuscript. This work was supported in part by DFG and NSFC through funds provided to the
Sino-German CRC 110 ``Symmetries and the Emergence of Structure in QCD'' (NSFC
Grant No.~12070131001, Project-ID 196253076 - TRR 110) and by BMBF (Grant No.~05P18PCFP1).



%
%
%
%
%
%
%
%
\bibliography{ref}

\providecommand{\href}[2]{#2}\begingroup\raggedright\begin{thebibliography}{10}

\bibitem{Luscher:1985dn}
M.~L{\"u}scher, \emph{{Volume dependence of the energy spectrum in massive
  quantum field theories. 1. stable particle states}},
  \href{https://doi.org/10.1007/BF01211589}{\emph{Commun. Math. Phys.}
  {\bfseries 104} (1986) 177}.

\bibitem{Luscher:1986pf}
M.~L{\"u}scher, \emph{{Volume dependence of the energy spectrum in massive
  quantum field theories. 2. scattering states}},
  \href{https://doi.org/10.1007/BF01211097}{\emph{Commun. Math. Phys.}
  {\bfseries 105} (1986) 153}.

\bibitem{Luscher:1990ux}
M.~L{\"u}scher, \emph{{Two particle states on a torus and their relation to the
  scattering matrix}},
  \href{https://doi.org/10.1016/0550-3213(91)90366-6}{\emph{Nucl. Phys. B}
  {\bfseries 354} (1991) 531}.

\bibitem{Rummukainen:1995vs}
K.~Rummukainen and S.~A. Gottlieb, \emph{{Resonance scattering phase shifts on
  a nonrest frame lattice}},
  \href{https://doi.org/10.1016/0550-3213(95)00313-H}{\emph{Nucl. Phys. B}
  {\bfseries 450} (1995) 397}
  [\href{https://arxiv.org/abs/hep-lat/9503028}{{\ttfamily hep-lat/9503028}}].

\bibitem{Kim:2005gf}
C.~H. Kim, C.~T. Sachrajda and S.~R. Sharpe, \emph{{Finite-volume effects for
  two-hadron states in moving frames}},
  \href{https://doi.org/10.1016/j.nuclphysb.2005.08.029}{\emph{Nucl. Phys. B}
  {\bfseries 727} (2005) 218}
  [\href{https://arxiv.org/abs/hep-lat/0507006}{{\ttfamily hep-lat/0507006}}].

\bibitem{Gockeler:2012yj}
M.~G{\"o}ckeler, R.~Horsley, M.~Lage, U.-G. Mei{\ss}ner, P.~E.~L. Rakow,
  A.~Rusetsky et~al., \emph{{Scattering phases for meson and baryon resonances
  on general moving-frame lattices}},
  \href{https://doi.org/10.1103/PhysRevD.86.094513}{\emph{Phys. Rev. D}
  {\bfseries 86} (2012) 094513}
  [\href{https://arxiv.org/abs/1206.4141}{{\ttfamily 1206.4141}}].

\bibitem{Luu:2011ep}
T.~Luu and M.~J. Savage, \emph{{Extracting scattering phase-shifts in higher
  partial-waves from lattice QCD calculations}},
  \href{https://doi.org/10.1103/PhysRevD.83.114508}{\emph{Phys. Rev. D}
  {\bfseries 83} (2011) 114508}
  [\href{https://arxiv.org/abs/1101.3347}{{\ttfamily 1101.3347}}].

\bibitem{Fu:2011xz}
Z.~Fu, \emph{{Rummukainen-Gottlieb's formula on two-particle system with
  different mass}},
  \href{https://doi.org/10.1103/PhysRevD.85.014506}{\emph{Phys. Rev. D}
  {\bfseries 85} (2012) 014506}
  [\href{https://arxiv.org/abs/1110.0319}{{\ttfamily 1110.0319}}].

\bibitem{Leskovec:2012gb}
L.~Leskovec and S.~Prelovsek, \emph{{Scattering phase shifts for two particles
  of different mass and non-zero total momentum in lattice QCD}},
  \href{https://doi.org/10.1103/PhysRevD.85.114507}{\emph{Phys. Rev. D}
  {\bfseries 85} (2012) 114507}
  [\href{https://arxiv.org/abs/1202.2145}{{\ttfamily 1202.2145}}].

\bibitem{Briceno:2014oea}
R.~A. Briceno, \emph{{Two-particle multichannel systems in a finite volume with
  arbitrary spin}},
  \href{https://doi.org/10.1103/PhysRevD.89.074507}{\emph{Phys. Rev. D}
  {\bfseries 89} (2014) 074507}
  [\href{https://arxiv.org/abs/1401.3312}{{\ttfamily 1401.3312}}].

\bibitem{Lage:2009zv}
M.~Lage, U.-G. Mei{\ss}ner and A.~Rusetsky, \emph{{A Method to measure the
  antikaon-nucleon scattering length in lattice QCD}},
  \href{https://doi.org/10.1016/j.physletb.2009.10.055}{\emph{Phys. Lett. B}
  {\bfseries 681} (2009) 439}
  [\href{https://arxiv.org/abs/0905.0069}{{\ttfamily 0905.0069}}].

\bibitem{He:2005ey}
S.~He, X.~Feng and C.~Liu, \emph{{Two particle states and the S-matrix elements
  in multi-channel scattering}},
  \href{https://doi.org/10.1088/1126-6708/2005/07/011}{\emph{JHEP} {\bfseries
  07} (2005) 011} [\href{https://arxiv.org/abs/hep-lat/0504019}{{\ttfamily
  hep-lat/0504019}}].

\bibitem{Hansen:2012tf}
M.~T. Hansen and S.~R. Sharpe, \emph{{Multiple-channel generalization of
  Lellouch-L\"uscher formula}},
  \href{https://doi.org/10.1103/PhysRevD.86.016007}{\emph{Phys. Rev. D}
  {\bfseries 86} (2012) 016007}
  [\href{https://arxiv.org/abs/1204.0826}{{\ttfamily 1204.0826}}].

\bibitem{Doring:2012eu}
M.~D{\"o}ring, U.-G. Mei{\ss}ner, E.~Oset and A.~Rusetsky, \emph{{Scalar mesons
  moving in a finite volume and the role of partial wave mixing}},
  \href{https://doi.org/10.1140/epja/i2012-12114-6}{\emph{Eur. Phys. J. A}
  {\bfseries 48} (2012) 114} [\href{https://arxiv.org/abs/1205.4838}{{\ttfamily
  1205.4838}}].

\bibitem{Feng:2004ua}
X.~Feng, X.~Li and C.~Liu, \emph{{Two particle states in an asymmetric box and
  the elastic scattering phases}},
  \href{https://doi.org/10.1103/PhysRevD.70.014505}{\emph{Phys. Rev. D}
  {\bfseries 70} (2004) 014505}
  [\href{https://arxiv.org/abs/hep-lat/0404001}{{\ttfamily hep-lat/0404001}}].

\bibitem{Briceno:2017max}
R.~A. Briceno, J.~J. Dudek and R.~D. Young, \emph{{Scattering processes and
  resonances from lattice QCD}},
  \href{https://doi.org/10.1103/RevModPhys.90.025001}{\emph{Rev. Mod. Phys.}
  {\bfseries 90} (2018) 025001}
  [\href{https://arxiv.org/abs/1706.06223}{{\ttfamily 1706.06223}}].

\bibitem{Mai:2021lwb}
M.~Mai, M.~D{\"o}ring and A.~Rusetsky, \emph{{Multi-particle systems on the
  lattice and chiral extrapolations: a brief review}},
  \href{https://doi.org/10.1140/epjs/s11734-021-00146-5}{\emph{Eur. Phys. J.
  ST} {\bfseries 230} (2021) 1623}
  [\href{https://arxiv.org/abs/2103.00577}{{\ttfamily 2103.00577}}].

\bibitem{Chen:2016qju}
H.-X. Chen, W.~Chen, X.~Liu and S.-L. Zhu, \emph{{The hidden-charm pentaquark
  and tetraquark states}},
  \href{https://doi.org/10.1016/j.physrep.2016.05.004}{\emph{Phys. Rept.}
  {\bfseries 639} (2016) 1} [\href{https://arxiv.org/abs/1601.02092}{{\ttfamily
  1601.02092}}].

\bibitem{Guo:2017jvc}
F.-K. Guo, C.~Hanhart, U.-G. Mei\ss{}ner, Q.~Wang, Q.~Zhao and B.-S. Zou,
  \emph{{Hadronic molecules}},
  \href{https://doi.org/10.1103/RevModPhys.90.015004}{\emph{Rev. Mod. Phys.}
  {\bfseries 90} (2018) 015004}
  [\href{https://arxiv.org/abs/1705.00141}{{\ttfamily 1705.00141}}].

\bibitem{Brambilla:2019esw}
N.~Brambilla, S.~Eidelman, C.~Hanhart, A.~Nefediev, C.-P. Shen, C.~E. Thomas
  et~al., \emph{{The $XYZ$ states: experimental and theoretical status and
  perspectives}},
  \href{https://doi.org/10.1016/j.physrep.2020.05.001}{\emph{Phys. Rept.}
  {\bfseries 873} (2020) 1} [\href{https://arxiv.org/abs/1907.07583}{{\ttfamily
  1907.07583}}].

\bibitem{Bedaque:2006yi}
P.~F. Bedaque, I.~Sato and A.~Walker-Loud, \emph{{Finite volume corrections to
  pi-pi scattering}},
  \href{https://doi.org/10.1103/PhysRevD.73.074501}{\emph{Phys. Rev. D}
  {\bfseries 73} (2006) 074501}
  [\href{https://arxiv.org/abs/hep-lat/0601033}{{\ttfamily hep-lat/0601033}}].

\bibitem{Sato:2007ms}
I.~Sato and P.~F. Bedaque, \emph{{Fitting two nucleons inside a box:
  Exponentially suppressed corrections to the L\"uscher's formula}},
  \href{https://doi.org/10.1103/PhysRevD.76.034502}{\emph{Phys. Rev. D}
  {\bfseries 76} (2007) 034502}
  [\href{https://arxiv.org/abs/hep-lat/0702021}{{\ttfamily hep-lat/0702021}}].

\bibitem{Jansen:2015lha}
M.~Jansen, H.~W. Hammer and Y.~Jia, \emph{{Finite volume corrections to the
  binding energy of the X(3872)}},
  \href{https://doi.org/10.1103/PhysRevD.92.114031}{\emph{Phys. Rev. D}
  {\bfseries 92} (2015) 114031}
  [\href{https://arxiv.org/abs/1505.04099}{{\ttfamily 1505.04099}}].

\bibitem{Doring:2011vk}
M.~D{\"o}ring, U.-G. Mei{\ss}ner, E.~Oset and A.~Rusetsky, \emph{{Unitarized
  Chiral Perturbation Theory in a finite volume: Scalar meson sector}},
  \href{https://doi.org/10.1140/epja/i2011-11139-7}{\emph{Eur. Phys. J. A}
  {\bfseries 47} (2011) 139} [\href{https://arxiv.org/abs/1107.3988}{{\ttfamily
  1107.3988}}].

\bibitem{Guo:2016zep}
Z.-H. Guo, L.~Liu, U.-G. Mei\ss{}ner, J.~A. Oller and A.~Rusetsky,
  \emph{{Chiral study of the $a_0(980)$ resonance and $\pi\eta$ scattering
  phase shifts in light of a recent lattice simulation}},
  \href{https://doi.org/10.1103/PhysRevD.95.054004}{\emph{Phys. Rev. D}
  {\bfseries 95} (2017) 054004}
  [\href{https://arxiv.org/abs/1609.08096}{{\ttfamily 1609.08096}}].

\bibitem{Guo:2018tjx}
Z.-H. Guo, L.~Liu, U.-G. Mei\ss{}ner, J.~A. Oller and A.~Rusetsky,
  \emph{{Towards a precise determination of the scattering amplitudes of the
  charmed and light-flavor pseudoscalar mesons}},
  \href{https://doi.org/10.1140/epjc/s10052-018-6518-1}{\emph{Eur. Phys. J. C}
  {\bfseries 79} (2019) 13} [\href{https://arxiv.org/abs/1811.05585}{{\ttfamily
  1811.05585}}].

\bibitem{Albaladejo:2012jr}
M.~Albaladejo, J.~A. Oller, E.~Oset, G.~Rios and L.~Roca, \emph{{Finite volume
  treatment of $\pi \pi$ scattering and limits to phase shifts extraction from
  lattice QCD}}, \href{https://doi.org/10.1007/JHEP08(2012)071}{\emph{JHEP}
  {\bfseries 08} (2012) 071} [\href{https://arxiv.org/abs/1205.3582}{{\ttfamily
  1205.3582}}].

\bibitem{Chen:2012rp}
H.-X. Chen and E.~Oset, \emph{{$\pi\pi$ interaction in the $\rho$ channel in
  finite volume}},
  \href{https://doi.org/10.1103/PhysRevD.87.016014}{\emph{Phys. Rev. D}
  {\bfseries 87} (2013) 016014}
  [\href{https://arxiv.org/abs/1202.2787}{{\ttfamily 1202.2787}}].

\bibitem{Hall:2013qba}
J.~M.~M. Hall, A.~C.~P. Hsu, D.~B. Leinweber, A.~W. Thomas and R.~D. Young,
  \emph{{Finite-volume matrix Hamiltonian model for a $\Delta \to N\pi$
  system}}, \href{https://doi.org/10.1103/PhysRevD.87.094510}{\emph{Phys. Rev.
  D} {\bfseries 87} (2013) 094510}
  [\href{https://arxiv.org/abs/1303.4157}{{\ttfamily 1303.4157}}].

\bibitem{Wu:2014vma}
J.-J. Wu, T.~S.~H. Lee, A.~W. Thomas and R.~D. Young, \emph{{Finite-volume
  Hamiltonian method for coupled-channels interactions in lattice QCD}},
  \href{https://doi.org/10.1103/PhysRevC.90.055206}{\emph{Phys. Rev. C}
  {\bfseries 90} (2014) 055206}
  [\href{https://arxiv.org/abs/1402.4868}{{\ttfamily 1402.4868}}].

\bibitem{Liu:2015ktc}
Z.-W. Liu, W.~Kamleh, D.~B. Leinweber, F.~M. Stokes, A.~W. Thomas and J.-J. Wu,
  \emph{{Hamiltonian effective field theory study of the $\mathbf{N^*(1535)}$
  resonance in lattice QCD}},
  \href{https://doi.org/10.1103/PhysRevLett.116.082004}{\emph{Phys. Rev. Lett.}
  {\bfseries 116} (2016) 082004}
  [\href{https://arxiv.org/abs/1512.00140}{{\ttfamily 1512.00140}}].

\bibitem{Li:2021mob}
Y.~Li, J.-j. Wu, D.~B. Leinweber and A.~W. Thomas, \emph{{Hamiltonian effective
  field theory in elongated or moving finite volume}},
  \href{https://doi.org/10.1103/PhysRevD.103.094518}{\emph{Phys. Rev. D}
  {\bfseries 103} (2021) 094518}
  [\href{https://arxiv.org/abs/2103.12260}{{\ttfamily 2103.12260}}].

\bibitem{Morningstar:2017spu}
C.~Morningstar, J.~Bulava, B.~Singha, R.~Brett, J.~Fallica, A.~Hanlon et~al.,
  \emph{{Estimating the two-particle $K$-matrix for multiple partial waves and
  decay channels from finite-volume energies}},
  \href{https://doi.org/10.1016/j.nuclphysb.2017.09.014}{\emph{Nucl. Phys. B}
  {\bfseries 924} (2017) 477}
  [\href{https://arxiv.org/abs/1707.05817}{{\ttfamily 1707.05817}}].

\bibitem{Li:2019qvh}
Y.~Li, J.-J. Wu, C.~D. Abell, D.~B. Leinweber and A.~W. Thomas, \emph{{Partial
  Wave Mixing in Hamiltonian Effective Field Theory}},
  \href{https://doi.org/10.1103/PhysRevD.101.114501}{\emph{Phys. Rev. D}
  {\bfseries 101} (2020) 114501}
  [\href{https://arxiv.org/abs/1910.04973}{{\ttfamily 1910.04973}}].

\bibitem{Lee:2021kfn}
F.~X. Lee, A.~Alexandru and R.~Brett, \emph{{Validation of the finite-volume
  quantization condition for two spinless particles}},
  \href{https://arxiv.org/abs/2107.04430}{{\ttfamily 2107.04430}}.

\bibitem{Lee:2020fbo}
F.~X. Lee, C.~Morningstar and A.~Alexandru, \emph{{Energy spectrum of
  two-particle scattering in a periodic box}},
  \href{https://doi.org/10.1142/S0129183120501314}{\emph{Int. J. Mod. Phys. C}
  {\bfseries 31} (2020) 2050131}.

\bibitem{Briceno:2013lba}
R.~A. Briceno, Z.~Davoudi and T.~C. Luu, \emph{{Two-Nucleon Systems in a Finite
  Volume: (I) Quantization Conditions}},
  \href{https://doi.org/10.1103/PhysRevD.88.034502}{\emph{Phys. Rev. D}
  {\bfseries 88} (2013) 034502}
  [\href{https://arxiv.org/abs/1305.4903}{{\ttfamily 1305.4903}}].

\bibitem{Morningstar:2013bda}
C.~Morningstar, J.~Bulava, B.~Fahy, J.~Foley, Y.~C. Jhang, K.~J. Juge et~al.,
  \emph{{Extended hadron and two-hadron operators of definite momentum for
  spectrum calculations in lattice QCD}},
  \href{https://doi.org/10.1103/PhysRevD.88.014511}{\emph{Phys. Rev. D}
  {\bfseries 88} (2013) 014511}
  [\href{https://arxiv.org/abs/1303.6816}{{\ttfamily 1303.6816}}].

\bibitem{Bernard:2008ax}
V.~Bernard, M.~Lage, U.-G. Mei{\ss}ner and A.~Rusetsky, \emph{{Resonance
  properties from the finite-volume energy spectrum}},
  \href{https://doi.org/10.1088/1126-6708/2008/08/024}{\emph{JHEP} {\bfseries
  08} (2008) 024} [\href{https://arxiv.org/abs/0806.4495}{{\ttfamily
  0806.4495}}].

\bibitem{dresselhaus2007group}
M.~Dresselhaus, G.~Dresselhaus and A.~Jorio, \emph{Group Theory: Application to
  the Physics of Condensed Matter}, SpringerLink: Springer e-Books. Springer
  Berlin Heidelberg, 2007.

\bibitem{Doring:2018xxx}
M.~D{\"o}ring, H.~W. Hammer, M.~Mai, J.~Y. Pang, t.~A. Rusetsky and J.~Wu,
  \emph{{Three-body spectrum in a finite volume: the role of cubic symmetry}},
  \href{https://doi.org/10.1103/PhysRevD.97.114508}{\emph{Phys. Rev. D}
  {\bfseries 97} (2018) 114508}
  [\href{https://arxiv.org/abs/1802.03362}{{\ttfamily 1802.03362}}].

\bibitem{Woss:2020cmp}
{\scshape Hadron Spectrum} collaboration, \emph{{Efficient solution of the
  multichannel L\"uscher determinant condition through eigenvalue
  decomposition}},
  \href{https://doi.org/10.1103/PhysRevD.101.114505}{\emph{Phys. Rev. D}
  {\bfseries 101} (2020) 114505}
  [\href{https://arxiv.org/abs/2001.08474}{{\ttfamily 2001.08474}}].

\bibitem{Epelbaum:2008ga}
E.~Epelbaum, H.-W. Hammer and U.-G. Mei{\ss}ner, \emph{{Modern Theory of
  Nuclear Forces}},
  \href{https://doi.org/10.1103/RevModPhys.81.1773}{\emph{Rev. Mod. Phys.}
  {\bfseries 81} (2009) 1773}
  [\href{https://arxiv.org/abs/0811.1338}{{\ttfamily 0811.1338}}].

\bibitem{Machleidt:2011zz}
R.~Machleidt and D.~R. Entem, \emph{{Chiral effective field theory and nuclear
  forces}}, \href{https://doi.org/10.1016/j.physrep.2011.02.001}{\emph{Phys.
  Rept.} {\bfseries 503} (2011) 1}
  [\href{https://arxiv.org/abs/1105.2919}{{\ttfamily 1105.2919}}].

\bibitem{Epelbaum:2019kcf}
E.~Epelbaum, H.~Krebs and P.~Reinert, \emph{{High-precision nuclear forces from
  chiral EFT: State-of-the-art, challenges and outlook}},
  \href{https://doi.org/10.3389/fphy.2020.00098}{\emph{Front. in Phys.}
  {\bfseries 8} (2020) 98} [\href{https://arxiv.org/abs/1911.11875}{{\ttfamily
  1911.11875}}].

\bibitem{Epelbaum:2014sza}
E.~Epelbaum, H.~Krebs and U.-G. Mei{\ss}ner, \emph{{Precision nucleon-nucleon
  potential at fifth order in the chiral expansion}},
  \href{https://doi.org/10.1103/PhysRevLett.115.122301}{\emph{Phys. Rev. Lett.}
  {\bfseries 115} (2015) 122301}
  [\href{https://arxiv.org/abs/1412.4623}{{\ttfamily 1412.4623}}].

\bibitem{Reinert:2017usi}
P.~Reinert, H.~Krebs and E.~Epelbaum, \emph{{Semilocal momentum-space
  regularized chiral two-nucleon potentials up to fifth order}},
  \href{https://doi.org/10.1140/epja/i2018-12516-4}{\emph{Eur. Phys. J. A}
  {\bfseries 54} (2018) 86} [\href{https://arxiv.org/abs/1711.08821}{{\ttfamily
  1711.08821}}].

\bibitem{Entem:2017gor}
D.~R. Entem, R.~Machleidt and Y.~Nosyk, \emph{{High-quality two-nucleon
  potentials up to fifth order of the chiral expansion}},
  \href{https://doi.org/10.1103/PhysRevC.96.024004}{\emph{Phys. Rev. C}
  {\bfseries 96} (2017) 024004}
  [\href{https://arxiv.org/abs/1703.05454}{{\ttfamily 1703.05454}}].

\bibitem{Epelbaum:2003xx}
E.~Epelbaum, W.~Gl{\"o}ckle and U.-G. Mei{\ss}ner, \emph{{Improving the
  convergence of the chiral expansion for nuclear forces. 2. Low phases and the
  deuteron}}, \href{https://doi.org/10.1140/epja/i2003-10129-8}{\emph{Eur.
  Phys. J. A} {\bfseries 19} (2004) 401}
  [\href{https://arxiv.org/abs/nucl-th/0308010}{{\ttfamily nucl-th/0308010}}].

\bibitem{Baru:2011bw}
V.~Baru, C.~Hanhart, M.~Hoferichter, B.~Kubis, A.~Nogga and D.~R. Phillips,
  \emph{{Precision calculation of threshold $\pi^-d$ scattering, $\pi$N
  scattering lengths, and the GMO sum rule}},
  \href{https://doi.org/10.1016/j.nuclphysa.2011.09.015}{\emph{Nucl. Phys. A}
  {\bfseries 872} (2011) 69} [\href{https://arxiv.org/abs/1107.5509}{{\ttfamily
  1107.5509}}].

\bibitem{Reinert:2020mcu}
P.~Reinert, H.~Krebs and E.~Epelbaum, \emph{{Precision determination of
  pion-nucleon coupling constants using effective field theory}},
  \href{https://doi.org/10.1103/PhysRevLett.126.092501}{\emph{Phys. Rev. Lett.}
  {\bfseries 126} (2021) 092501}
  [\href{https://arxiv.org/abs/2006.15360}{{\ttfamily 2006.15360}}].

\bibitem{Epelbaum:2003gr}
E.~Epelbaum, W.~Gl{\"o}ckle and U.-G. Mei{\ss}ner, \emph{{Improving the
  convergence of the chiral expansion for nuclear forces. 1. Peripheral
  phases}}, \href{https://doi.org/10.1140/epja/i2003-10096-0}{\emph{Eur. Phys.
  J. A} {\bfseries 19} (2004) 125}
  [\href{https://arxiv.org/abs/nucl-th/0304037}{{\ttfamily nucl-th/0304037}}].

\bibitem{Stoks:1993tb}
V.~G.~J. Stoks, R.~A.~M. Klomp, M.~C.~M. Rentmeester and J.~J. de~Swart,
  \emph{{Partial wave analaysis of all nucleon-nucleon scattering data below
  350-MeV}}, \href{https://doi.org/10.1103/PhysRevC.48.792}{\emph{Phys. Rev. C}
  {\bfseries 48} (1993) 792}.

\bibitem{Mai:2019fba}
M.~Mai, M.~D{\"o}ring, C.~Culver and A.~Alexandru, \emph{{Three-body unitarity
  versus finite-volume $\pi^+\pi^+\pi^+$ spectrum from lattice QCD}},
  \href{https://doi.org/10.1103/PhysRevD.101.054510}{\emph{Phys. Rev. D}
  {\bfseries 101} (2020) 054510}
  [\href{https://arxiv.org/abs/1909.05749}{{\ttfamily 1909.05749}}].

\bibitem{Fachruddin:2001az}
I.~Fachruddin, C.~Elster and W.~Gl{\"o}ckle, \emph{{Nucleon-nucleon scattering
  in a three-dimensional approach}},
  \href{https://doi.org/10.1016/S0375-9474(01)00892-2}{\emph{Nucl. Phys. A}
  {\bfseries 689} (2001) 507}
  [\href{https://arxiv.org/abs/nucl-th/0104027}{{\ttfamily nucl-th/0104027}}].

\bibitem{Stoks:1990us}
V.~G.~J. Stoks and J.~J. De~Swart, \emph{{The Magnetic moment interaction in
  nucleon-nucleon phase shift analyses}},
  \href{https://doi.org/10.1103/PhysRevC.42.1235}{\emph{Phys. Rev. C}
  {\bfseries 42} (1990) 1235}.

\bibitem{Dudek:2012gj}
J.~J. Dudek, R.~G. Edwards and C.~E. Thomas, \emph{{S and D-wave phase shifts
  in isospin-2 pi pi scattering from lattice QCD}},
  \href{https://doi.org/10.1103/PhysRevD.86.034031}{\emph{Phys. Rev. D}
  {\bfseries 86} (2012) 034031}
  [\href{https://arxiv.org/abs/1203.6041}{{\ttfamily 1203.6041}}].

\bibitem{Cheung:2020mql}
{\scshape Hadron Spectrum} collaboration, \emph{{DK I = 0,$ D\overline{K} $I =
  0, 1 scattering and the $ {D}_{s0}^{\ast } $(2317) from lattice QCD}},
  \href{https://doi.org/10.1007/JHEP02(2021)100}{\emph{JHEP} {\bfseries 02}
  (2021) 100} [\href{https://arxiv.org/abs/2008.06432}{{\ttfamily
  2008.06432}}].

\bibitem{Baru:2015ira}
V.~Baru, E.~Epelbaum, A.~A. Filin and J.~Gegelia, \emph{{Low-energy theorems
  for nucleon-nucleon scattering at unphysical pion masses}},
  \href{https://doi.org/10.1103/PhysRevC.92.014001}{\emph{Phys. Rev. C}
  {\bfseries 92} (2015) 014001}
  [\href{https://arxiv.org/abs/1504.07852}{{\ttfamily 1504.07852}}].

\bibitem{Baru:2016evv}
V.~Baru, E.~Epelbaum and A.~A. Filin, \emph{{Low-energy theorems for
  nucleon-nucleon scattering at $M_\pi=450$ MeV}},
  \href{https://doi.org/10.1103/PhysRevC.94.014001}{\emph{Phys. Rev. C}
  {\bfseries 94} (2016) 014001}
  [\href{https://arxiv.org/abs/1604.02551}{{\ttfamily 1604.02551}}].

\bibitem{Bernard:1991zc}
V.~Bernard, N.~Kaiser and U.-G. Mei{\ss}ner, \emph{{Chiral perturbation theory
  in the presence of resonances: Application to pi pi and pi K scattering}},
  \href{https://doi.org/10.1016/0550-3213(91)90586-M}{\emph{Nucl. Phys. B}
  {\bfseries 364} (1991) 283}.

\bibitem{Bruns:2004tj}
P.~C. Bruns and U.-G. Mei{\ss}ner, \emph{{Infrared regularization for spin-1
  fields}}, \href{https://doi.org/10.1140/epjc/s2005-02118-0}{\emph{Eur. Phys.
  J. C} {\bfseries 40} (2005) 97}
  [\href{https://arxiv.org/abs/hep-ph/0411223}{{\ttfamily hep-ph/0411223}}].

\bibitem{Woloshyn:1973mce}
R.~M. Woloshyn and A.~D. Jackson, \emph{{Comparison of three-dimensional
  relativistic scattering equations}},
  \href{https://doi.org/10.1016/0550-3213(73)90626-3}{\emph{Nucl. Phys. B}
  {\bfseries 64} (1973) 269}.

\bibitem{Castillejo:1955ed}
L.~Castillejo, R.~H. Dalitz and F.~J. Dyson, \emph{{Low's scattering equation
  for the charged and neutral scalar theories}},
  \href{https://doi.org/10.1103/PhysRev.101.453}{\emph{Phys. Rev.} {\bfseries
  101} (1956) 453}.

\bibitem{Oller:2000ug}
J.~A. Oller, E.~Oset and J.~E. Palomar, \emph{{Pion and kaon vector
  form-factors}}, \href{https://doi.org/10.1103/PhysRevD.63.114009}{\emph{Phys.
  Rev. D} {\bfseries 63} (2001) 114009}
  [\href{https://arxiv.org/abs/hep-ph/0011096}{{\ttfamily hep-ph/0011096}}].

\bibitem{Gasser:1984gg}
J.~Gasser and H.~Leutwyler, \emph{{Chiral Perturbation Theory: Expansions in
  the Mass of the Strange Quark}},
  \href{https://doi.org/10.1016/0550-3213(85)90492-4}{\emph{Nucl. Phys. B}
  {\bfseries 250} (1985) 465}.

\bibitem{Ecker:1988te}
G.~Ecker, J.~Gasser, A.~Pich and E.~de~Rafael, \emph{{The Role of Resonances in
  Chiral Perturbation Theory}},
  \href{https://doi.org/10.1016/0550-3213(89)90346-5}{\emph{Nucl. Phys. B}
  {\bfseries 321} (1989) 311}.

\bibitem{Fischer:2020fvl}
{\scshape Extended Twisted Mass, ETM} collaboration, \emph{{The
  $\rho$-resonance from $N_f$ = 2 lattice QCD including the physical pion
  mass}}, \href{https://doi.org/10.1016/j.physletb.2021.136449}{\emph{Phys.
  Lett. B} {\bfseries 819} (2021) 136449}
  [\href{https://arxiv.org/abs/2006.13805}{{\ttfamily 2006.13805}}].

\bibitem{Estabrooks:1974vu}
P.~Estabrooks and A.~D. Martin, \emph{{pi pi Phase Shift Analysis Below the K
  anti-K Threshold}},
  \href{https://doi.org/10.1016/0550-3213(74)90488-X}{\emph{Nucl. Phys. B}
  {\bfseries 79} (1974) 301}.

\bibitem{Protopopescu:1973sh}
S.~D. Protopopescu, M.~Alston-Garnjost, A.~Barbaro-Galtieri, S.~M. Flatte,
  J.~H. Friedman, T.~A. Lasinski et~al., \emph{{Pi pi Partial Wave Analysis
  from Reactions pi+ p ---\ensuremath{>} pi+ pi- Delta++ and pi+ p
  ---\ensuremath{>} K+ K- Delta++ at 7.1-GeV/c}},
  \href{https://doi.org/10.1103/PhysRevD.7.1279}{\emph{Phys. Rev. D} {\bfseries
  7} (1973) 1279}.

\bibitem{data}
{\scshape ETM} collaboration, \emph{Data repository,
  https://github.com/urbach/datarhonf2}, .

\bibitem{James:1994vla}
F.~James, \emph{{MINUIT function minimization and error analysis:
  referencemanual version 94.1}}, .

\bibitem{Filin:2019eoe}
A.~A. Filin, V.~Baru, E.~Epelbaum, H.~Krebs, D.~M{\"o}ller and P.~Reinert,
  \emph{{Extraction of the neutron charge radius from a precision calculation
  of the deuteron structure radius}},
  \href{https://doi.org/10.1103/PhysRevLett.124.082501}{\emph{Phys. Rev. Lett.}
  {\bfseries 124} (2020) 082501}
  [\href{https://arxiv.org/abs/1911.04877}{{\ttfamily 1911.04877}}].

\bibitem{Filin:2020tcs}
A.~A. Filin, D.~M{\"o}ller, V.~Baru, E.~Epelbaum, H.~Krebs and P.~Reinert,
  \emph{{High-accuracy calculation of the deuteron charge and quadrupole form
  factors in chiral effective field theory}},
  \href{https://doi.org/10.1103/PhysRevC.103.024313}{\emph{Phys. Rev. C}
  {\bfseries 103} (2021) 024313}
  [\href{https://arxiv.org/abs/2009.08911}{{\ttfamily 2009.08911}}].

\bibitem{NPLQCD:2011naw}
{\scshape NPLQCD} collaboration, \emph{{The Deuteron and Exotic Two-Body Bound
  States from Lattice QCD}},
  \href{https://doi.org/10.1103/PhysRevD.85.054511}{\emph{Phys. Rev. D}
  {\bfseries 85} (2012) 054511}
  [\href{https://arxiv.org/abs/1109.2889}{{\ttfamily 1109.2889}}].

\bibitem{Inoue:2011ai}
{\scshape HAL QCD} collaboration, \emph{{Two-Baryon Potentials and H-Dibaryon
  from 3-flavor Lattice QCD Simulations}},
  \href{https://doi.org/10.1016/j.nuclphysa.2012.02.008}{\emph{Nucl. Phys. A}
  {\bfseries 881} (2012) 28} [\href{https://arxiv.org/abs/1112.5926}{{\ttfamily
  1112.5926}}].

\bibitem{Yamazaki:2012hi}
T.~Yamazaki, K.-i. Ishikawa, Y.~Kuramashi and A.~Ukawa, \emph{{Helium nuclei,
  deuteron and dineutron in 2+1 flavor lattice QCD}},
  \href{https://doi.org/10.1103/PhysRevD.86.074514}{\emph{Phys. Rev. D}
  {\bfseries 86} (2012) 074514}
  [\href{https://arxiv.org/abs/1207.4277}{{\ttfamily 1207.4277}}].

\bibitem{Yamazaki:2015asa}
T.~Yamazaki, K.-i. Ishikawa, Y.~Kuramashi and A.~Ukawa, \emph{{Study of quark
  mass dependence of binding energy for light nuclei in 2+1 flavor lattice
  QCD}}, \href{https://doi.org/10.1103/PhysRevD.92.014501}{\emph{Phys. Rev. D}
  {\bfseries 92} (2015) 014501}
  [\href{https://arxiv.org/abs/1502.04182}{{\ttfamily 1502.04182}}].

\bibitem{Orginos:2015aya}
K.~Orginos, A.~Parreno, M.~J. Savage, S.~R. Beane, E.~Chang and W.~Detmold,
  \emph{{Two nucleon systems at $m_\pi\sim 450~{\rm MeV}$ from lattice QCD}},
  \href{https://doi.org/10.1103/PhysRevD.92.114512}{\emph{Phys. Rev. D}
  {\bfseries 92} (2015) 114512}
  [\href{https://arxiv.org/abs/1508.07583}{{\ttfamily 1508.07583}}].

\bibitem{Horz:2020zvv}
B.~H{\"o}rz et~al., \emph{{Two-nucleon S-wave interactions at the $SU(3)$
  flavor-symmetric point with $m_{ud}\simeq m_s^{\rm phys}$: A first lattice
  QCD calculation with the stochastic Laplacian Heaviside method}},
  \href{https://doi.org/10.1103/PhysRevC.103.014003}{\emph{Phys. Rev. C}
  {\bfseries 103} (2021) 014003}
  [\href{https://arxiv.org/abs/2009.11825}{{\ttfamily 2009.11825}}].

\bibitem{Davoudi:2020ngi}
Z.~Davoudi, W.~Detmold, K.~Orginos, A.~Parre\~no, M.~J. Savage, P.~Shanahan
  et~al., \emph{{Nuclear matrix elements from lattice QCD for electroweak and
  beyond-Standard-Model processes}},
  \href{https://doi.org/10.1016/j.physrep.2020.10.004}{\emph{Phys. Rept.}
  {\bfseries 900} (2021) 1} [\href{https://arxiv.org/abs/2008.11160}{{\ttfamily
  2008.11160}}].

\bibitem{NPLQCD:2012mex}
{\scshape NPLQCD} collaboration, \emph{{Light Nuclei and Hypernuclei from
  Quantum Chromodynamics in the Limit of SU(3) Flavor Symmetry}},
  \href{https://doi.org/10.1103/PhysRevD.87.034506}{\emph{Phys. Rev. D}
  {\bfseries 87} (2013) 034506}
  [\href{https://arxiv.org/abs/1206.5219}{{\ttfamily 1206.5219}}].

\bibitem{Barnea:2013uqa}
N.~Barnea, L.~Contessi, D.~Gazit, F.~Pederiva and U.~van Kolck,
  \emph{{Effective Field Theory for Lattice Nuclei}},
  \href{https://doi.org/10.1103/PhysRevLett.114.052501}{\emph{Phys. Rev. Lett.}
  {\bfseries 114} (2015) 052501}
  [\href{https://arxiv.org/abs/1311.4966}{{\ttfamily 1311.4966}}].

\bibitem{Drischler:2019xuo}
C.~Drischler, W.~Haxton, K.~McElvain, E.~Mereghetti, A.~Nicholson, P.~Vranas
  et~al., \emph{{Towards grounding nuclear physics in QCD}},
  \href{https://doi.org/10.1016/j.ppnp.2021.103888}{\emph{Prog. Part. Nucl.
  Phys.} {\bfseries in press} (2021) }
  [\href{https://arxiv.org/abs/1910.07961}{{\ttfamily 1910.07961}}].

\end{thebibliography}\endgroup
\end{document}